%% file: main.tex
\PassOptionsToPackage{table}{xcolor}
\documentclass[acmsmall]{acmart}
\setcopyright{none}
\usepackage{booktabs}   %% For formal tables:
\usepackage{subcaption} %% For complex figures with subfigures/subcaptions
\usepackage{stackengine}
\usepackage{lipsum}
\usepackage{tabularx}
\usepackage{color, colortbl}
\usepackage{xcolor}
\usepackage{listings}
\usepackage[inline]{enumitem}
\usepackage{mathpartir}
\usepackage{mathtools}
\usepackage{amssymb}
\usepackage{pifont}
\usepackage{hyperref}
\usepackage[noend]{algpseudocode}
\usepackage{makecell}
\usepackage{amsfonts}
\usepackage{accents}
\usepackage{mdframed}
\usepackage{verbatim}
\usepackage[linesnumbered,noend]{algorithm2e}
\usepackage{wrapfig}
\usepackage{tikz-cd}
\usepackage{booktabs}
\usepackage{longtable}
\usepackage{makecell}
\usepackage{cellspace}
\usepackage{tcolorbox}
\usepackage{multirow}
\DeclareMathSizes{8}{8}{5}{4}

\SetKwRepeat{Do}{do}{while}
\begin{document}

%% File conatining newly defined commands
\input{commands}

%% Title information
\title[\tool]{\tool: \papername}
                                        %% [Short Title] is optional;
                                        %% when present, will be used in
                                        %% header instead of Full Title.
%\titlenote{Sisyphus: (n) Greek legend: a king in ancient Greece who offended Zeus and whose punishment was to roll a huge boulder to the top of a steep hill; each time the boulder neared the top it rolled back down and Sisyphus was forced to start again}            %% \titlenote is optional;
                                        %% can be repeated if necessary;
                                        %% contents suppressed with 'anonymous'
%\subtitle{Subtitle}                    %% \subtitle is optional
%\subtitlenote{with subtitle note}      %% \subtitlenote is optional;
                                        %% can be repeated if necessary;
                                        %% contents suppressed with 'anonymous'

%% Author information
%% Contents and number of authors suppressed with 'anonymous'.
%% Each author should be introduced by \author, followed by
%% \authornote (optional), \orcid (optional), \affiliation, and
%% \email.
%% An author may have multiple affiliations and/or emails; repeat the
%% appropriate command.
%% Many elements are not rendered, but should be provided for metadata
%% extraction tools.

% 1
%% Author with single affiliation.
\author{Kia Rahmani}
%% \authornote{with author1 note}          %% \authornote is optional;
\orcid{0000-0001-9064-0797}
\affiliation{
  \position{Research Assistant}
%  \department{Department of Computer Science}              %% \department is recommended
  \institution{Purdue University}            %% \institution is required
%  \streetaddress{305 N. University Street}
%  \city{West Lafayette}
%  \state{Indiana}
%  \postcode{47906}
%  \country{USA}                    %% \country is recommended
}
\email{rahmank@purdue.edu}          %% \email is recommended

% 2
\author{Kartik Nagar}
\affiliation{
  \position{Post-doctoral Research Associate}
%  \department{Department of Computer Science}             %% \department is recommended
  \institution{Purdue University}           %% \institution is required
 % \streetaddress{305 N. University Street}
 % \city{West Lafayette}
 % \state{Indiana}
 % \postcode{47906}
  \country{USA}                    %% \country is recommended
}
\email{nagark@purdue.edu}         %% \email is recommended

% 3
\author{Benjamin Delaware}
\affiliation{
  \position{Assistant Professor}
%  \department{Department of Computer Science}             %% \department is recommended
  \institution{Purdue University}           %% \institution is required
 % \streetaddress{305 N. University Street}
 % \city{West Lafayette}
 % \state{Indiana}
 % \postcode{47906}
  \country{USA}                    %% \country is recommended
}
\email{bendy@purdue.edu}         %% \email is recommended

% 4
\author{Suresh Jagannathan}
\affiliation{
  \position{Samuel D. Conte Professor}
%  \department{Department of Computer Science}             %% \department is recommended
  \institution{Purdue University}           %% \institution is required
 % \streetaddress{305 N. University Street}
 % \city{West Lafayette}
 % \state{Indiana}
 % \postcode{47906}
  \country{USA}                    %% \country is recommended
}
\email{suresh@cs.purdue.edu}         %% \email is recommended

%% Abstract
%% Note: \begin{abstract}...\end{abstract} environment must come
%% before \maketitle command
\begin{abstract}
\input{sections/abstract}
\end{abstract}

%% 2012 ACM Computing Classification System (CSS) concepts
%% Generate at 'http://dl.acm.org/ccs/ccs.cfm'.

% default
%\begin{CCSXML}
%<ccs2012>
%<concept>
%<concept_id>10011007.10011006.10011008</concept_id>
%<concept_desc>Software and its engineering~General programming languages</concept_desc>
%<concept_significance>500</concept_significance>
%</concept>
%<concept>
%<concept_id>10003456.10003457.10003521.10003525</concept_id>
%<concept_desc>Social and professional topics~History of programming languages</concept_desc>
%<concept_significance>300</concept_significance>
%</concept>
%</ccs2012>
%\end{CCSXML}

%\ccsdesc[500]{Software and its engineering~General programming languages}
%\ccsdesc[300]{Social and professional topics~History of programming languages}
% end of default

%% End of generated code

%% Keywords
%% comma separated list
  %% \keywords are mandatory in final camera-ready submission

%% \maketitle
%% Note: \maketitle command must come after title commands, author
%% commands, abstract environment, Computing Classification System
%% environment and commands, and keywords command.
\maketitle

%%%%%%%%%%%%%%%%%%%%%%%%%%%%%%%%%%%%%%%%%%%%%%%%%%%%%%
\section{Introduction}
\label{sec:intro}
\input{sections/intro}

\input{sections/overview}
\section{Abstract Model}
\label{sec:model}
\input{sections/model}
\input{sections/ser}

\section{Static Anomaly Generation}
\label{sec:encoding}
\input{sections/encoding}
%
%
%\newpage
\section{Implementation}
\label{sec:impl}
\input{sections/impl}
%
%
%\newpage
\section{Evaluation}
\label{sec:eval}
\input{sections/eval}

\subsection{Comparison with Random Testing}
\label{sec:blackbox}
\input{sections/blackbox}
\section{Related Work}
\label{sec:related}
\input{sections/related}

\section{
Conclusions
  and 
  Future Works 
}
\label{sec:conclusion}
\input{sections/conclusion}

%%%%%%%%%%%%%%%%%%%%%%%%%%%%%%%%%%%%%%%%%%%%%%%%%%%%%%

%% %% Acknowledgments
%% \begin{acks}                            %% acks environment is optional
%%                                         %% contents suppressed with 'anonymous'
%%   %% Commands \grantsponsor{<sponsorID>}{<name>}{<url>} and
%%   %% \grantnum[<url>]{<sponsorID>}{<number>} should be used to
%%   %% acknowledge financial support and will be used by metadata
%%   %% extraction tools.
%%   This material is based upon work supported by the
%%   \grantsponsor{GS100000001}{National Science
%%     Foundation}{http://dx.doi.org/10.13039/100000001} under Grant
%%   No.~\grantnum{GS100000001}{nnnnnnn} and Grant
%%   No.~\grantnum{GS100000001}{mmmmmmm}.  Any opinions, findings, and
%%   conclusions or recommendations expressed in this material are those
%%   of the author and do not necessarily reflect the views of the
%%   National Science Foundation.
%% \end{acks}

%% Bibliography
\bibliography{bib}

%% Appendix
%\newpage
%\appendix
%\section{Full Operational semantics}
%\label{app:semantics}
%\input{sections/full_operational_semantics}
%\label{app:encoding}
%\input{sections/full_fol_encoding}

\end{document}

%% file: commands.tex
%% Paper Specific Commands
\newcommand{\tool}[0]{\textsc{clotho}}        % name of the tool
\newcommand{\papername}[0]{
Directed Test Generation for Weakly Consistent Database Systems}
%  Bug Detection and Test Administration in Cloud Database Applications}                          % title of the paper

%%  Commenats
\definecolor{DOrange}{RGB}{145,20,20}
\definecolor{DPurple}{RGB}{79,21,95}
\newcommand\BD[1]{\textcolor{DOrange}{[\texttt{BD: #1}]}} % Ben's Comments
\newcommand\KR[1]{\textcolor{DOrange}{[KR: #1]}} % Kia's Comments
%\newcommand\SJ[1]{\textcolor{DOrange}{[SJ: #1]}} % Suresh's Comments

%% OPERATIONAL SEMANTICS
\newcommand{\rc}[0]{\Psi_\textsc{rc}}
\newcommand{\rr}[0]{\Psi_\textsc{rr}}
\newcommand{\lin}[0]{\Psi_\textsc{lin}}
\newcommand{\ser}[0]{\Psi_\textsc{ser}}
\newcommand{\folphi}[1]{\varphi_{\text{\textsc{#1}}}}
\newcommand{\cv}[0]{\Psi_\textsc{cv}}
\newcommand{\cc}[0]{\Psi_\textsc{cc}}

\newcommand{\rec}[0]{\mathfrak{r}}

\newcommand{\rulelabelinlinec}[1]{\fcolorbox{black}{Main-Theme-Complement}{\textrm{\sc{\color{black} #1}}}
}

\newcommand{\rulelabel}[1]{
  \vspace{4mm}
  \begin{flushleft}
    \hspace{15mm}\noindent\fcolorbox{black}{Main-Theme-Lighter}{\textrm{\sc{\color{black} #1}}}
  \end{flushleft}
}
\newcommand{\RULE}[2]{
\footnotesize
\frac{\begin{array}{c}#1\end{array}}
     {\begin{array}{c}#2\end{array}}
}

\newcommand{\TEXTRULE}[2]{\footnotesize\frac{\begin{array}{c}#1\end{array}}{#2}}
\newcommand{\FOLRULE}[3]{
  \footnotesize
  \frac{\begin{array}{c}#1\end{array}}{\let\scriptstyle\textstyle
  \substack{#2\\#3}}}
\newcommand{\step}[3]{#1\xrightarrow{#2} #3}
\newcommand{\hstep}[3]{#1\hookrightarrow #3}
% functions
\newcommand{\F}[0]{\mathcal{F}}    % function f used in full operational semantics
\newcommand{\B}[2]{#2, #1 \Downarrow } 
\newcommand{\folB}[2]{\stx{[\![#1]\!]^\mathbb{B}_{#2}}}
\newcommand{\I}[2]{#2, #1 \Downarrow } 
\newcommand{\folI}[2]{\stx{[\![#1]\!]^\mathbb{Z}_{#2}}}
\newcommand{\R}[1]{\stx{[\![#1]\!]^\mathbb{R}_{\theta}}}
% notations
\newcommand{\vis}[0]{\mathtt{vis}}
\newcommand{\store}[0]{\mathtt{str}}
\newcommand{\st}[0]{\mathtt{st}}
\newcommand{\ar} [0]{\mathtt{ar}}
\newcommand{\txnR} [0]{\mathtt{txn}}
\newcommand{\cmdR} [0]{\mathtt{cmd}}
\newcommand{\wrr}[1]{\textcolor{purple}{\mathsf{wr}#1}}
\newcommand{\wrrp}[1]{\textcolor{purple}{\mathsf{wr^+}#1}}
\newcommand{\inn}[1]{\textcolor{purple}{\mathsf{in}#1}}
\newcommand{\rdd}[1]{\textcolor{purple}{\mathsf{rd}#1}}
\newcommand{\rddp}[1]{\textcolor{purple}{\mathsf{rd^+}#1}}
\newcommand{\pk} [0]{\mathsf{PK}}
\newcommand{\MAX} [0]{\mathtt{MAX}}
\newcommand{\txninstance}[0]{\mathcal{T}}
\newcommand{\txntype}[0]{\mathbb{T}}
\newcommand{\set}[1]{\mathcal{P}(#1)}

\newcommand{\fol}[1]{#1^{\triangleright}}
\newcommand{\folversionsub}[2]{\mathtt{ver}_{#2}^{\mathtt{#1}}}

\newcommand{\rulelabelc}[1]{
  \vspace{3mm}
  \begin{flushleft}
    \hspace{15mm}\noindent\fcolorbox{black}{Main-Theme-Complement}{\textrm{\sc{\color{black} #1}}}
  \end{flushleft}
}

%% FOL ENCODING
\newcommand{\txnsort}[0]{\mathfrak{t}}              % uninterpreted sort txns for encoding
\newcommand{\opsort}[0]{\mathfrak{q}}               % uninterpreted sort operations  for encoding
\newcommand{\partsort}[0]{\mathfrak{p}}               % uninterpreted sort operations  for encoding
\newcommand{\rowsort}[0]{\mathfrak{r}}              % uninterpreted sort operations  for encoding
\newcommand{\varproj}[0]{\rho}                    % variable projection function
\newcommand{\folww}[0]{\mathtt{WW}}
\newcommand{\folD}[0]{\mathtt{D}}
\newcommand{\folst}[0]{\mathtt{ST}}
\newcommand{\folstp}[0]{\mathtt{ST^+}}
\newcommand{\folX}[0]{\mathtt{X}}
\newcommand{\folSib}[0]{\mathtt{sibling}}
\newcommand{\folwr}[0]{\mathtt{WR}}
\newcommand{\folrw}[0]{\mathtt{RW}}
\newcommand{\folR}[0]{\mathtt{R}}
\newcommand{\folvis}[0]{\mathtt{vis}}
\newcommand{\folar} [0]{\mathtt{ar}}
\newcommand{\folalive} [0]{\mathtt{Alive}}
\newcommand{\foltime}[0]{\mathtt{time}}
\newcommand{\folval}[0]{\mathtt{val}}
\newcommand{\folTxn}[0]{\mathtt{txn}}
\newcommand{\folinit}[0]{\mathtt{init}}
\newcommand{\folversion}[1]{\mathtt{ver^{#1}}}
\newcommand{\folparam}[1]{\mathtt{param^{#1}}}
\newcommand{\folproj}[1]{\mathtt{proj^{#1}}}
\newcommand{\folabs}[0]{\mathtt{abs}}
\newcommand{\folconf}[1]{\mathtt{conf^{#1}}}
\newcommand{\folpart}[0]{\mathtt{part}}
\newcommand{\folparent}[0]{\mathtt{txn}}
\newcommand{\denote}[3]{[\![#1]\!]^{#2}_{#3}}
\newcommand{\V}[2]{\stx{[\![#1]\!]^\mathbb{V}_{#2}}}
\newcommand{\folstep}[3]{#1\rightsquigarrow{#2} #3}
\newcommand{\folpath}[1]{\phi^{p}_{#1}}
\newcommand{\foldval}[0]{\psi}
\newcommand{\foldcond}[0] {\varphi}

%% SYNTAX
\newcommand{\inv}[1]{{\boldmath$\mathsf{#1}$}}
\newcommand{\ir}[0]{$\mathcal{AR}$} % Intermediate abstract representation
\newcommand{\stx}[1]{{#1}}   % generic syntax
\newcommand{\fld}[1]{\mathtt{#1}}   % fields
\newcommand{\sql}[1]{\textcolor{Sql-color}{\uppercase{\mathtt{#1}}\;}}
\newcommand{\intt}[0]{\mathbb{Z}}   % integers
\newcommand{\bvv}[0]{\mathbb{BV}}   % bitvectors
\newcommand{\natt}[0]{\mathbb{N}}   % natural numbers
\newcommand{\booll}[0]{\mathbb{B}}   % natural numbers
\newcommand{\ALT}{~\mid~}

%% SQL APPLICATION REPRESENTATION
\newcommand{\txn}[1]{\texttt{#1}}
\newcommand{\tabl}[1]{\textsc{#1}}
\newcommand{\ms}[0]{\tiny{ms}}               %tiny milliseconds
\newcommand{\crr}[1]{\textsc{cr\Small{#1}}}  %consistency requirement
\newcommand{\ncr}[1]{\textsc{ncr\Small{#1}}} %new consistency requirement
\newcommand{\passed}[0]{\color{weborange}\ding{51}\color{black}}
\newcommand{\failed}[0]{\color{red}\ding{53}\color{black}}

%% COLORS
\definecolor{dark-red}{HTML}{000000}
\definecolor{very-dark-red}{HTML}{000000}
\definecolor{dark-yellow}{HTML}{000000}
\definecolor{light-gray}{HTML}{FAFAFA}
\definecolor{dark-green}{HTML}{000000}
\definecolor{very-dark-green}{HTML}{000000}
\definecolor{dark-blue}{HTML}{154360}

\definecolor{Main-Theme-Complement}{HTML}{F0F3FF}
\definecolor{Main-Theme-Complement-Dark}{HTML}{D7DFFF}
\definecolor{Main-Theme-Triadic}   {HTML}{FFF7F9}
\definecolor{Main-Theme-Lightest}  {HTML}{FFFCF0}
\definecolor{Main-Theme-Lighter}   {HTML}{FFF8D6}
\definecolor{Main-Theme-Light}     {HTML}{F4EBBB}
\definecolor{Main-Theme-Dark}      {HTML}{FFDD96}
\definecolor{Main-Theme-Darker}    {HTML}{D95F0E}
\definecolor{purple}    {HTML}{330066}
\definecolor{yelloww}    {HTML}{7A4300}
\definecolor{Main-Background}      {rgb}{0.95,0.95,0.95 }
\definecolor{Sql-color}{RGB}{0, 0, 120}

%% MISC
\newcommand{\aws}[1]{\textup{\color{blue}#1}}   % aws ec2 instance types
\newcommand{\rulesep}{\unskip\ \vrule width 0.1pt\ }
\newcommand*\circledA[0]{\raisebox{0.9pt}[0pt][0pt]{\tikz[baseline=(char.base)]{
\node[shape=circle,draw,fill=black,text=white,text depth = 0.2 mm,inner
sep=0.8pt] (char) {\small a};}}}
\newcommand*\circledB[0]{\raisebox{0.1pt}[0pt][0pt]{\tikz[baseline=(char.base)]{
\node[shape=circle,draw,fill=black,text=white,text depth = 0.2 mm,inner
sep=0.2pt] (char) {\small b};}}}

%% file: sections/abstract.tex
Relational database applications are notoriously difficult to test and
debug.  Concurrent execution of database transactions may violate
complex structural invariants that constraint how changes to the
contents of one (shared) table affect the contents of
another. Simplifying the underlying concurrency model is one way to
ameliorate the difficulty of understanding how concurrent accesses and
updates can affect database state with respect to these sophisticated
properties.  Enforcing serializable execution of all
transactions achieves this simplification, but it comes at a
significant price in performance, especially at scale, where database
state is often replicated to improve latency and availability.

To address these challenges, this paper presents a novel testing
framework for detecting serializability violations in (SQL)
database-backed Java applications executing on weakly-consistent
storage systems.  We manifest our approach in a tool named \tool, that
combines a static analyzer and a model checker to generate abstract
executions, discover serializability violations in these executions,
and translate them back into concrete test inputs suitable for
deployment in a test environment.  To the best of our knowledge,
\tool\, is the first automated test generation facility for
identifying serializability anomalies of Java applications intended to
operate in geo-replicated distributed environments. An experimental
evaluation on a set of industry-standard benchmarks demonstrates the
utility of our approach.

%% file: sections/intro.tex
%% 1
%% Intro to testing frameworks
%%

Realistic SQL-based databases typically have sophisticated structural
relationships (schemas), and clients must ensure that every possible
use of a database operation preserves these relationships. Testing
that client applications do so is particularly challenging, as the
operations they perform depends on the control structure and initial
state of the application. Generating appropriate test inputs to
discover violations of these relationships necessarily requires
reasoning about the behaviors of these operations in the context of
the program's execution. Further complicating matters is that the
initial state of the database itself needs to be chosen so that 
tests expose interesting integrity
and assertion violations.

One important simplification that helps
reduce the complexity of developing a useful testing strategy for
database applications is to treat database transactions
as \emph{serializable}, restricting the set of interleavings that must
be considered to those that maintain transaction atomicity and
isolation.

As an example, consider the transactions
\texttt{txnWrite} and \texttt{txnRead} from the simple pseudocode application
presented in \autoref{fig:intro_example}.  Transaction \texttt{txnWrite} updates rows
from \texttt{table1} and \texttt{table2} with the value \texttt{val}
where both rows are accessed using the given \texttt{id}.  Similarly,
transaction \texttt{txnRead} reads rows from both tables that are
associated with the given \texttt{id} and asserts that they are equal.
This is a very common pattern in database applications, e.g. when
backup versions of tables are maintained.

\input{figures/intro_example}

Under the guarantees
afforded by serializability, an automated testing framework need only
consider interleavings of entire transactions in order to completely
cover the state of possible valid concurrent executions.  This is
because interleavings that expose intermediate transaction states
(failure of \emph{atomicity}), or which allow an already executing
transaction to observe the effects of other completed transactions
(failure of \emph{isolation}) are prohibited, and hence the space of executions that
must be considered to validate an application's correctness is greatly
reduced.  
While conceptually simple to reason about, maintaining
serializability imposes strong constraints on the underlying database
and storage system that hinders performance at scale.  In large
distributed environments, especially those use replication to improve
latency, the cost of enforcing serializability can be prohibitive,
requiring global synchronization to enforce atomic updates and
maintain a single global view of database state.  Consequently,
modern-day cloud systems often implement and support storage
abstractions that provide significantly weaker consistency and
isolation guarantees than those necessary to enforce serializable
executions of database programs.

For example, when the transactions in the above example are executed
on a replicated database systems offering eventual
consistency (EC), each individual read and write operation
can be executed independently making the space of possible executions
significantly larger than before.  Not surprisingly, new anomalous
executions are also introduced under such weak semantics. These
executions can trigger assertion violations that would otherwise not
happen, for example, if both read operations occur between the two
writes and vice versa.  To make matters worse, these violations are
only triggered if the initial state of the database for both rows
differs from what the \texttt{txnWrite} transaction is writing.  Given
these complexities, the obvious question that arises is whether a
database application, originally written assuming the underlying
storage system provides strong consistency and isolation guarantees,
will still operate correctly (i.e., behave as though serializability
were still maintained) in weaker environments where these guarantees
are no longer assured.

Given these concerns, the need for a practical automated testing
framework for database applications executing in weakly consistent
environments becomes is particularly acute.  This paper takes an
important first step towards addressing this need by presenting a tool
(\tool) that systematically and efficiently explores abstract
executions of database-backed Java programs for serializability
violations, assuming a weakly-consistent storage abstraction.  To do
so, we employ a bounded model-checking serializability violation
detector  that is parameterized by a specification of the
underlying storage model.  Abstract executions capture various
visibility and ordering relations among read and write operations
on the database generated by queries; the structure of these
relations is informed by the underlying data consistency model.
Potential serializability violations in an abstract execution manifest
as cycles in a dependency graph induced by these relations.  When such
violations are discovered, \tool\ synthesizes concrete tests that can
be used to drive executions that manifest the problem for assessment
by application developers.

Through our experimental evaluation, we demonstrate that \tool\, can
efficiently, automatically, and reliably detect and concretely
manifest serializability anomalies of database programs executing on
top of weakly-consistent data stores.  An important by-product of its
design is that \tool\, does not bake-in any specific consistency or
isolation assumptions about the underlying storage infrastructure,
allowing users to strengthen visibility and ordering constraints as
desired.

The contributions of this paper are as follows:
\begin{enumerate}
\item We present an abstract representation and a novel encoding of
  database applications capable of modeling a diverse range of weak
  guarantees offered by real-world distributed databases.
\item We develop a search algorithm capable of effectively identifying
  unique serializability anomalies accompanied by a complete execution
  history that leads to that anomaly.
\item We bridge the gap between abstract and concrete executions by
  developing a front-end compiler and a test administration
  environment for database-backed Java applications that admits a wide range distributed
  database system behaviors and features.
\item We demonstrate the utility of our approach on a set of standard
  benchmarks and present a comprehensive empirical study comparing
  \tool's effectiveness against a tuned dynamic (random) testing
  approach.
\end{enumerate}

The remaining of this paper is organized as follows.  
A brief overview of our solution using a motivational example is presented in 
Section~\ref{sec:overview}.  
Sections~\ref{sec:model} and~\ref{sec:ser}
respectively define the operational semantics of our abstract
programming model and the notion anomalies in that model.
Section~\ref{sec:encoding}
presents our precise encoding of anomalies in database 
applications.  
Section~\ref{sec:impl} introduces our effective search algorithm and the 
implementation details of \tool.  
We present two sets of empirical results in
Section~\ref{sec:eval} showing applicability and effectiveness of our
approach and its performance compared to a state of the art dynamic
testing framework.  Lastly we summarize related works in
section~\ref{sec:related} before concluding the paper in Section~\ref{sec:conclusion}.

%% file: figures/intro_example.tex
%%-----------------
% settings

\definecolor{pgrey}{rgb}{0.26,0.25,0.28}
\definecolor{background}{rgb}{0.92,0.92,0.92 }
\definecolor{javared}{rgb}{0.2,0.2,0.7} % for strings
\definecolor{javagreen}{rgb}{0.2,0.45,0.3} % comments
\definecolor{javapurple}{rgb}{0.5,0,0.35} % keywords
\definecolor{javadocblue}{rgb}{0.25,0.35,0.75} % javadoc
\definecolor{weborange}{RGB}{0,75,0}
\definecolor{annotateColor}{RGB}{175,25,20}

\lstset{
  %language=C,
  basicstyle=\ttfamily\scriptsize,
  breaklines=true,
  backgroundcolor=\color{Main-Background},
  emph={read, if, write, assert},
  emphstyle={\bfseries},
  frame=single,
  %rulecolor=\color{pgrey},
  %keywordstyle=\color{javapurple}\bfseries,
  %stringstyle=\color{javared},
  %commentstyle=\color{javagreen},
  %morecomment=[s][\color{javadocblue}]{/**}{*/},
  numbers=none,
  %title=\footnotesize{A1\_Ins1.java},
  %xleftmargin=1.9em,
  %framexleftmargin=1.25em,
  tabsize=1,
  showspaces=false,
  showstringspaces=false,
  captionpos=b,
  abovecaptionskip=-0pt,
  %classoffset=2, % starting new class
  %morekeywords={stmt,rs},
  %keywordstyle=\color{weborange},
  %moredelim=[is][\textcolor{red}]{\%\%}{\%\%},
}

%%-----------------
% code figure
\begin{wrapfigure}{r}{0.48\textwidth}  
%\begin{minipage}{0.5\textwidth}
%
  \vspace{-1mm}
  \centering
\begin{center}
  \begin{subfigure}[b]{0.23\textwidth}
  \begin{lstlisting}[title=\scriptsize{(a) two writes}]
txnWrite(id,val):
 write(table1,id,val)
 write(table2,id,val)
  \end{lstlisting}
\end{subfigure}
\hfill
\begin{subfigure}[b]{0.21\textwidth}
  \begin{lstlisting}[title=\scriptsize{(b) two reads}]
txnRead(id):
 v1=read(table1,id)
 v2=read(table2,id)
 assert:(v1=v2)
\end{lstlisting}
\end{subfigure}
\end{center}
\vspace{-6mm}
\caption{}
\label{fig:intro_example}
\vspace{-4mm}
%\end{minipage}
\end{wrapfigure}

%% file: sections/overview.tex
\vspace{4mm}
\input{figures/motivating_example}
\vspace{-1mm}

\section{Overview}
\label{sec:overview}

%%
%%
% introduce TPC-C
%%%%
TPC-C is a canonical Online Transaction Processing (OLTP) benchmark
application that emulates a warehouse management and order
entry/delivery system. The benchmark was originally designed to
validate a mixture of requirements (including transactional safety)
expected from relational database systems~\cite{tpcc}.
To illustrate our approach, consider
\mbox{\autoref{fig:motivating} (left)} that presents a pruned code snippet implementing
a transaction from TPC-C named $\txn{payment}$.
The snippet shows a procedure that updates the total number of successful
payments from a customer, where
the customer's associated row from $\mathsf{CUST}$ table is initially retrieved
\mbox{(lines 3-5)} and is then rewritten with an incremented value~\mbox{(lines 10-13)}.
%
%%
%%
% introduce an anomaly and what it depends on
%%%%
Now, consider the deployment of TPC-C on a replicated cloud database
presented in \mbox{\autoref{fig:motivating} (right)} where two $\txn{payment}$
transactions are concurrently executed
on weakly-consistent replicas \texttt{A} and \texttt{B} that do not enforce
serializability.
The initial database state in this execution consists
of the table $\mathsf{CUST(id,c\_pay\_cnt)}$
with a single row $\mathsf{r_0\!:=\!(10,50)}$.
Additionally, the transaction instances  {\sf txn1} and {\sf txn2} are
given arguments such that both transactions internally access and update
$\mathsf{r_0}$.
Both transactions concurrently read the initial value of $\mathsf{r_0}$
at different replicas (depicted as incoming labeled arrows to the transaction)
and then both locally replace it  with updated values $\mathsf{r_0^{A}\!:=\!(10,51)}$ and
$\mathsf{r_0^{B}\!:=\!(10,51)}$ (depicted by outgoing labeled arrows).
Each replica then propagates the locally updated row to
the other replica (depicted by dashed arrows), where the
written value at \texttt{B} supersedes value written at
\texttt{A}, leaving the database with the final
state~of~$\mathsf{r_0\!:=\!(10,51)}$.
Because the above scenario is not equivalent to any sequential
execution of the two transaction instances, this behavior is properly
classified as a \emph{serializability anomaly}.
%
%Serializability anomalies are effective pathways to detect
%concurrency bugs in database applications[+elaborate]
%
Unfortunately, buggy executions like this are often permitted by
modern cloud databases which drop support for atomic and isolated
transactions in favor of fault tolerance, scalability, and
availability.
This foists a significant challenge onto OLTP application developers,
requiring them to detect the potential anomalies that can occur in a
particular application and validate them against different database
systems in order to diagnose and remedy undesired behaviors.

% random testing and its challenges
%%%
Developing a testing framework that discovers anomalies of this kind
is challenging, however.  Part of the problem is due to the large set
of possible interleaved executions that are possible.  For example,
the serializability anomaly described above is crucially dependent on
the exact order in which operations are transmitted to different
replicas and occurs only when the same customer (out of 30000
customers) is accessed by both instances.  Combined with the
possibility that operations within a single transaction instance can
also be unreachable in some program paths, and that the execution
paths taken are highly dependent on the initial database state, the
chance of randomly encountering conflicts of this kind becomes even
smaller.

In addition to the above challenges, the high computational cost of
detecting serializability anomalies at
runtime\cite{Papadimitriou:1979:Ser} means that practical testing
methods often leverage a set of user-provided application-level
invariants and assertions to check for serializability failure-induced
violations.  But, these assertions are often underspecified.  For
example, the anomaly depicted in \autoref{fig:motivating}, which
most developers would classify as a bug, does not directly violate any of the
twelve officially specified invariants of the TPC-C benchmark.

\input{figures/pipeline}
\subsection*{\tool}

To overcome these challenges, we have developed a
principled test construction and administration framework called
\tool\ that targets database applications intended to be deployed
on weakly-consistent storage environments.
Our approach reasons over \emph{abstract executions} of a database
application.  Serializability anomalies manifest as cycles in such
executions.  \tool\ translates the discovered abstract anomalies
back to concrete executions which can be then automatically replayed and validated.

%
%% \tool\ is backed by a powerful encoding of database applications and serializability
%% anomalies which is free of false positives by construction and naturally
%% provides it with specific details that can be incorporated to automatically map
%% abstract anomalies back to concrete executions.
%
The test configuration details produced by \tool\ include
\begin{enumerate*}[label=(\roman*)]
\item concrete transaction instances and parameters,
\item concurrent execution schedules,
\item network partitionings throughout the execution,
\item session management, and
\item initial database state.
\end{enumerate*}
Our test administration framework allows arbitrary distributed or centralized
databases to be plugged into its managed containers.
Given a set of test configurations, \tool\, automatically executes multiple application instances
according to each test configuration, in order to effectively manifest and validate the intended
serializability anomaly against a variety of actual database systems.

%%
%%
% briefly introduce the encoding
%%%
\autoref{fig:pipeline} presents an overview of \tool's design.
\tool's  static analysis backend is based on an intermediate abstract
representation of database programs,
automatically generated from input Java source code via a front-end compiler.
The abstract program representation is passed to an encoding engine (\ding{202})
that constructs FOL formulae that capture the necessary conditions under which a
dependency cycle forms over instances of database operations.
This fine-grained SAT representation of the problem is then passed to an off-the-shelf
theorem prover  (\ding{203}); responses with a satisfying solution (\ding{204})
are recorded and re-encoded into the
context to be passed to the solver again with the intention of finding another
anomaly.
This iterative approach allows
for a complete search of the space
of anomalies within a configurable cycle-detection length.
Once the bounded space is completely covered (i.e., there are no more satisfying
solutions), test configuration files (\ding{206}) are generated from the collected abstract anomalies
that provide details about concrete executions that can potentially manifest the intended
anomaly.
Additionally, the source code of the program is passed through a code annotator
(\ding{205},\ding{207}) which returns multiple instances of the source code
including concrete function parameters and annotations on all database
access operations.

For example, \textsf{A1.conf} in \autoref{fig:annotated_code} is the
configuration file that specifies the initial state of the database
needed to manifest the previously discussed anomaly on the $\mathsf{CUST}$
table.
It also specifies the execution order of operations from the two \txn{payment}
transaction instances and the network status at each logical step (\texttt{T1} to
\texttt{T4}).
The configuration is completed with annotated source code
that specifies the input parameters to the \txn{payment} transaction
and marks the database operations with their relative local orders
(\textsf{A\!1\_Ins1.java} and \textsf{A\!1\_Ins2.java}).

%
%
%
\input{figures/annotated_code}
%
%
%

% 8
% introduce the test administrator
%%%
\tool\, additionally offers a fully automated test administration framework,
which consists of centrally managed wrappers for arbitrary database
drivers which can temporarily block database access requests in order to
enforce a specific execution order (\ding{208}).
Similarly, database replicas are deployed in managed containers, communication
among which can be throttled (\ding{209}) in order to induce temporary
network partitionings.
This enables \tool\, to efficiently administer a wide range of
statically constructed potential serializability anomalies,
allowing developers to witness and study undesired application behaviors that
frequently occur in the wild but which are extremely difficult to catch using
existing testing frameworks.

%% file: figures/motivating_example.tex
%%-----------------
% settings

\definecolor{pgrey}{rgb}{0.26,0.25,0.28}
\definecolor{background}{rgb}{0.92,0.92,0.92 }
\definecolor{javared}{rgb}{0.2,0.2,0.7} % for strings
\definecolor{javagreen}{rgb}{0.2,0.45,0.3} % comments
\definecolor{javapurple}{rgb}{0.5,0,0.35} % keywords
\definecolor{javadocblue}{rgb}{0.25,0.35,0.75} % javadoc
\definecolor{weborange}{RGB}{0,75,0}

\lstset{language=Java,
  basicstyle=\ttfamily\scriptsize,
  breaklines=true,
  backgroundcolor=\color{Main-Background},
  frame=single,
  rulecolor=\color{pgrey},
  keywordstyle=\color{javapurple}\bfseries,
  stringstyle=\color{javared},
  commentstyle=\color{javagreen},
  morecomment=[s][\color{javadocblue}]{/**}{*/},
  numbers=left,
  xleftmargin=1.9em,
  framexleftmargin=1.25em,
  numberstyle=\tiny\color{pgrey},
  stepnumber=1,
  numbersep=5pt,
  tabsize=1,
  showspaces=false,
  showstringspaces=false,
  classoffset=1, % starting new class  
  morekeywords={stmt,rs},
  keywordstyle=\color{weborange},
}

%%-----------------
% code figure
\begin {figure}[h]
\begin{minipage}[b]{0.47\textwidth}
\begin{lstlisting}[]
public void payment(int c_id) 
        throws CustNotFoundException{
	stmt = prepareStatement("SELECT C_PAY_CNT FROM CUST WHERE C_ID=?");
	stmt.setInt(1, c_id);
	rs = stmt.executeQuery();
 if(!rs.next())
  throw new CustNotFoundException(c_id);
	int c_pay_cnt = rs.getInt("C_PAY_CNT");
	c_pay_cnt++;
 stmt = prepareStatement("UPDATE CUST SET C_PAY_CNT=? WHERE C_ID=?");
	stmt.setInt(1, c_pay_cnt);
	stmt.setInt(2, c_id);
	stmt.executeUpdate();
}
\end{lstlisting}
\end{minipage}
\hfill
\begin{minipage}[t][][t]{0.51\textwidth}
  \raisebox{0.2\height}{\includegraphics[scale=0.595]{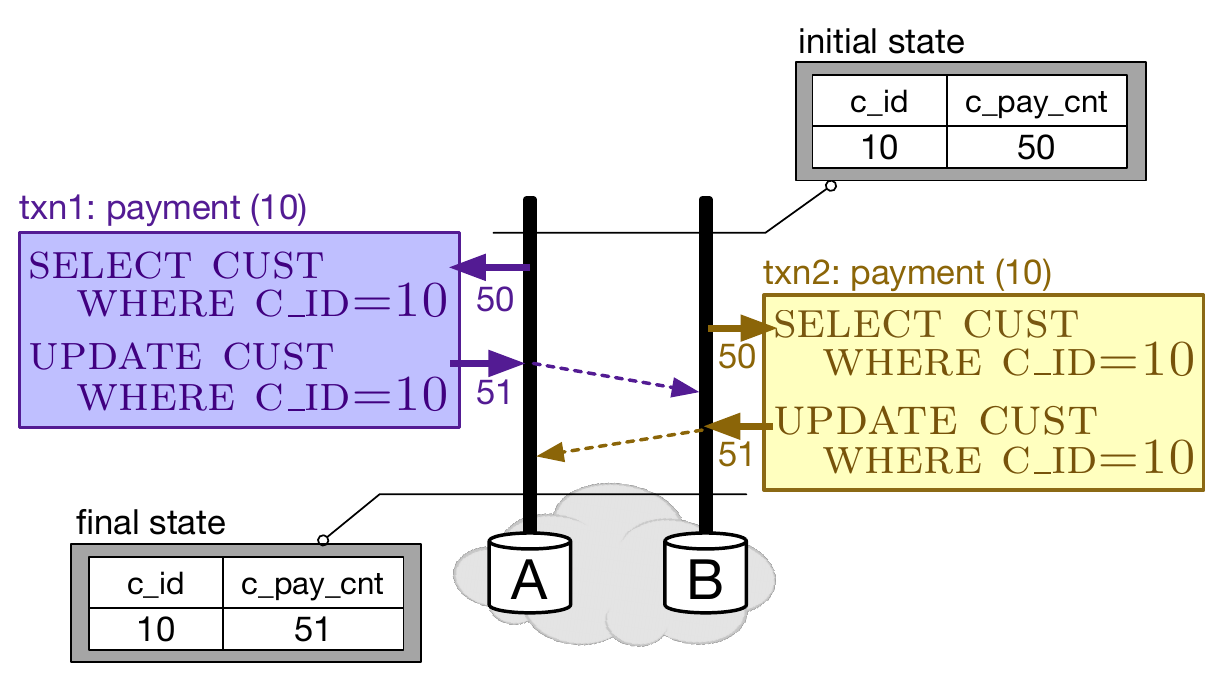}}
\end{minipage}

%\hfill
%\lstset{language=Java,
%  basicstyle=\ttfamily\scriptsize,
%  breaklines=true,
%  backgroundcolor=\color{background},
%  frame=single,
%  rulecolor=\color{pgrey},
%  keywordstyle=\color{javapurple}\bfseries,
%  stringstyle=\color{javared},
%  commentstyle=\color{javagreen},
%  morecomment=[s][\color{javadocblue}]{/**}{*/},
%  numbers=left,
%  xleftmargin=1.9em,
%  framexleftmargin=1.25em,
%  numberstyle=\tiny\color{pgrey},
%  stepnumber=1,
%  numbersep=5pt,
%  tabsize=1,
%  showspaces=false,
%  showstringspaces=false,
%  classoffset=2, % starting new class  
%  morekeywords={stmt,rs},
%  keywordstyle=\color{weborange},
%}
%\begin{minipage}[b]{0.49\textwidth}\centering
%\begin{lstlisting}[title={Delivery},captionpos=b]
%public void delivery(int c_id){
%	stmt = prepareStatement("SELECT C_DELIVERY_CNT FROM CUSTOMER WHERE C_ID=?");
%	stmt.setInt(1, c_id);
%	rs = stmt.executeQuery();
%	if (!rs.next())
%		throw new CustNotFoundException;
%	int c_delivery_cnt = 
%			rs.getInt("C_DELIVERY_CNT");
%	c_delivery_cnt++;
%	stmt = prepareStatement("UPDATE CUSTOMER SET C_DELIVERY_CNT=? WHERE C_ID=?");
%	stmt.setInt(1, c_delivery_cnt);
%	stmt.setInt(2, c_id);
%	stmt.executeUpdate();
%}
%\end{lstlisting}
%\end{minipage}
\caption{A transaction from TPC-C benchmark in Java (left) and an anomalous
execution (right)}
\label{fig:motivating}
\end{figure}

%% file: figures/pipeline.tex
\begin {figure}[b]
    \includegraphics[scale=0.455]{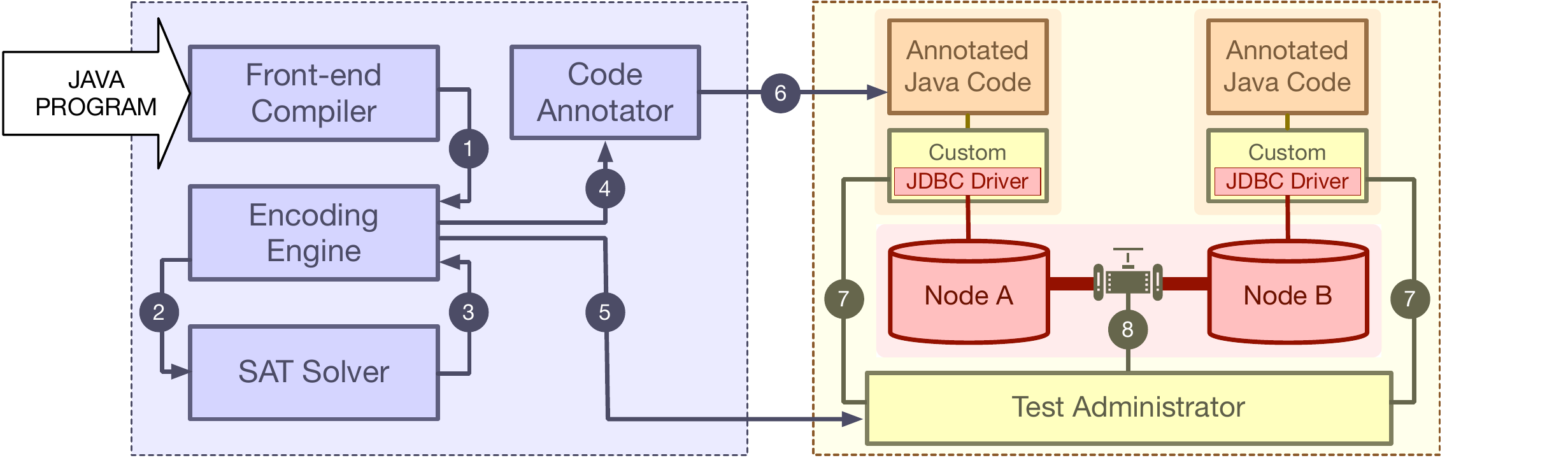}
\vspace{1mm}
  \caption{Pipeline of \tool}
\vspace{-2mm}
\label{fig:pipeline}
\end{figure}

%% file: figures/annotated_code.tex
%%-----------------
% settings

\definecolor{pgrey}{rgb}{0.26,0.25,0.28}
\definecolor{background}{rgb}{0.92,0.92,0.92 }
\definecolor{javared}{rgb}{0.2,0.2,0.7} % for strings
\definecolor{javagreen}{rgb}{0.2,0.45,0.3} % comments
\definecolor{javapurple}{rgb}{0.5,0,0.35} % keywords
\definecolor{javadocblue}{rgb}{0.25,0.35,0.75} % javadoc
\definecolor{weborange}{RGB}{0,75,0}
\definecolor{annotateColor}{RGB}{175,25,20}

\lstset{language=Java,
  basicstyle=\ttfamily\scriptsize,
  breaklines=true,
  backgroundcolor=\color{Main-Background},
  frame=single,
  rulecolor=\color{pgrey},
  keywordstyle=\color{javapurple}\bfseries,
  stringstyle=\color{javared},
  commentstyle=\color{javagreen},
  morecomment=[s][\color{javadocblue}]{/**}{*/},
  numbers=left,
  title=\footnotesize{A1\_Ins1.java},
  xleftmargin=1.9em,
  framexleftmargin=1.25em,
  numberstyle=\tiny\color{pgrey},
  stepnumber=1,
  numbersep=5pt,
  tabsize=1,
  captionpos=b
  showspaces=false,
  showstringspaces=false,
  classoffset=2, % starting new class  
  morekeywords={stmt,rs},
  keywordstyle=\color{weborange},
  moredelim=[is][\textcolor{red}]{\%\%}{\%\%},
}

%%-----------------
% code figure
\begin {figure}[t]
\begin{minipage}[b]{0.32\textwidth}
\begin{lstlisting}[]
%%@Parameters(10)%%
public void payment ... {
 ... 
 %%@Sched(node="A", order=1)%%
 rs = stmt.executeQuery();
 ...
 %%@Sched(node="A", order=2)%%
	stmt.executeUpdate();
}
\end{lstlisting}
\end{minipage}
\lstset{language=Java,
  basicstyle=\ttfamily\scriptsize,
  breaklines=true,
  backgroundcolor=\color{Main-Background},
  frame=single,
  rulecolor=\color{pgrey},
  keywordstyle=\color{javapurple}\bfseries,
  stringstyle=\color{javared},
  commentstyle=\color{javagreen},
  morecomment=[s][\color{javadocblue}]{/**}{*/},
  numbers=left,
  title=\footnotesize{A1\_Ins2.java},
  xleftmargin=1.9em,
  framexleftmargin=1.25em,
  numberstyle=\tiny\color{pgrey},
  stepnumber=1,
  numbersep=5pt,
  tabsize=1,
  captionpos=b
  showspaces=false,
  showstringspaces=false,
  classoffset=1, % starting new class
  morekeywords={stmt,rs},
  keywordstyle=\color{weborange},
  moredelim=[is][\textcolor{red}]{\%\%}{\%\%},
}
\begin{minipage}[b]{0.32\textwidth}
\begin{lstlisting}[]
%%@Parameters(10)%%
public void payment ... {
 ... 
 %%@Sched(node="B", order=1)%%
 rs = stmt.executeQuery();
 ...
 %%@Sched(node="B", order=2)%%
	stmt.executeUpdate();
}
\end{lstlisting}
\end{minipage}
\lstset{language=ruby,
  basicstyle=\ttfamily\scriptsize,
  breaklines=true,
  backgroundcolor=\color{Main-Background},
  frame=single,
  rulecolor=\color{pgrey},
  keywordstyle=\color{javapurple}\bfseries,
  stringstyle=\color{javared},
  commentstyle=\color{javagreen},
  morecomment=[s][\color{javadocblue}]{/**}{*/},
  numbers=none,
  title=\footnotesize{A1.conf},
  captionpos=b,
  xleftmargin=0.9em,
  framexleftmargin=0.25em,
  numberstyle=\tiny\color{pgrey},
  stepnumber=1,
  numbersep=5pt,
  tabsize=1,
  showspaces=false,
  showstringspaces=false,
  classoffset=2, % starting new class  
  morekeywords={stmt,rs},
  keywordstyle=\color{weborange},
  moredelim=[is][\textcolor{red}]{\%\%}{\%\%},
}
\begin{minipage}[b]{0.32\textwidth}
\begin{lstlisting}[]
# initialize:
INSERT INTO 
  CUST(c_id,c_pay_cnt) 
  VALUES (10,50);
# schedule: 
@T1@partitions{A,B}: Ins1-O1
@T2@partitions{A,B}: Ins2-O1
@T3@partitions{A,B}: Ins1-O2
@T4@partitions{A,B}: Ins2-O2
\end{lstlisting}
\end{minipage}
%\begin{minipage}[b]{0.15\textwidth}
%\begin{lstlisting}[]
%Anomaly#1 
%---------
%@T1@C1:1 {A,B}
%@T2@C2:1 {A,B}
%@T3@C1:2 {A,B}
%@T4@C2:2 {A,B}
%\end{lstlisting}
%\end{minipage}
\vspace{0mm}
\caption{Annotated code of two transaction instances and the test configuration 
file}
\vspace{0mm}
\label{fig:annotated_code}
\end{figure}

%% file: sections/model.tex
\ir\ is the target language of \tool's front-end compiler and is
designed to capture key features of database-backed applications,
including schemas, transactions, and data retrieval and modification
operations.  We begin by introducing the syntax of \ir\ programs and
presenting an overview of its semantics. A complete account of the
semantics of \ir\ can be found in~\cite{Rahmani:Clotho:2019}.

\subsection{Abstract Representation}
\label{subsec:ar}
\input{figures/ar_syntax}

\ir\ programs are parameterized over a fixed schema that describes a
set of database tables.  Each table is defined as an ordered list of
field names from the set $\mathtt{Fields}$.  In order to uniquely
identify individual records, each table is assigned a non-empty subset
of its fields that acts as a \emph{primary key}.
For simplicity, our presentation assumes a single
table with only integer-valued fields, but \tool\ relaxes this
restriction and supports an arbitrary number of tables and other Java
primitive types.

The syntax of \ir\ is given in \autoref{fig:syntax}.
Programs in \ir\ consist of a set of parameterized transactions, each
of which has a unique name drawn from the set $\mathtt{TxnNames}$.
The body of a transaction is a command, denoted by $c$, which can be
either a database query, a guarded command, a loop, or a sequence of
commands.
Database queries, denoted by $q$, retrieve ($\sql{select}\!\!$) or
modify~($\sql{update}\!\!$, $\sql{insert}\!$and $\sql{delete}\!\!\!\;$)
records in the database.  The result of each $\sql{select}\!$query is
stored as an ordered list of records in a unique variable $v$.

Boolean expressions of \ir, denoted by $\phi$, consist of standard arithmetic
comparison and boolean~operators.

Arithmetic expressions, denoted by $e$, include integer constants,
transaction arguments, arithmetic operations, non-deterministic
values, iteration counters, the number of results held in a variable,
and field accesses, denoted as \texttt{proj}. For example,
$\mathtt{proj}(age,v,3)$ returns the value of $age$ field in the third
record stored in the variable $v$.
The expression $\mathtt{any}(\phi)$ can evaluate to any value
satisfying the predicate $\stx{\phi}$ and is used to model source
language features such as unknown function calls and updates to program
variables within non-deterministic loops.
The iteration counter $\mathtt{iter}$ represents the number of times a
loop has been repeated during a program execution.
Field names can also appear in an arithmetic expression as $\mathtt{this}.f$,
which can only be used in conditionals in the $\sql{where}$clauses of queries and 
denote the field values of records that are being evaluated in that~clause.

\lstset{basicstyle=\ttfamily\small} As an example,
\autoref{fig:compile_example} presents a Java code snippet and its
\ir\ encoding.  The program scans a table and retrieves all records
representing employees younger than 35 and then increases their
salaries within a\;\lstinline{while} loop.  \lstset{language=Java,
basicstyle=\ttfamily\scriptsize}

\input{figures/compile_example}

%

%%%%%%%%%%%%%%%%%%%%%%%%%%%%%%%%%%%%%%%%%%%%%%%%%%%%%%%%%%%%%%%%%%%%%%%%%%%%%%%%%%%%%%%%%%%%%%%%%%%%

%
\subsection{System Configurations}
\label{subsec:sys_conf}
The operational semantics of \ir\ is defined by a small-step reduction
relation, $\rightarrow~\subseteq~\Sigma\times\Gamma\times\Sigma\times\Gamma$, on
the current \emph{system state} of the underlying database, denoted as
$\Sigma$, and a set of concurrently running transaction instances,
denoted as $\Gamma$.
Mimicking the execution of database-backed programs on replicated
stores, \ir\ programs are interpreted on a finite number of \emph{partitions},
each of which has its own copy of the database.
System states are therefore represented as a triple,
$(\store, \ar, \vis)$, which holds a history of database reads and
writes executed at each partition and a pair of orderings on these
events. Executing a database operation generates a set of \emph{read
  effects}, $\rdd{(r,f)}$, witnessing that the field $f$ of record $r$
was accessed, and \emph{write effects}, $\wrr{(r,f,n)}$, recording
that the field $f$ of record $r$ was set to the value $n$. Read
effects may optionally include the value read,
e.g. $\rdd{(r,f,n)}$. Each effect is implicitly tagged with a unique
identifier, so that we can distinguish, for example, between the same
field being read by multiple queries.
Given a finite set $\mathcal{P}$ of partition names, $\store$
represents the set of effects generated on each partition as a mapping
from partition names to disjoint sets of effects. When used without an
argument, $\store$ refers to the complete set of effects from all
partitions, i.e.
$\store:=
\underset{\scriptscriptstyle{p}\in\mathcal{P}}{\bigcup}\store(p)$.

The $\ar$ component of the system state records the exact
sequence of database operations that have been executed as an 
order between effects called an \emph{arbitration relation}.
This relation is however too coarse-grained to capture causal relationships
between effects: an effect $\eta_1$ created at partition $p_1$, does
not necessarily influence the execution of a query at partition $p_2$
that creates another effect $\eta_2$, even though $\eta_1$ may be
arbitrated before $\eta_2$.
The system state therefore maintains a more refined \emph{visibility}
relation, $\vis\subseteq\ar$, between effects, which only relates two
effects if one actually witnesses the other at the time of creation.

It is possible to construct a current copy, or \emph{local view}, of
the database at each partition $p$ from a system state by ``applying''
the effects of query operations stored in $\store(p)$ according to the
order in $\ar$. Such local views are denoted as $\sigma$ and are
modeled as functions from primary keys to records. To ensure $\sigma$
is total, all tables include a special field,
$alive \in \mathtt{Fields}$, whose value determines if a record is
present in the table or not.

%%%%%%%%%%%%%%%%%%%%%%%%%%%%%%%%%%%%%%%%%%%%%%%%%%%%%%%%%%%%%%%%%%%%%%%%%%%%%%%%%%%%%%%%%%%%%%%%%%%%

%
\input{figures/rule_example}
\subsection{Operational Semantics}
\label{subsec:op}
\autoref{fig:rule_example} presents the rules defining the operational
semantics of \ir\ programs and commands. These rules are parameterized
over an \ir\ program, $\mathbb{P}$. Each step either adds a new
instance to the set of currently running transactions
via \textsc{e-spawn}\footnote{For
  simplicity of presentation, we assume the spawned transaction has
  only one argument} 
or executes the body of one
of the currently running transactions via \textsc{e-step}. 
The reduction rules for commands non-deterministically select a
partition $p$ to execute the command on. This partition is used by an
auxiliary function~$\Delta$ to construct a local view of the
database. These local views are used by evaluation relation for
boolean and arithmetic expressions, $\Downarrow$. The rules for
non-query commands are straightforward outside for the
\textsc{(e-iter)} rule, which uses the $\mathtt{concat}(n,c)$ function
that sequences $n$ copies of the command $c$, with any occurrences of
\texttt{iter} being instantiated as expected.

\input{figures/aux_rules}

The operational semantics for queries, presented in
\autoref{fig:aux_rules}, are more interesting. The key component of
each of these rules is how they construct the set of new effects
generated by each database operation, denoted by $\varepsilon_1$ and
$\varepsilon_2$. 
In the rule for $\sql{select}\!\!$, $\varepsilon_1$ includes appropriate read effects
for each record satisfying the operation's $\sql{where}$condition.
Any occurrence of $\mathtt{this}.f$ construct in the conditional $\phi$,
is substituted with the value of the
corresponding field in each record instance under evaluation.
The rule for $\sql{update}$similarly includes a write
effect in $\varepsilon_1$ for each record satisfying the operation's
$\sql{where}$condition. 
To model the deletion of records, the rule for $\sql{delete}$creates
write effects with a constant value 0 for the $alive$ fields of
all records that satisfy its $\sql{where}$condition.

In order to capture the data accesses that occur during database-wide
scans, all three rules include new read effects in $\varepsilon_2$ for every field that
is predicated in the operation's  $\sql{where}$clause.
To do so, these rules use the auxiliary function $\F{}$ which extracts
any such fields from boolean expressions. $\F{}$ additionally always
includes the $alive$ field. As an example, the result of this function
for the where clause from the query in \autoref{fig:compile_example}
is, \mbox{$\F{(age<35)}=\{age,alive\}$}.

Finally, each rule updates the $\ar$ (resp. $\vis$) relation to reflect
the relationship between the newly created effects in
$\varepsilon_1\cup\varepsilon_2$ and already existing effects in $\store$ (resp.
$\store(p)$).

\input{figures/execution}

\autoref{fig:execution} depicts a concrete example of two consecutive
execution steps of the \ir\ program
from~\autoref{fig:compile_example}. 
We depict an execution in which arithmetic expressions in the update 
operation ($q_2$)
have been fully evaluated and both operations execute on the same
partition. The example furthermore simplifies $\store$ to be just
the set of effects in that particular partition.
The initial system state encodes a database with three employee
records, only two of which are alive.  Because the $\sql{select}$query
does not modify any database record, the local views
$\sigma$ and $\sigma'$ are identical.

The $\sql{select}$query constructs a set of new effects in
$\varepsilon_2$, including three read effects on the $age$ field of
all records (since $age$ is constrained by its $\sql{where}$clause)
and three read effects on all $alive$ fields.  
Since only one of the
witnessed records is both alive and satisfies the
$\sql{where}$condition (i.e. $age<35$), a single read effect on that
record, $\rdd{(1,sal)}$, is also created and included in $\varepsilon_1$.
Similarly, $\varepsilon_1'$ is defined for the execution of
$\sql{update}$query, which includes only a single write effect,
$\wrr{(1,sal,86)}$, capturing the modified salary value for the
previously selected employee record.
$\varepsilon_2'$ is empty, because the $\sql{update}$query uses
the primary key of the table to access the record, without any database-wide
scans.
Note that both steps update the arbitration and visibility relations
identically, since the example assumes only a single~partition.

%% file: figures/ar_syntax.tex
\begin {figure}[t]
 \begin{minipage}{.85\textwidth}
   \begin{mdframed}[
  backgroundcolor=Main-Theme-Lightest,
     ]
  \begin{mathpar}
  \footnotesize
  \begin{array}{lcl}
  \multicolumn{3}{c}{
    \stx{a} \in \mathtt{Arguments} \qquad
    \stx{v} \in \mathtt{Variables} \qquad
    \stx{f} \in \mathtt{Fields} \qquad
    \stx{A} \in \{\tt{min, max}\} } \\
  \multicolumn{3}{c}{
    \stx{T} \in \mathtt{TxnNames} \qquad
    \oplus \in \{+, -, \times, /\} \qquad
    \odot \in \{<, \leq, =, >, \geq\} \qquad
    \circ \in \{\wedge, \vee\} }\\
\\
\phi & \coloneqq & \stx{e\odot e}
\ALT
\neg\phi \ALT \phi\circ\phi
\ALT \mathtt{true} \ALT \mathtt{false}
\\
\stx{e} & \coloneqq &
\intt
\ALT \stx{a}
\ALT \stx{e\oplus e}
\ALT \stx{\mathtt{any(\phi)}}
\ALT \stx{\mathtt{iter}}
\ALT \stx{\mathtt{size}(v)}
\ALT \stx{\mathtt{proj}(f,v,e)}
\ALT \stx{\mathtt{this}.f}
\\[0.3mm]
\stx{q} & \coloneqq & \sql{select} \stx{f}\; \sql{as} \stx{v}\;
\sql{where} \stx{\phi}
\ALT
\sql{select} \stx{A(f)}\; \sql{as} \stx{v}\;
\sql{where} \stx{\phi}
\ALT
\\[-0.3mm]
        & & \sql{update} \sql{set}
        \stx{f=e} \; \sql{where} \stx{\phi}
\ALT
\sql{insert} \sql{values} \stx{\overline{f}=\overline{e}}
\ALT
 \sql{delete} \sql{where} \stx{\phi}
 \\[0.2mm]
\stx{c} & \coloneqq &
\stx{q}
\ALT
\mathtt{if(\phi)}\{c\}
\ALT
\mathtt{iterate}(e)\{c\}
\ALT
c;c
\ALT
\mathtt{skip}
\\
\mathbb{P} & \coloneqq & \stx{\overline{T(\overline{a})\{c\}}}

\end{array}
  \end{mathpar}
\end{mdframed}
\end{minipage}
\caption{Syntax of transactional programs written in the abstract
\label{fig:syntax}
representaiton (\ir)}
\end{figure}

%% file: figures/compile_example.tex
\definecolor{pgrey}{rgb}{0.26,0.25,0.28}
\definecolor{background}{rgb}{0.92,0.92,0.92 }
\definecolor{javared}{rgb}{0.2,0.2,0.7} % for strings
\definecolor{javagreen}{rgb}{0.2,0.45,0.3} % comments
\definecolor{javapurple}{rgb}{0.5,0,0.35} % keywords
\definecolor{javadocblue}{rgb}{0.25,0.35,0.75} % javadoc
\definecolor{weborange}{RGB}{0,75,0}

\lstset{language=Java,
  basicstyle=\ttfamily\scriptsize, %\tiny,
  breaklines=true,
  backgroundcolor=\color{Main-Background},
  title={},
  caption={},
  frame=none,
  rulecolor=\color{pgrey},
  keywordstyle=\color{javapurple}\bfseries,
  stringstyle=\color{javared},
  commentstyle=\color{javagreen},
  morecomment=[s][\color{javadocblue}]{/**}{*/},
  numbers=none,
  xleftmargin=0em,
  xrightmargin=0em,
  framexleftmargin=0em,
  numberstyle=\tiny\color{pgrey},
  stepnumber=1,
  numbersep=5pt,
  tabsize=1,
  showspaces=false,
  showstringspaces=false,
  classoffset=1, % starting new class  
  morekeywords={stmt,rs},
  keywordstyle=\color{weborange},
}

\begin {figure}[t]
\begin{minipage}[t]{.95\textwidth}
\begin{mdframed}[
  innerleftmargin=0cm,
  innerrightmargin=0cm,
  innertopmargin=0pt,
  innerbottommargin=0pt,
  linewidth=0.6bp,
  linecolor=black,%
  usetwoside=false,
  backgroundcolor=Main-Background,]
~
  \begin{minipage}{0.62\textwidth}
    \vspace{1mm}  
    \center \scriptsize
      \textsc{java code}
    \vspace{0.5mm}  
  \end{minipage}
  \vrule 
   \setlength{\fboxsep}{0pt}\colorbox{Main-Theme-Lightest}{%
  \begin{minipage}{0.37\textwidth}
    \center \scriptsize
    \vspace{0.25mm}
    \textsc {encoding in} \ir 
    \vspace{0.75mm}  
\end{minipage}}
  \hrule  
%%%%%%%%%%%%%%%%%%%%%%%%%%%%%%%%%%%%%%%%%%%%%%%%%%%%%%%%%%%%%%%%%%%%%%%%%%%%%%%%%%%%%%%%%%%%%%%%%%%
  \hrule
~
  \begin{minipage}{0.62\textwidth}
    \lstset{basicstyle=\linespread{0.1}\ttfamily\scriptsize}
      \begin{lstlisting}[]
stmt=prepareStatement("SELECT sal WHERE age<35");
rs = stmt.executeQuery();
while(rs.next()){
  int emp_id=rs.getInt("id");
  int old_sal=rs.getInt("sal");
  stmt=prepareStatement("UPDATE SET sal=? WHERE id=?"); 
  stmt.setInt(1,old_sal+1);
  stmt.setInt(2,emp_id);
  stmt.executeUpdate();
}
      \end{lstlisting}
  \end{minipage}
  \vrule
	  \setlength{\fboxsep}{0pt}\colorbox{Main-Theme-Lightest}{%
  \begin{minipage}{0.37\textwidth}
    \vspace{5.6mm}
    
    $ \scriptstyle
    \;\;\sql{select}{sal}\;\sql{AS}\; \stx{v}\;\;\sql{where}
    \mathtt{this}.{age}<35;$
    \\[-0.4mm]
    $
  \scriptstyle
  \stx{\;\; \mathtt{iterate(size(}v))\{}
    \\[-1mm]
  \stx{\;\;\;\;\;\sql{update}\sql{set}} \stx{{sal} =
  \mathtt{proj}({{sal}},v,\mathtt{iter}) + 1}
  \scriptstyle
  $\\[-1mm]
  $
  \scriptstyle
  \stx{\;\;\;\;\;
    \sql{where} \mathtt{this}.{id}=\mathtt{proj}({id},v,\mathtt{iter})
  }
  $\\[-2mm]
  $
  \scriptstyle
\stx{\;\;\}}
  $
  \vspace{6mm}
  \end{minipage}}
\end{mdframed}
\end{minipage}
%
%%%%%%%%%%%%%%%%%%%%%%%%%%%%%%%%%%%%%%%%%%%%%%%%%%%%%%%%%%%%%%%%%%%%%%%%%%%%%%%%%%%%%%%%%%%%%%%%%%%
%
\vspace{1mm}
\caption{ A Java code block and its encoding  in \ir}
\label{fig:compile_example}
\end{figure}

%% file: figures/rule_example.tex
\begin{figure}[t]
  \begin{minipage}{\textwidth}
  \begin{mdframed}[backgroundcolor=Main-Theme-Lightest]

    \vspace{-2mm}
\begin{minipage}[t]{.23\textwidth}
  \vspace{3mm}
  \begin{flushleft}
    {\scriptsize
    \textrm{\sc{\color{black} (e-spawn)}}}
  \end{flushleft}
$$
\RULE{
  T(a)\{c\} \in \mathbb{P}
\qquad
n\in \intt

}
%% ------------------------------------------------------------------------------------------------
{
\stx{
  \step{\Sigma,\Gamma}
    {}% ->
    {\Sigma,\Gamma \!\cup\! \{c[a/n]\}}
  }
}
$$

%\caption{A simplified rule capturing execution of an $\sql{update}$ operation}
    \label{fig:rule_example}
\end{minipage}
\hfill
\begin{minipage}[t]{.23\textwidth}
  \vspace{3mm}
  \vspace{0mm}
  \begin{flushleft}
    {\scriptsize
    \textrm{\sc{\color{black} (e-step)}}}
  \end{flushleft}
$$
\RULE{
\step{\Sigma, c}{}{\Sigma', c'}

}
%% ------------------------------------------------------------------------------------------------
{
\stx{
  \step{\Sigma,\{c\}\!\cup\!\Gamma}
    {}% ->
    {\Sigma',\{c'\}\!\cup\!\Gamma}
  }
}
$$

%\caption{A simplified rule capturing execution of an $\sql{update}$ operation}
    \label{fig:rule_example}
\end{minipage}
\hfill
\begin{minipage}[t]{.18\textwidth}
  \vspace{3mm}
  \begin{flushleft}
    %\hspace{-2mm}\fcolorbox{black}{Main-Theme-Lighter}
    {\scriptsize
    \textrm{\sc{\color{black} (e-skip)}}}
  \end{flushleft}
$$
\RULE{
\\
}
%% ------------------------------------------------------------------------------------------------
{
\stx{
  \step{\Sigma, \mathtt{skip}; c}
    {}% ->
    {\Sigma, c}
  }
}
$$
    \label{fig:rule_example}
\end{minipage}
\hfill
\begin{minipage}[t]{.19\textwidth}
  \vspace{3mm}
  \begin{flushleft}
    {\scriptsize
    \textrm{\sc{\color{black} (e-seq)}}}
  \end{flushleft}
$$
\RULE{
  \stx{
    \step{\Sigma, c}
    {}% ->
    {\Sigma', c''}
  }

}
{
\stx{
  \step{\Sigma,c; c'}
    {}% ->
    {\Sigma', c''; c'}
  }
}
$$
    \label{fig:rule_example}
\end{minipage}
%
%

%%%%%%%%%%%%%%%%%%%%%%%%%%%%%%%%%%%%%%%%%%%%%%%%%%%%%%%%%%%%%%%%%%%%%%%%%%%%%%%%%%%%%%%%%%%%%%%%%%%%

\begin{minipage}[t]{.31\textwidth}
  \vspace{3mm}
  \begin{flushleft}
    {\scriptsize
    \textrm{\sc{\color{black} (e-cond-t)}}}
  \end{flushleft}
$$
\RULE{
  \Sigma \equiv (\store,\ar,\vis) 
  \quad 
  p\in\mathcal{P}  \\
  \sigma = \Delta(\ar,\store(p)) 
  \quad
  \B{\phi}{\sigma} \mathsf{true}

}
{
\stx{
  \step{\Sigma, \mathtt{if}~\phi~\{c\}}
    {}% ->
    {\Sigma, c}
  }
}
$$
    \label{fig:rule_example}
\end{minipage}
\hfill
\begin{minipage}[t]{.31\textwidth}
  \vspace{3mm}
  \begin{flushleft}
    {\scriptsize
    \textrm{\sc{\color{black} (e-cond-f)}}}
  \end{flushleft}
$$
\RULE{
  \Sigma \equiv (\store,\ar,\vis) 
  \quad
  p\in\mathcal{P}  \\
  \sigma = \Delta(\ar,\store(p)) 
  \quad
  \B{\phi}{\sigma} \mathsf{false}

}
{
\stx{
  \step{\Sigma, \mathtt{if}~\phi~\{c\}}
    {}% ->
    {\Sigma, \mathtt{skip}}
  }
}
$$
\end{minipage}
\hfill
\begin{minipage}[t]{.34\textwidth}
  \vspace{3mm}
  \begin{flushleft}
    {\scriptsize
    \textrm{\sc{\color{black} (e-iter)}}}
  \end{flushleft}
$$
\RULE{
  \Sigma \equiv (\store,\ar,\vis)
  \quad 
  p\in\mathcal{P}
  \\
  \sigma = \Delta(\ar,\store(p))
  \quad
  \I{e}{\sigma} n

}
{
\stx{
  \step{\Sigma, \mathtt{iterate}(e)\{c\} }
    {}% ->
    {\Sigma, \mathtt{concat}(n,c)}
  }
}
$$
\end{minipage}
\end{mdframed}
\end{minipage}
\caption{Operational semantics of \ir\ programs and commands}
\label{fig:rule_example}
\end{figure}

%% file: figures/aux_rules.tex
\begin{figure}[h]

%\eta

  \begin{minipage}[b]{.95\textwidth}
    \begin{mdframed}[backgroundcolor=Main-Theme-Lightest]
  \vspace{0mm}
  \begin{flushleft}
    \hspace{-2mm}\fcolorbox{black}{Main-Theme-Lighter}{\scriptsize
    \textrm{\sc{\color{black} e-select}}}
  \end{flushleft}
  \vspace{-7mm}
$$
\RULE{
 q\equiv \sql{select} f\; \sql{as} x\; \sql{where} \phi\
 \qquad
 {p}\in \mathcal{P}
 \qquad 
 \sigma = \Delta(\ar,\store({p})) \\

  \stx{\varepsilon_1}  =
  \{\rdd{(r,f)} \;|\;\B{\phi[\mathtt{this}.f'\mapsto r(f')]}{\sigma} \texttt{true}\}
  \qquad
 \stx{\varepsilon_2}  =  
  \stx{\{\rdd{(r,f')}\;|\;f'\in\F(\phi)\}}
  \qquad 
  \varepsilon = \varepsilon_1\cup\varepsilon_2
  \\
    \vis' = \vis~\cup~\{(\eta,\eta')\ALT \eta'\in\varepsilon \wedge \eta \in
    \store({p})\}  \qquad \;\;
    \ar' = \ar~\cup~\{(\eta,\eta')\ALT \eta'\in\varepsilon \wedge \eta \in\store\}

}
{
\stx{
  \step{(\store,\vis,\ar), 
  q
}
    {}% ->
    {(\store[{p} {\mapsto}\store({p})\cup\varepsilon],\vis',\ar'), \texttt{skip}}
  }
}
$$

\vspace{-1mm}
\end{mdframed}
\end{minipage}

%
%
%
%
%
%
%
%
%
%
%
%pp
%
%

\begin{minipage}[b]{.95\textwidth}
    \begin{mdframed}[backgroundcolor=Main-Theme-Lightest]
  \vspace{0mm}
  \begin{flushleft}
    \hspace{-2mm}\fcolorbox{black}{Main-Theme-Lighter}{\scriptsize
    \textrm{\sc{\color{black} e-update}}}
  \end{flushleft}
  \vspace{-7mm}
$$
\RULE{
  q\equiv \sql{update} \sql{set} f=v\; \sql{where} \phi
  \qquad
    {p}\in \mathcal{P} \qquad
  \sigma = \Delta(\ar,\store({p}))
  \qquad 
  \I{v}{\sigma} n 
  \\
  \stx{\varepsilon_1} =
  \{\wrr{(r,f, n)} \;|\;\B{\phi[\mathtt{this}.f'\mapsto r(f')]}{\sigma} \texttt{true}\}
  \qquad
  \stx{\varepsilon_2} =
  \{\rdd{(r,f')}\;|\; f'\in\F(\phi)\}
  \qquad 
  \varepsilon = \varepsilon_1 \cup \varepsilon_2
  \\
    \vis' =
    \vis~\cup~\{(\eta,\eta')\ALT
    \eta'\in\varepsilon
    \wedge
    \eta \in \store({p})
    \}
    \qquad
    \ar' =
    \ar~\cup~\{(\eta,\eta')\ALT \eta'\in\varepsilon
    \wedge \eta \in \store\}
}
{
  \stx{
    \step{(\store,\vis,\ar), q
  }
    {}% ->
    {(\store[{p}{\mapsto}\store({p})\!~\cup~\!\varepsilon],\vis',\ar'), \texttt{skip}}
  }
}
$$
  \end{mdframed}
\end{minipage}

\begin{minipage}[b]{.95\textwidth}
    \begin{mdframed}[backgroundcolor=Main-Theme-Lightest]
  \vspace{0mm}
  \begin{flushleft}
    \hspace{-2mm}\fcolorbox{black}{Main-Theme-Lighter}{\scriptsize
    \textrm{\sc{\color{black} e-delete}}}
  \end{flushleft}
  \vspace{-7mm}
$$
\RULE{
    q\equiv \sql{delete} \sql{where} \phi 
    \qquad
    {p}\in \mathcal{P} \qquad
  \sigma = \Delta(\ar,\store({p}))
  \\
  \stx{\varepsilon_1} =
  \{\wrr{(r, {alive},0)} \;|\; \B{\phi[\mathtt{this}.f'\mapsto r(f')]}{\sigma} \texttt{true}\}
  \qquad 
  \varepsilon_2 =
  \{\rdd{(r,f')}\;|\; f'\in\F(\phi)\}
  \qquad 
  \varepsilon = \varepsilon_1 \cup \varepsilon_2
  \\
    \vis' =
    \vis~\cup~\{(\eta,\eta')\ALT
    \eta'\in\varepsilon
    \wedge
    \eta \in \store({p})
    \}
    \qquad
    \ar' =
    \ar~\cup~\{(\eta,\eta')\ALT \eta'\in\varepsilon
    \wedge \eta \in \store\}
}
%% ------------------------------------------------------------------------------------------------
{
  \stx{
    \step{(\store,\vis,\ar),q
  }
    {}% ->
    {(\store[{p}{\mapsto}\store({p})\!~\cup~\!\varepsilon],\vis',\ar'), \texttt{skip}}
  }
}
$$
  \end{mdframed}
\end{minipage}

\caption{Operational semantics of \ir\ queries.}
\label{fig:aux_rules}

\end{figure}

%% file: figures/execution.tex
\begin {wrapfigure}[23]{r}{0.42\textwidth}
\centering
\vspace{-2mm}
\begin{subfigure}[b]{.4\textwidth}
	\includegraphics[width=\textwidth]{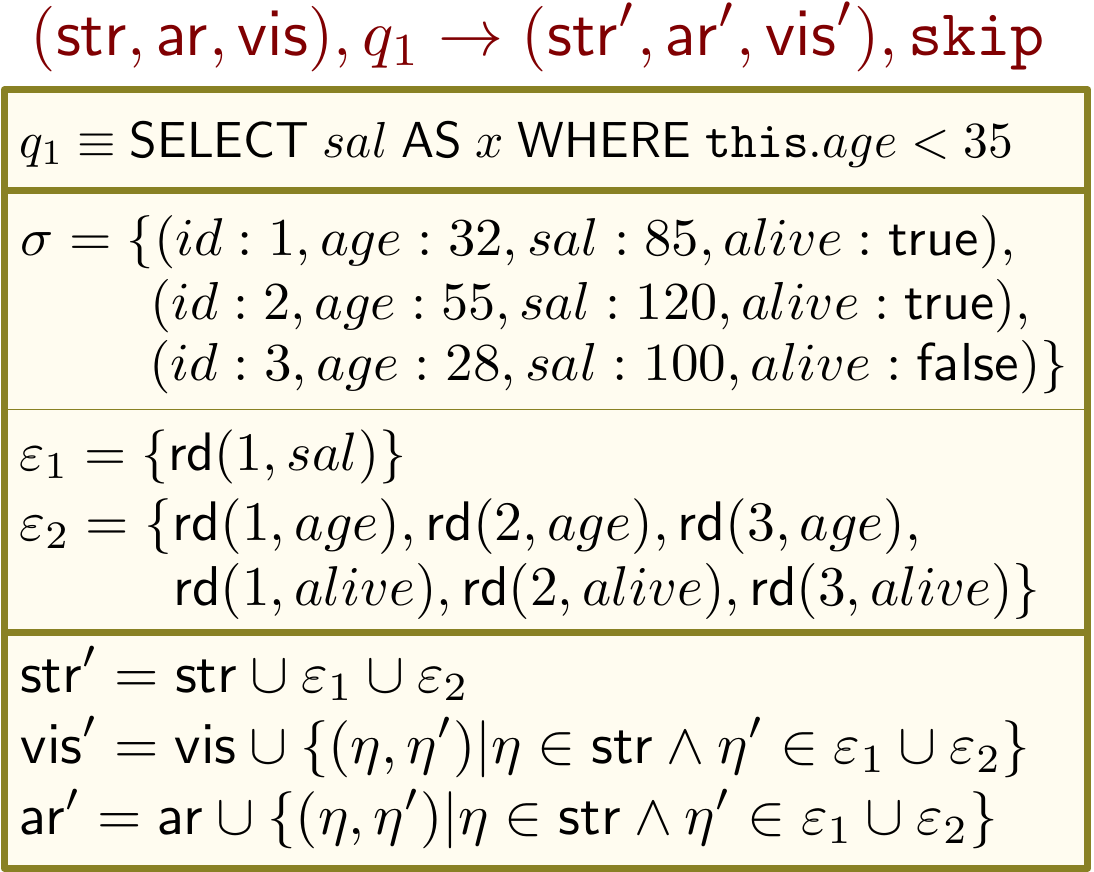}
\end{subfigure}
\vspace{2mm}

\begin{subfigure}[b]{.4\textwidth}
	\includegraphics[width=\textwidth]{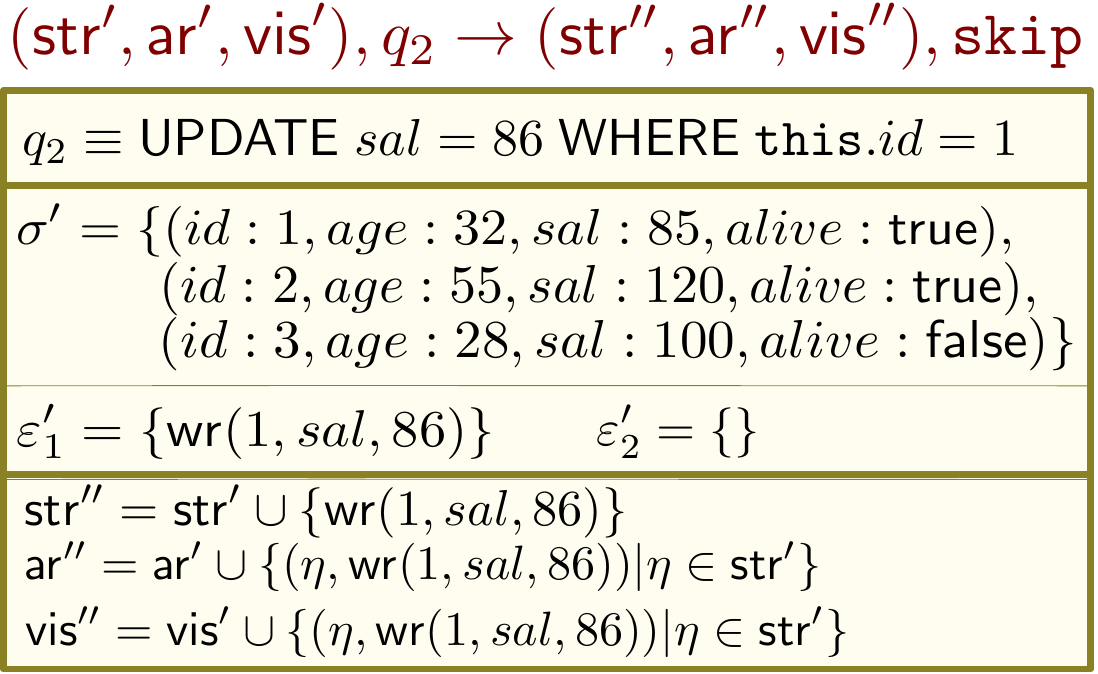}
\end{subfigure}
\caption{Execution of the program in \autoref{fig:compile_example}}
  \label{fig:execution}
\end{wrapfigure}

%% file: sections/ser.tex
\section{Serializability Anomalies}
\label{sec:ser}

We now turn to the question of identifying and statically detecting
undesirable executions in \ir~programs.

\subsection{Serial(izable) Executions}
\label{subsec:serializability_anomalies}
An \emph{execution history} of an \ir\ program is simply a finite
sequence of system configurations allowed by the reduction relation:
$H\equiv (\Sigma_0,\Gamma_0) \rightarrow (\Sigma_1,\Gamma_1)
\rightarrow \dots \rightarrow (\Sigma_k,\Gamma_k)$.
An execution history is said to be \emph{valid}, if all the spawned
transactions are fully executed, i.e. $\Gamma_0$ and $\Gamma_k$ are
both empty. A valid history furthermore requires $\Sigma_0$ to be
consistent with any user-defined constraints on the initial state of a database,
e.g. that certain tables are initially empty.
%

%
\input{tables/consistency_guarantees}

Following the literature on weak consistency and isolation
semantics~\cite{burckhardt14} we can specify
constraints on $\vis$ and $\ar$ relations which identify when system
states are consistent with a particular database's consistency guarantees.
\autoref{table:consistency_guarantees} presents some well-known instances
of such guarantees and their specifications, where the predicate $\st$ simply
relates effects created by queries from the {\bf s}ame {\bf
  t}ransaction instance.
System states can be thought of as finite models for these guarantees.
For example, a system state $\Sigma$ is said to be \emph{strictly serial} if
and only if $\ser$ is satisfied by $\Sigma$, denoted as
\mbox{$\Sigma\models\ser$}.
Since both $\vis$ and $\ar$ only grow during the execution of a
program, if the final system state satisfies a particular
specification, so will every system state in the execution
history. We correspondingly
extend this notion to execution histories and define a history $H$ to
conform to a guarantee~$\Psi$, denoted by $H\models\Psi$, if and only
if $\Psi$ is satisfied in the final system state of $H$.

Although strictly serial executions are easily comprehensible and
desirable, forcing the database to execute transactions one-by-one is
unnecessarily restrictive and must usually rely on poorly scalable pessimistic
locking mechanisms to enforce~\cite{pldi15}.
We instead define an execution history $H$ to be \emph{serializable}
if there exists another strictly serial history $H'$, that is
constructed by reordering execution steps of $H$, such that the final
set of effects in both histories are equivalent.

\input{figures/serializability_example}

For example, \autoref{fig:serializability_example} depicts the system states
from a history $H$~(left),
and another serial history~$H'$ (right) which is constructed by reordering
steps \ding{173} and \ding{174} of $H$.
Although $H$ is not strictly serial (since both \hyperref[rc]{$\rc$}
and \hyperref[rr]{$\rr$} are violated in $\Sigma_4$), its final set of
events, $\store_4$, is the same as $\store_4'$, the final set of events in $H'$,
modulo renaming of their implicit unique identifiers,
and is therefore~serializable.

Given an \ir\ program, our goal is to statically find non-serializable
execution histories that are consistent with the guarantees of the
underlying database, i.e. to detect any \emph{serializability
  anomalies} in the program.
As a first step, observe that execution histories that contain a
serializability anomaly can be decomposed into a serial execution
history followed by a (usually smaller) non-serializable history.
For instance, a non-serializable execution of ten transactions,
may be decomposed into eight serially executed transactions
which insert appropriate records into
initially empty tables, needed for the remaining two (concurrently executed)
transactions that actually manifest the serializability anomaly.
Following this observation, \tool\
systematically prunes the space of all possible interleavings of
transactions involved in lengthy anomalous executions
and instead searches for \emph{independent}
serializability anomalies, defined as two disjoint
histories $H_1\rightarrow H_2$, where $H_1$ is a valid serial execution
and $H_2$ is a non-serializable execution history.

%%%%%%%%%%%%%%%%%%%%%%%%%%%%%%%%%%%%%%%%%%%%%%%%%%%%%%%%%%%%%%%%%%%%%%%%%%%%%%%%%%%%%%%%%%%%%%%%%%%%

%
%
\subsection{Internal and External Serializability}
\label{subsec:internal}

During the course of our experiments, we observed that a subset of
serializability anomalies could be considered benign, in that they
always leave the final state of the database indistinguishable from
the state after some serializable execution of the same instances.
As an example, consider the executions $H$ and $H'$ presented in the
left and right of \autoref{fig:abstract_anomaly2}. $H$ consists of two
instances of the $\mathsf{upd}$ transaction, which reads the variable
$\mathsf{x}$ and then updates it to a fresh value.  Similarly, $H'$
includes two instances of the $\mathsf{inc}$ transaction which reads the
value of $\mathsf{x}$ and then \emph{increments} it by 10. Both of
these executions are non-serializable as they manifest the classic
\emph{lost update} anomaly~\cite{adyaphd}. However, only $H'$ leaves
the database in a state that is inconsistent with any serializable
execution. We dub the sort of benign anomalies seen in $H$
\emph{external serializability anomalies}.

\input{figures/abstract_anomaly2}

The key observation is that it is possible for some of the read effects in a
non-serializable execution to not impact a transaction's control
flow or any of later write effects.
As a result, those read events 
%(denoted by $\rddp{}$) 
could be
excluded from the correctness analyses, in order to focus on
\emph{internal serializability anomalies}. Such anomalies are more
harmful in that they leave the database state permanently diverged
from a serializable execution state and should be carefully studied
and addressed by developers.  We denote such \emph{unused} read
effects as~$\rddp{}$, which can be detected through a straightforward
analysis of the source~program. For these reasons, by default,
\tool\ only identifies internal serializabilty anomalies.

%%%%%%%%%%%%%%%%%%%%%%%%%%%%%%%%%%%%%%%%%%%%%%%%%%%%%%%%%%%%%%%%%%%%%%%%%%%%%%%%%%%%%%%%%%%%%%%%%%%%

%
%
\subsection{Dependency Cycles}
\label{subsec:dep}
Following the approach of~\citet{AD00}, we reduce the problem of
determining serializability of an execution history to the detection
of cycles in the \emph{dependency graph} of its final state.
To this end, we first define three \emph{dependency relations} over the set of
effects in an execution state:
\begin{enumerate*}[label=(\roman*)]
      \item Read dependency, $\folwr$, which relates two effects
        if one witnesses a value in a record's field that is
        written by the other.
      \item Write dependency, $\folww$, which relates two effects if one overwrites the
        value written by the other.
      \item Read anti-dependency, $\folrw$, which relates two effects if one witnesses
        a value in a field that is later overwritten by the other
        effect.
    \end{enumerate*}
For example, in \autoref{fig:abstract_anomaly2} (left) both dependency
relations~$\folww(\wrr{(x,f,1)},
\wrr{(x,f,2)})$ and $\folrw(\rddp{(x,f,0)},
\wrr{(x,f,2)})$ hold.

Recalling that serializability is defined by a reordering of database
operations, we lift the definition of the above dependency relations
(and previously defined $\vis$ and $\ar$) from effects to query
instances.
Given an execution history $H$ and for all relations $\folR\in\{\ar,\vis,\folrw,\folwr,\folww\}$,
if effects $\eta$ and $\eta'$ are created respectively by queries $q$ and $q'$ in $H$,
then $\folR(\eta,\eta')\Rightarrow\folR(q,q')$.
This lifting is necessary because in order to determine serializability of a
history $H$,
the execution order of queries in $H$ must be altered such that the new history
$H'$ is serial (see section \ref{subsec:serializability_anomalies}).
Intuitively, such reordering cannot not exist when some queries in $H$
are cyclically dependent to each other.
\input{figures/cycles}

We are now ready to formally define the dependency graph $G$ of a system state
$\Sigma$ as a directed multigraph whose nodes are query instances (that created
the effects in $\Sigma$) and
edges are in $\{\folrw,\folwr,\folww,\folst\}$.
The relation $\folst$ is simply defined to relate query instances
originated from the same transaction.
We further define a valid dependency cycle as a 2-regular subgraph of $G$, that
includes at least two dependency edges (i.e. $\folwr,\folrw,\folww$)
which are connected by an $\folst$ edge.
We do not allow dependency edges between query instances from the same
transaction and w.l.o.g assume they are replaced with $\folst$ edges.
\autoref{fig:cycles} presents schematic examples of valid and invalid cycle structures in
dependency~graphs.

We conclude this section by describing the connection between the dependency
graph of the final state in an execution to its serializability.
To this end, an execution history $(\Sigma_0,\Gamma_0)\rightarrow\dots\rightarrow(\Sigma_k,\Gamma_k)$
is serializable if and only if $\Sigma_k$'s dependency graph is acyclic~\cite{AD00}.
As we will explain in the next section, this connection enables \tool\ to
encode potential dependency cycles in a given \ir\ program 
into a decidable first-order formula and 
statically construct an execution history that manifests a serializability
anomaly.

%%%%%%%%%%%%%%%%%%%%%%%
%%%%%%%%%%%%%%%%%%%%%%%
\begin{comment}
%%%%%%%%%%%%%%%%%%%%%%%
%%%%%%%%%%%%%%%%%%%%%%%

We conclude this section by showing that our particular choice of
\hyperref[fig:rule_example]{operational semantics}
is permissive enough to produce any valid dependency cycle that 
an \ir\ program may contain.
%
In other words, given a \emph{test configuration} consisted of concurrent
execution schedules and concrete input arguments 
to transactions, \tool's replaying mechanism 
is capable of replaying all serializability anomalies induced by that schedule.

\begin{theorem}
      For any  execution state $S$ which manifests a valid dependency cycle $C$,
the necessary constraints imposed by $C$ do not violate causal visibility.
\proof \cite{Rahmani:Clotho:2019}
    \end{theorem}

  \end{comment}

%% file: tables/consistency_guarantees.tex
% settings
\setlength{\arrayrulewidth}{0.5pt}
\setlength{\tabcolsep}{5pt}

\def\arraystretch{1}
% table
\begin{table}[b!]
\centering
\begin{footnotesize}
{\rowcolors{1}{Main-Theme-Light}{Main-Theme-Lighter}
  \begin{tabular}{|l | l |} 
 \hline
 \rowcolor{Main-Theme-Dark}
 \multicolumn{1}{|c|}{Guarantee}
    & 
    \multicolumn{1}{c|}{Specification} 
   \\
 \hline
    Causal Visibility & $\cv \equiv \forall \eta_1\eta_2\eta_3.\;\vis(\eta_1,\eta_2)\wedge\vis(\eta_2,\eta_3)
    \Rightarrow \vis(\eta_1,\eta_3)$ \\
     Causal Consistency & $\cc \equiv \forall \eta_1\eta_2.\; \cv\wedge
     (\st(\eta_1,\eta_2) \Rightarrow \vis(\eta_1,\eta_2)\vee \vis(\eta_2,\eta_1))$ \\
     Read Committed  & $\rc  \equiv \forall
     \eta_1\eta_2\eta_3.\;\st(\eta_1,\eta_2) \label{rc}
     \wedge \vis(\eta_1,\eta_3) \Rightarrow
    \vis(\eta_2,\eta_3) $\\
     Repeatable Read &  $\rr  \equiv \forall \eta_1\eta_2\eta_3.\; \label{rr}
     \st(\eta_1,\eta_2) \wedge \vis(\eta_3,\eta_1) \Rightarrow
    \vis(\eta_3,\eta_2) $ \\
     Linearizable & $ \lin  \equiv \ar \subseteq \vis $ \\
    Strictly Serial & $ \ser \equiv \rc \wedge \rr \wedge \lin$ \\
    [0.3ex] 
 \hline
\end{tabular}
}
\end{footnotesize}
\vspace{2mm}
  \caption{Consistency and isolation guarantees}
\label{table:consistency_guarantees}
\end{table}

%% file: figures/serializability_example.tex
\begin{figure}[t]
  \begin{subfigure}[b]{.22\textwidth}
    \centering
    \includegraphics[scale=0.26]{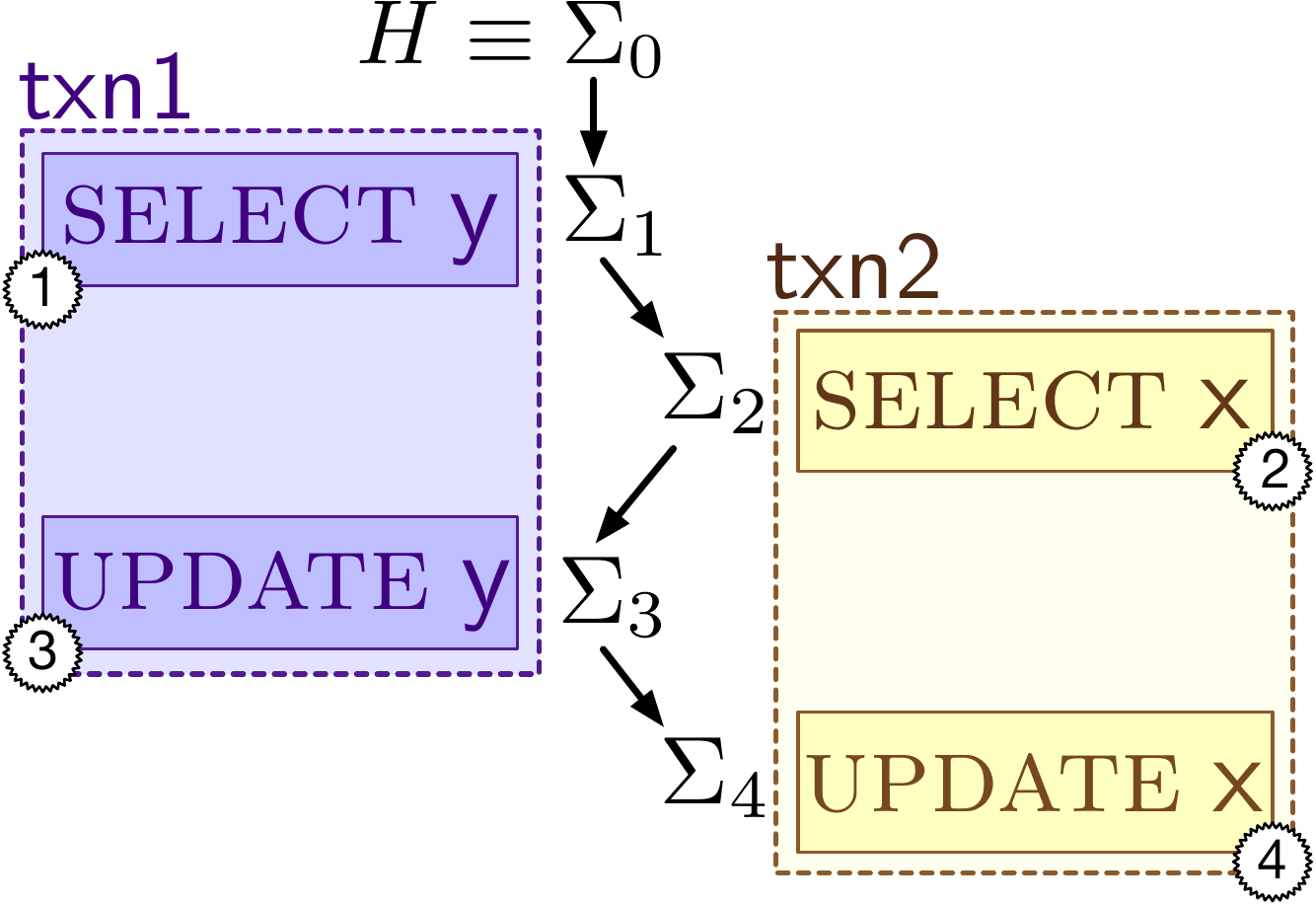}
  \end{subfigure}
  ~\;\;\;\;\;
  \begin{subfigure}[b]{.185\textwidth}
    $\scriptstyle \store_0= \{\wrr{(\mathsf{x},f,0)},$ \\[-1mm]
  $\hspace*{7.7mm} \scriptstyle \wrr{(\mathsf{y},f,0)}\}$ \\
  $\scriptstyle \store_1=\store_0\cup \{\rdd{(\mathsf{y},f,0)}\}$ \\
  $\scriptstyle \store_2=\store_1\cup \{\rdd{(\mathsf{x},f,0)}\}$\\
  $\scriptstyle \store_3=\store_2\cup \{\wrr{(\mathsf{y},f,1)}\}$\\
  $\scriptstyle \store_4=\store_3\cup \{\wrr{(\mathsf{x},f,1)}\}$
  \end{subfigure}
  ~\;\;\; \vline\vline \;\;\;\;
  \begin{subfigure}[b]{.22\textwidth}
    \centering
    \includegraphics[scale=0.26]{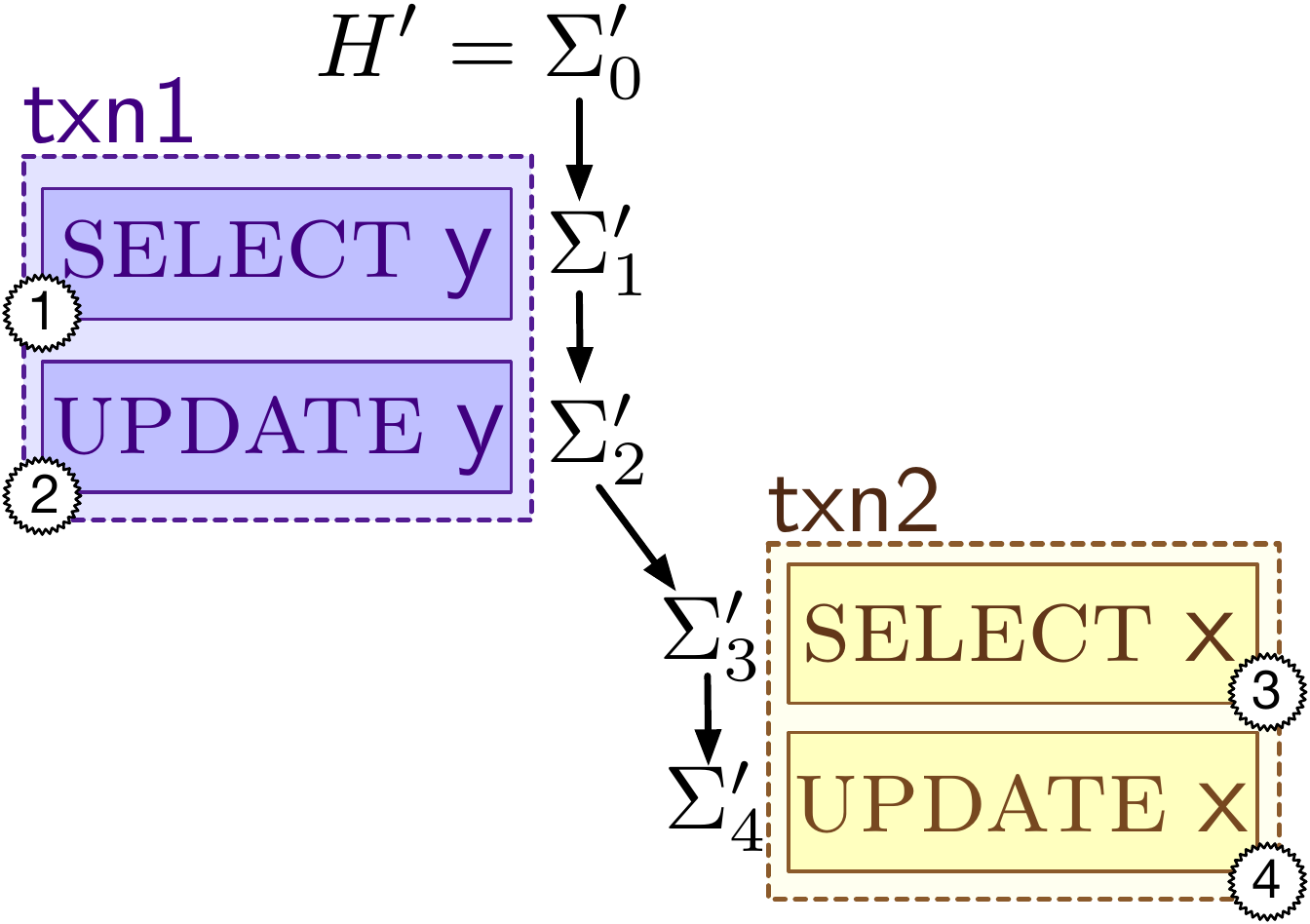}
  \end{subfigure}
  ~\;\;\;\;\;
  \begin{subfigure}[b]{.185\textwidth}
  $\scriptstyle \store_0'= \{\wrr{(\mathsf{x},f,0)},$\\[-1mm]
  $\hspace*{7.7mm}\scriptstyle \wrr{(\mathsf{y},f,0)}\}$ \\
  $\scriptstyle \store_1'=\store_0'\cup \{\rdd{(\mathsf{y},f,0)}\}$ \\
  $\scriptstyle \store_2'=\store_1'\cup \{\wrr{(\mathsf{y},f,1)}\}$\\
  $\scriptstyle \store_3'=\store_2'\cup \{\rdd{(\mathsf{x},f,0)}\}$\\
  $\scriptstyle \store_4'=\store_3'\cup \{\wrr{(\mathsf{x},f,1)}\}$
  \end{subfigure}

  \caption{A serializable execution (left) and its equivalent serial execution
  (right)}
    \label{fig:serializability_example}
  \end{figure}

%% file: figures/abstract_anomaly2.tex
\begin{figure}[t]
%\begin{mdframed}
  \begin{subfigure}[c]{.22\textwidth}
    \centering
    \includegraphics[scale=0.225]{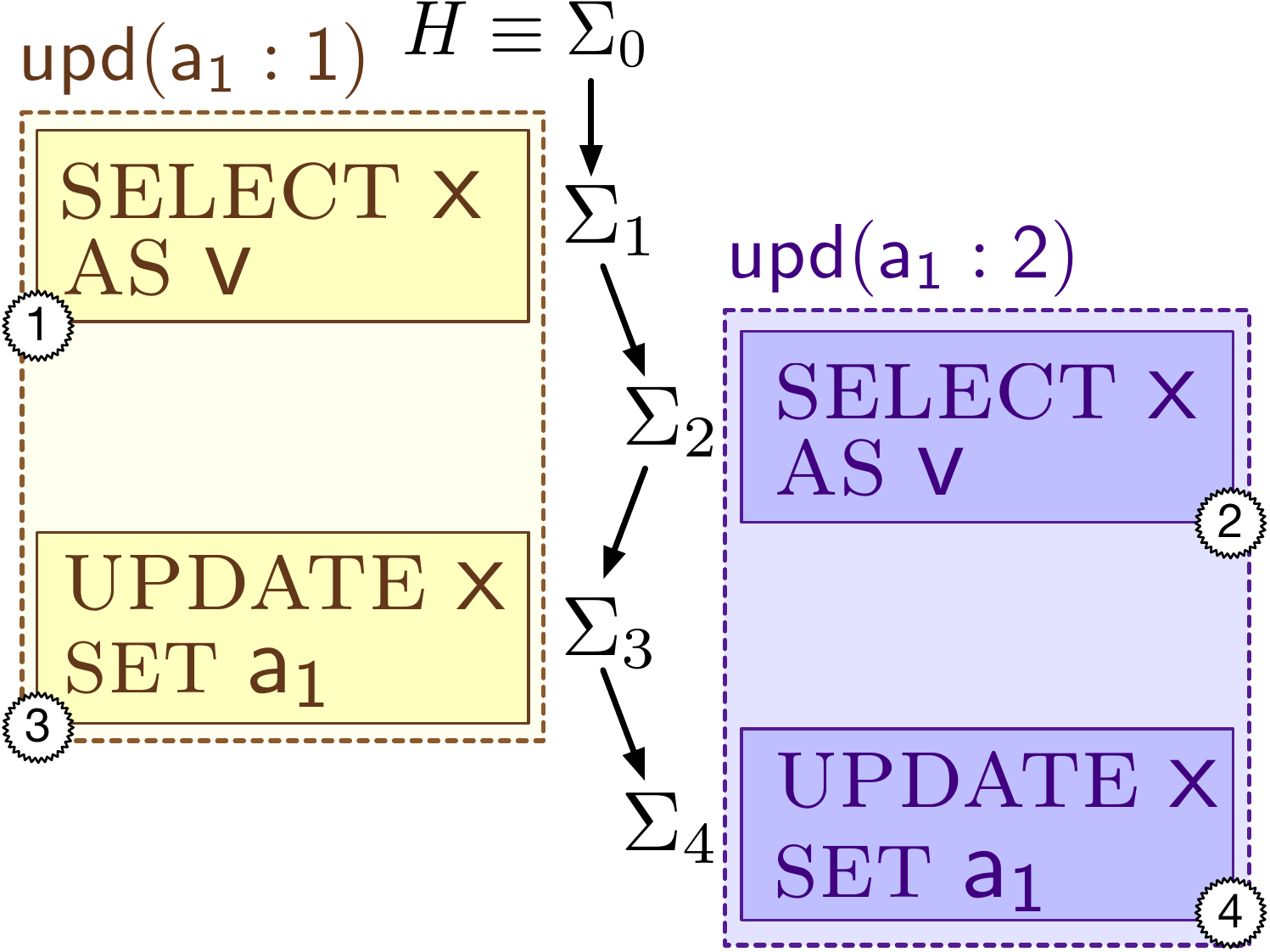}
  \end{subfigure}
  ~\;\;\;\;\;
  \begin{subfigure}[c]{.195\textwidth}
    $\scriptstyle \store_0= \{\wrr{(\mathsf{x},f,0)}\}$ 
  \\
  $\scriptstyle \store_1=\store_0\cup \{\rddp{(\mathsf{x},f,0)}\}$ \\
  $\scriptstyle \store_2=\store_1\cup \{\rddp{(\mathsf{x},f,0)}\}$\\
  $\scriptstyle \store_3=\store_2\cup \{\wrr{(\mathsf{x},f,1)}\}$\\
  $\scriptstyle \store_4=\store_3\cup \{\wrr{(\mathsf{x},f,2)}\}$
  \end{subfigure}
  ~\;\;\; \vline\vline \;\;\;\;\;
  \begin{subfigure}[c]{.225\textwidth}
    \centering
    \includegraphics[scale=0.225]{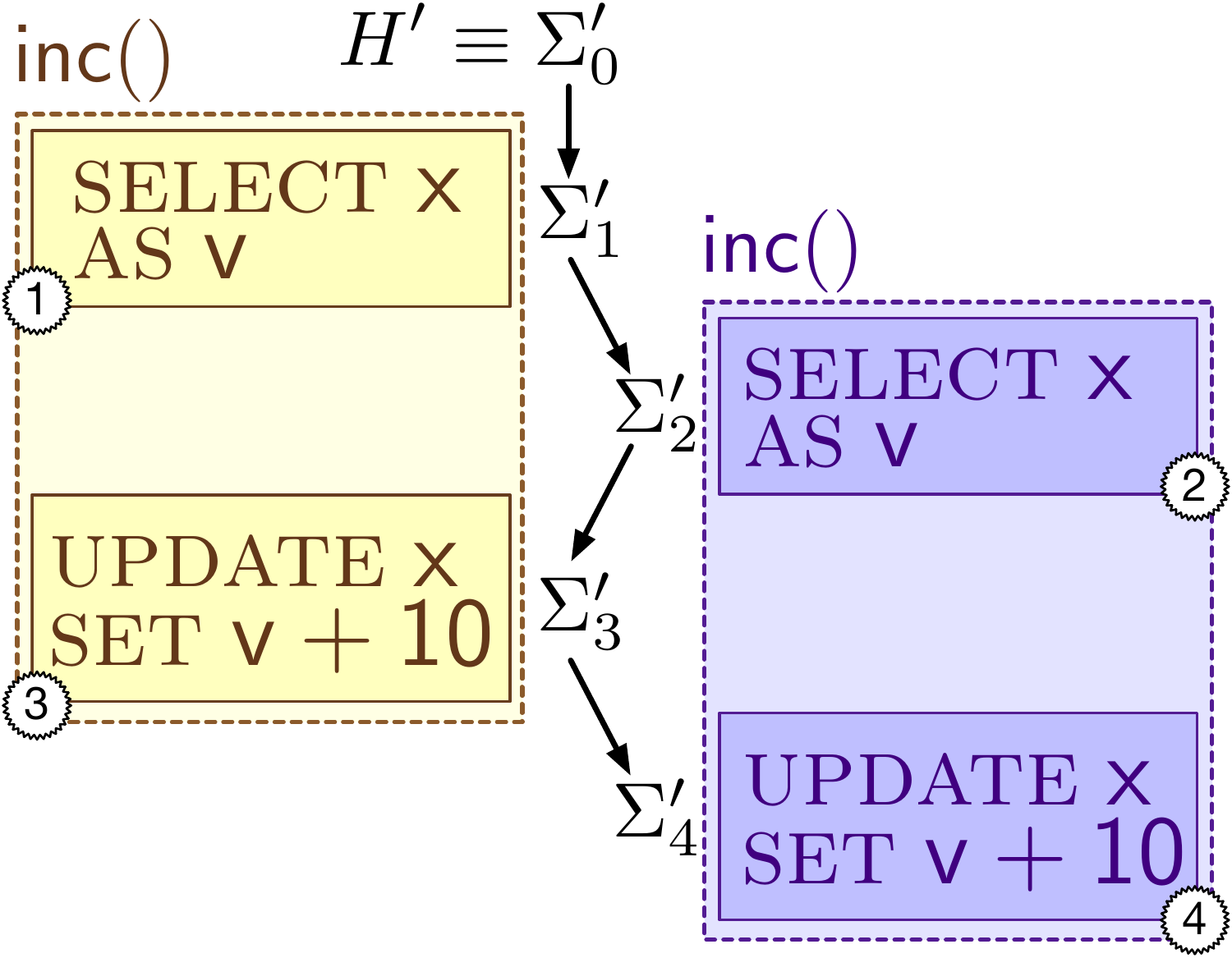}
  \end{subfigure}
  ~\;\;\;\;
  \begin{subfigure}[c]{.195\textwidth}
    $\scriptstyle \bar{S}_0'= \{\wrr{(\mathsf{x},f,0)}\}$ \\
  $\scriptstyle \store_1'=\store_0'\cup \{\rdd{(\mathsf{x},f,0)}\}$ \\
  $\scriptstyle \store_2'=\store_1'\cup \{\rdd{(\mathsf{x},f,0)}\}$\\
  $\scriptstyle \store_3'=\store_2'\cup \{\wrr{(\mathsf{x},f,10)}\}$\\
  $\scriptstyle \store_4'=\store_3'\cup \{\wrr{(\mathsf{x},f,10)}\}$
  \end{subfigure}

  \caption{External (left) and internal (right)  lost update anomalies.}
    \label{fig:abstract_anomaly2}
  \end{figure}

%% file: figures/cycles.tex
\begin{figure}[t]
%\begin{mdframed}
  \begin{subfigure}[b]{.23\textwidth}
    \centering
    \includegraphics[scale=0.245]{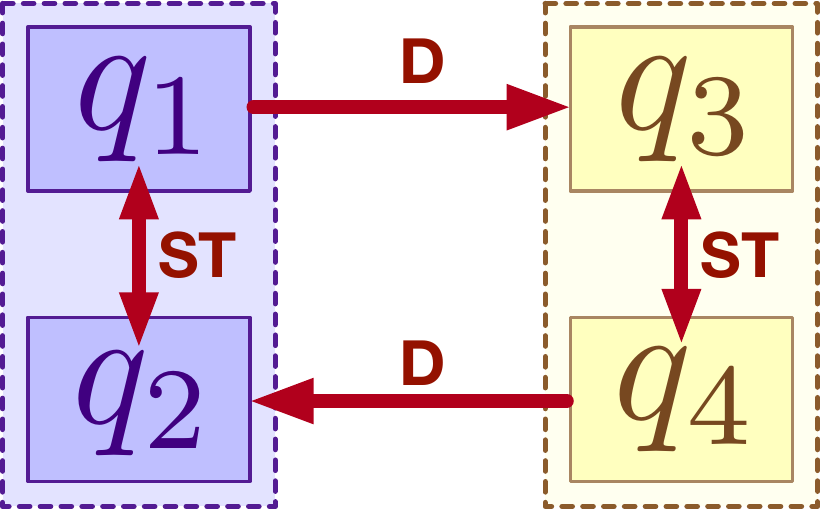}
    \caption{}
  \end{subfigure}
  ~
  \begin{subfigure}[b]{.23\textwidth}
    \centering
    \includegraphics[scale=0.245]{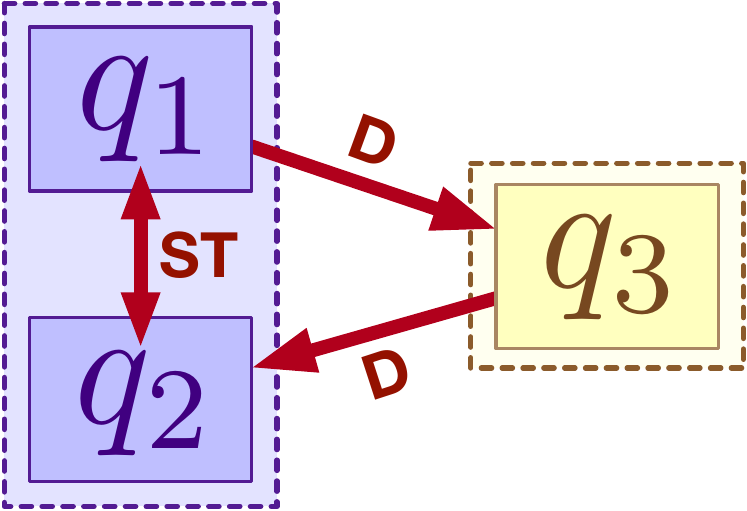}
    \caption{}
  \end{subfigure}
  ~
  \begin{subfigure}[b]{.23\textwidth}
    \centering
    \includegraphics[scale=0.245]{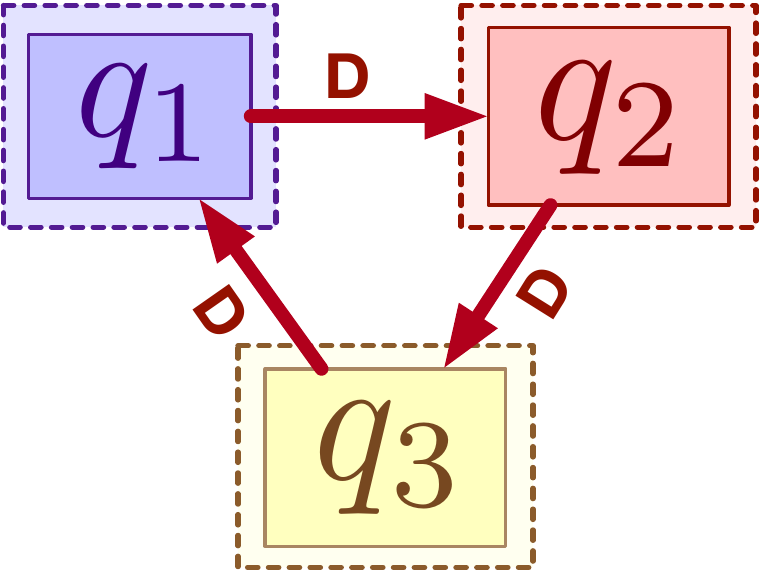}
    \caption{}
  \end{subfigure}
  ~
  \begin{subfigure}[b]{.23\textwidth}
    \centering
    \includegraphics[scale=0.245]{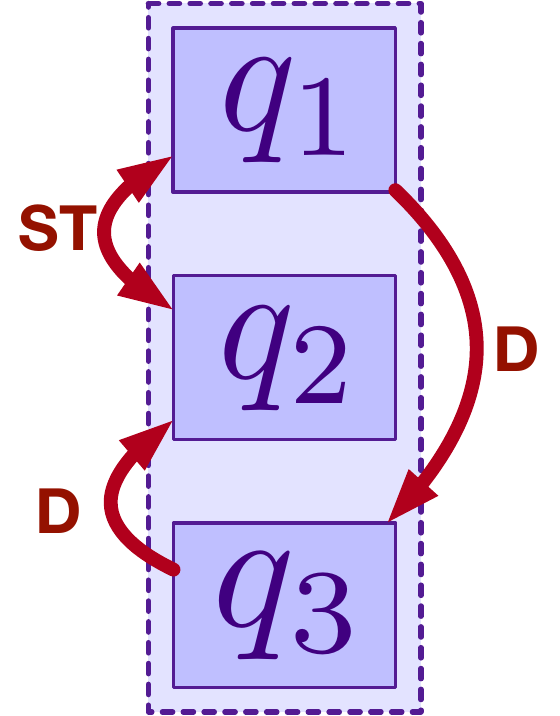}
    \caption{}
  \end{subfigure}
  \caption{Valid (a,b) and invalid (c,d) cycles where
  $\folD\in\{\folwr,\folrw,\folww\}$}
    \label{fig:cycles}
  \end{figure}

%% file: sections/encoding.tex
We now turn to the question of how to statically detect dependency cycles and construct
independent serializability anomalies in \ir\ programs.
We adopt the approach presented in~\cite{KR18} and~\cite{BR18}
and reduce (the bounded version of) our problem to
checking the satisfiability of an FOL formula
$\varphi_{\mathbb{C}}$, constructed from a given \ir\ program.
This formula includes variables for each of the dependency, visibility, and arbitration constraints
that can appear during the program's execution, and is designed such that the assignments to
these variables in any satisfying model can be used
to reconstruct an anomalous execution of the original program.
This allows us to use an off-the-shelf SMT solver to efficiently check for
anomalies in the given program.
Given bounds $max_p$, $max_t$ and $max_c$,
the shape of the full formula is a conjunction of five clauses,
each encoding a different aspect of the program:
\begin{equation}
\varphi_{\mathbb{C}}\overset{\Delta}{\equiv}
\folphi{context}
\wedge
\folphi{db}
\wedge
\folphi{dep$\rightarrow$}
\wedge \folphi{$\rightarrow$dep}
\wedge
\bigvee_{\substack{0\leq i\leq max_p\\2\leq j\leq max_t\\3\leq k\leq max_c}}
\!\!\!\!\!\folphi{anomaly}^{i,j,k}
\label{eq:C}
\end{equation}

In the above formula, $\folphi{context}$ represents a set of constraints on
variables and functions,
which ensures that a satisfying assignment corresponds to a
valid execution of any database program.
Clause $\folphi{DB}$ enforces a set of user-defined validity constraints on database records.
Clauses
$\folphi{dep$\rightarrow$}$ and $\folphi{$\rightarrow$dep}$
capture the necessary and sufficient conditions for establishing dependency
relations between queries in the given program.

\input{figures/anomaly}

Following our discussion of execution histories which manifest an independent
serializability anomaly in
section \ref{subsec:serializability_anomalies}, $\folphi{anomaly}^{i,j,k}$ forces the solver
to instantiate a serial execution of $i$ transactions that lead to
concurrent execution of $j$ transactions within which
a dependency cycle of length $k$ is formed.
\autoref{fig:anomaly} presents an example of such an instantiation where $i=3$, $j=2$ and $k=4$.
The remainder of this section lays the formal groundwork for the encoding and explains each of
the above clauses in more detail.
The complete details of the encoding can be found in the extended version of
this paper~\cite{Rahmani:Clotho:2019}.

\subsection{Components of FOL Encoding}
We assume uninterpreted sorts for
transactions ($\txnsort$), queries ($\opsort$),
partitions ($\partsort$), and records ($\rowsort$).
We define functions ($\mathsf{txn}:\opsort\rightarrow \txnsort$) and
(${\tau}:\opsort\rightarrow \partsort$)
to relate a query to the transaction instance
calling it and the partition it is executed on.
We use an auxiliary function
$\folalive{}{}:\rowsort\times \opsort\rightarrow \booll$
to capture the liveness of a record in a particular query instance.
Predicates $\vis$, $\ar$, $\folst$, $\folwr$, $\folrw$ and $\folww$,
all with the signature $\opsort\times\opsort\rightarrow\booll$ are also defined to
capture relations discussed in \autoref{sec:model}.
The disjunction of dependency relations $\folww$, $\folwr$ and $\folrw$ is
denoted by $\folD$.

\paragraph{\normalfont $\folphi{anomaly}$}

This clause specifies that an independent serializability
anomaly exists in the program, that consistes of
$i$ serially executed transactions
and
$j$ concurrent transactions that manifest a dependency cycle of
length $k$.
It is also required that
all serializable transactions are executed before
the transactions involved in the cycle.

$$\folphi{anomaly}^{i,j,k} \overset{\Delta}{\equiv}
\exists t_1,...,t_{i}.\!\!\!\!\bigwedge_{1\leq m\leq i}\!\!\!\folphi{ser}(t_m)
\wedge
\exists t'_{1},...,t'_{j}.\;
\folphi{cycle}^{k}(t'_1,...,t'_{j})
\wedge
\bigwedge_{\substack
  {1\leq m \leq i
    \\
   1\leq n\leq j}
 }
\folphi{order}(t_m,t'_n)
$$

In the above formula,
$\folphi{ser}(t)$
encodes the serializability of $t$ with respect to all other transaction
instances and
$\folphi{order}(t,t')$ enforces that all queries from $t$ are arbitrated before
the queries of $t'$.
Proposition $\folphi{cycle}^{k}(t_1,...,t_{j})$ forces the solver
to
instantiate $k$ queries within the
given transactions,
such that a cycle of length $k$ exists between
them,
at least two of its edges are dependency relations in $\{\folwr,\folrw,\folww\}$
and
one is an $\folst$ relation.
The rest of the edges can be of either kind.
\color{black}
  \begin{align*}
\folphi{cycle}^{k}(t_1,...,t_{j})
 \overset{\Delta}{\equiv}\;
 &
\exists q_1,...,q_{k}. \!\!
\bigwedge_{1\leq m\leq k}\!\!
\txnR(q_m)\in\{t_1,...,t_{j}\}
\,\wedge
\\[-2mm]
 &
 \folD(q_1,q_2)\wedge \folD(q_{k-1},q_{k})
 \wedge
 \folst(q_1,q_{k})
 \wedge
\!\bigwedge_{2\leq
m<{k-1}}\!(\folst(q_m,q_{m+1})\vee\folD(q_m,q_{m+1}))
\end{align*}

\autoref{fig:abstract_anomaly} presents concrete examples of dependency cycles detected
by \tool.
Figure \ref{fig:abstract_anomaly_1} depicts a scenario where the
$\sql{select}$query from
$\txn{oneRead}$ transaction witnesses the intermediate value written by
$\txn{twoWrites}$ transaction. \tool\ automatically determines the
execution order of queries (shown as circled numbers)
and values of input arguments required for manifestation of the anomaly. 
In this example, it suffices that all queries access the record with the same
key, i.e. $\mathsf{k_1=k_2=k_3}$.
The anomaly depicted in
figure \ref{fig:abstract_anomaly_2} occurs when an update
from transaction $\txn{twoWrites}'$ is executed at a node which
then becomes disconnected from the network to form partition 1.
Assuming that clients can access the rest of the network on
partition 2, the remaining queries need to be
submitted to available nodes in the specified order in order to
trigger the depicted serializability anomaly.
This anomaly does not occur when network partitioning is not possible.
Figures \ref{fig:abstract_anomaly_3} and \ref{fig:abstract_anomaly_4}
respectively present the
cycles detected for external and internal lost update anomaly from the example discussed in
section \ref{subsec:internal}.
Predicate $\folstp$ is defined similarly to $\folst$ but also requires that
no data dependency exists between the predicated query instances.
\tool\ users may optionally choose to
only detect internal serializability anomalies in their programs,
in which case $\folst$ will be replaced with $\folstp$ in $\folphi{cycle}$.

\input{figures/abstract_anomaly}
\paragraph{\normalfont$\folphi{context}$}
This clause ensures that a satisfying assignment corresponds to a
valid execution of database programs. It uses a collection of
functions $\mathtt{init}^f\!:\!\rowsort\rightarrow\intt$ and
$\folval^f\!:\!\opsort\times\rowsort\rightarrow \intt$ which are
defined for all fields $f$ in the given schema.  These functions are
used to identify the initial values of records and their values at
each query operation, respectively.
These functions are constrained so that the values read by operations
are from the initial state if the fields have not yet been written to. We
make this explicit via the following formula:
\begin{equation}
%  \tag{$\folphi{context}$}
  \bigwedge_{q_1,q_2 \in \opsort} \;\bigwedge_{r \in \rowsort} \neg\folwr(q_2,q_1)  \Rightarrow
  \folval^f(q_1,r)=\mathtt{init}^f(r)
  \label{eqn:init}
\end{equation}

Next, $\folphi{context}$ constrains the values of all records stored
in a variable $v$ by a $\sql{select}$query
to satisfy the
$\sql{where}$condition of that query.

The next set of restrictions encoded in $\folphi{context}$ constrains
the fields of records involved in a dependency cycle in order to make
the corresponding anomaly more understandable.
As a concrete example, consider the $\mathsf{twoWrites}$ transaction in
figure \ref{fig:abstract_anomaly_1}.
Without explicit constraints on the updated values in this
transaction, a SMT solver could trivially instantiate parameters
$\mathsf{a_1=a_2=0}$, making the dirty value read by $\mathsf{oneRead}$
transaction the same as the value read in a serializable execution of
the two transactions.  This could falsely appear benign to
developers, so $\folphi{context}$ constrains the
values of records to avoid such scenarios.
$\folphi{context}$ forces the values of field $f$ used in queries
$q_1$ and $q_2$ to be \textbf {identical} if there is a read
dependency between them, i.e. $\folwr(q_1,q_2)$ holds.  Similarly, if
there is an anti-dependency ($\folrw$) or a write dependency
($\folww$) between two queries, the values of any fields written or
read by those queries are forced to be \textbf {different}.

Finally, we constrain the visibility relation to capture how effects in a
partition are witnessed by any later query executed in that partition, i.e. we
enforce \emph{causal visibility within each partition}: 
\begin{equation}
%   \tag{$\folphi{context}$}
  \forall q_1,q_2.\;\ar(q_1,q_2)\wedge
  ({\tau}(q_1)={\tau}(q_2))\Rightarrow \vis(q_1,q_2)
\end{equation}

\paragraph{\normalfont $\folphi{db}$}
This clause enforces any consistency or isolation guarantees
(e.g. those from \autoref{table:consistency_guarantees}) provided by
the database under test, as well as any user-defined constraints on
record instances, e.g. requiring a table to be initially empty or
requiring $age$ field to be always greater than
21.

%%
%%
%% DEP->
%%
\paragraph{$\normalfont \folphi{dep$\rightarrow$}$}

This clause ensures that every database operation captures the effects of
any query it is related to by the dependency relation $\folD$. This is
accomplished via a proposition,
$\mu^{\folD\rightarrow}_{\stx{q,q'}}$, which asserts that
\begin{enumerate*}[label=(\alph*)]
\item there exists a concrete record instance that both queries access and
\item both queries are reached by the control flow of their transactions.
\end{enumerate*}

To this end, we first introduce function $\folB{}{t}:\phi\rightarrow \booll$
which constructs an FOL formula corresponding a boolean expression in~\ir, alongside any
additional conditions that must be satisfied for a successful construction.
When the input is a $\sql{where}$condition of a query, a record instance (that
must be checked if satisfies the condition) is also
passed to the function as an extra argument, e.g. $\folB{}{t,r}$.
We also assume a function $\Lambda: \opsort\rightarrow \phi$ which
returns the conjunction of all conditionals which must be satisfied in
order for a query to be reached by the program's control flow.

\input{figures/dep_then_rule_example}

\autoref{fig:dep_then_rule_example} gives a rule defining a formula that
enforces these conditions when an anti-dependency relation
$\folrw(\stx{q,q'})$ is established between a $\sql{select}$query $\stx{q}$ and
an $\sql{update}$query $\stx{q'}$ which access the same field
$\stx{f}$.
The transactions containing the two query are distinct.
The rule enforces that
there is a record which satisfies the $\sql{WHERE}$clause of both
queries
and
%($\exists r. \B{\phi}{t,r} \wedge~\B{\phi'}{t',r}$) and
is perceived alive by both of them.
%($\folalive{_{rd}}(r,q)\wedge\folalive{_{rd}}(r,q')$).
Both queries are also forced to be reachable in their corresponding
transactions.
%($\B{\Lambda(q)}{t}\wedge \B{\Lambda(q')}{t'}$).
Assuming the full set of rules defining $\mu^{\folD\rightarrow}_{\stx{}}$ for a
given program,
the $\folphi{dep$\rightarrow$}$ clause is simply defined as their conjunction:
$$\folphi{dep$\rightarrow$} \overset{\Delta}{\equiv}
\bigwedge_{\folD\in\{\folwr,\folrw,\folww\}}\bigwedge_{q,q'\in\opsort}
(\folD(q,q')\Rightarrow\mu^{\folD\rightarrow}_{q,q'})$$

%%
%% -> DEP
%%
%%
%%
\paragraph{$\folphi{$\rightarrow$dep}$}
\input{figures/delivery}
Examining the values of
$\ar$, $\tau$, $\mathtt{init}$ and $\folval$
in a satisfying assignment to $\varphi_{\mathbb{C}}$, gives us enough information to
recover a concrete execution of all the queries involved in a dependency cycle.
This suffices for constructing valid test cases for all the
examples discussed so far, all of which consist of transactions
that only include queries involved in the dependency cycle.
In practice, though, we observed that for the majority of detected
anomalies, transactions include operations which occur before the
start of the cycle.  Such operations may cause the database's
pre-cycle state to diverge from the state built by the conjunction of
the clauses previously described, preventing \tool\ from constructing
a valid test configuration that realizes the anomaly.

As a concrete example, consider the $\txn{delivery}$ transaction from the TPC-C benchmark,
where records from the tables $\mathsf{n\_order}$ and $\mathsf{cust}$ are retrieved
and updated.
As presented in \autoref{fig:delivery}~(left),
a lost update anomaly on $\mathsf{cust}$ table may occur if two instances of
$\txn{delivery}$ transaction concurrently select and update the same
$\mathsf{cust}$ record (depicted by red $\folrw$ edges).
Note that, the primary key for the $\mathsf{cust}$ record involved in this
anomaly, is determined based
on the $\mathsf{n\_order}$ records previously retrieved during steps \ding{172} and \ding{174}.

Mapping this cycle to a non-serializable execution requires
finding transaction arguments that force the retrieval of $\mathsf{n\_order}$
records at steps \ding{172} and \ding{174} that refer to the same
$\mathsf{cust}$ record.
Following proposition~(\ref{eqn:init}), 
our encoding thus far constrains the values of the
$\mathsf{n\_order}$ records  (including values in the $alive$ field) retrieved by
the queries at \ding{172}
and~\ding{174} to be the same.
Both queries will witness the initial state of the records, 
since there does not exist any $\folwr$ pointing at them.
As a result, a satisfying assignment could select the same value as
the argument to both transactions, i.e.~$\mathsf{a_1=a_2}$.
When replaying of this execution according to the specified arbitration
orders is attempted, however,
the $\mathsf{n\_order}$ record retrieved at step \ding{172} is deleted at
step \ding{173} and cannot be retrieved at step \ding{174}.
Consequently, the execution will either terminate at step \ding{174}
or fail to form the desired dependency cycle,
since two different $\mathsf{cust}$ records would be selected in steps \ding{176} and
\ding{177}.
%\BD{Might want to remind readers that ST edges exist between all the
%  operations in a transaction. }

The $\folphi{$\rightarrow$dep}$ clause remedies this problem by
forcing $\folwr$ edges to exist between
operations that are not part of the dependency cycle.
This clause forces the
$\folwr$ edge depicted by the green arrow in
\autoref{fig:delivery} (right) to be established.
When combined with the previous arbitration ordering, this
prevents the operation at step \ding{174} from selecting the same $\mathsf{n\_order}$
record as the one deleted at \ding{173} and eliminates 
the spurious dependency cycle between steps \ding{176}, \ding{177}, \ding{178} and
\ding{179}.
A satisfying assignment must now ensure that \emph{two}
records exist in the $\mathsf{n\_order}$ table,
both of which refer to the same $\mathsf{cust}$~record,
in order to subsequently manifest a valid 
lost update anomaly on that record.

The clause
$\folphi{$\rightarrow$dep}$,
using  predicates
$\mu^{\rightarrow\folD}_{\stx{q,q'}}$, defines
sufficient conditions under which a dependency relation $\folD$ between $q$ and $q'$
must exist.
\autoref{fig:then_dep_rule_example} presents the rule that defines the
predicate to force a $\folwr$ dependency when $\stx{q}$ is a
$\sql{select}$query and $\stx{q'}$ is an $\sql{update}$such that:
\begin{enumerate*}
  \item \color{dark-red} the update query is visible to the select query,
  \item \color{dark-green} both queries access the same alive row, \color{black} and
  \item  \color{dark-yellow} both $q$ and $q'$ will be executed.
\end{enumerate*}

\input{figures/then_dep_rule_example}

Using the complete set of rules for $\mu^{\rightarrow\folD}$,
we can give the full definition of the clause that forces the solver to
establish dependency edges also between operations outside of a cycle:
$$\folphi{$\rightarrow$dep} \overset{\Delta}{\equiv}
\bigwedge_{\folD\in\{\folwr,\folrw,\folww\}}\bigwedge_{q_1,q_2\in\opsort}
(\mu^{\rightarrow\folD}_{q_1,q_2} \Rightarrow \folD(q_1,q_2))
$$

We describe refinements to this encoding scheme in the following
section that allow \tool\ to manifest serializability anomalies for
realistic database programs, as evidenced by our experimental results.
These results support our contention that our encoding is sufficiently
precise to map abstract executions to realizable non-serializable
concrete ones.

%% We conclude this section with the following theorem which connects
%% satisfying assignments to $\varphi_{\mathbb{C}}$ to non-serializable
%% execution histories. 

%% \begin{theorem}
%%   Any satisfying assignment to the formula $\varphi_{\mathbb{C}}$,
%%   corresponds to a concrete test configuration (transaction instances,
%%   input arguments, execution schedule and partitioning) manifesting
%%   an independent serializability anomaly.
%% \end{theorem}

%
%
%

%% file: figures/anomaly.tex
\begin{wrapfigure}[6]{r}{0.47\textwidth}
  \vspace{-4.5mm}
  \centering
  \includegraphics[width=0.43\textwidth]{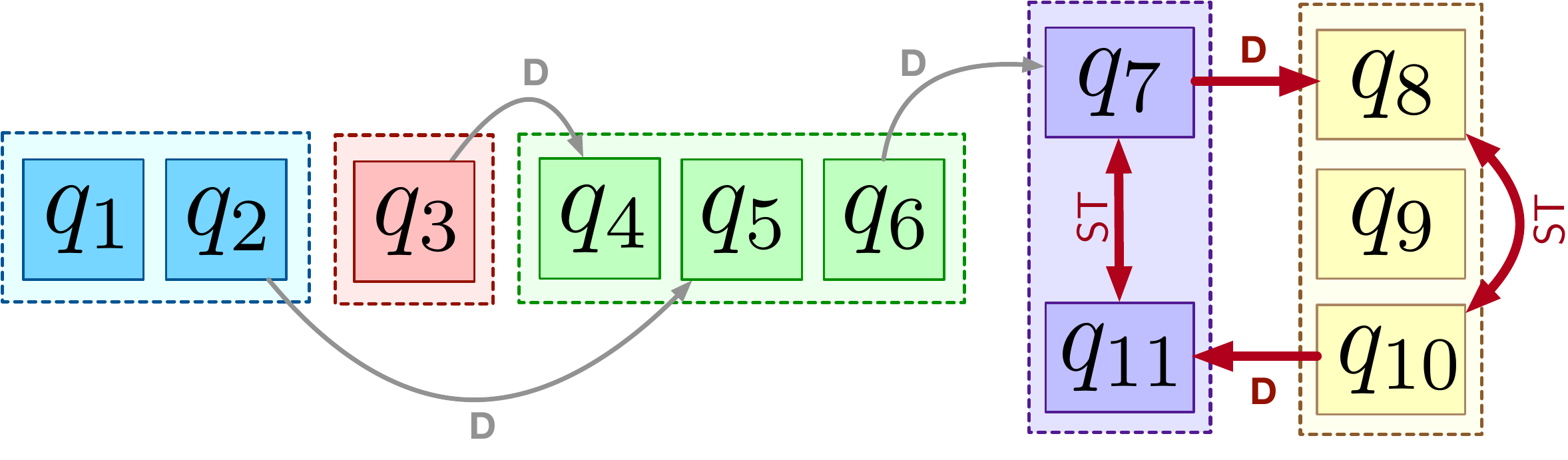}
  \vspace{-0mm}
  \caption{An anomaly constructed by \tool}
  \label{fig:anomaly}
  %\vspace{-1mm}
\end{wrapfigure}

%% file: figures/abstract_anomaly.tex
\begin{figure}[t]
  \begin{minipage}{.96\textwidth}
\begin{mdframed}[
  backgroundcolor=light-gray]
  \begin{subfigure}[b]{.49\textwidth}
    \centering
\includegraphics[scale=0.25]{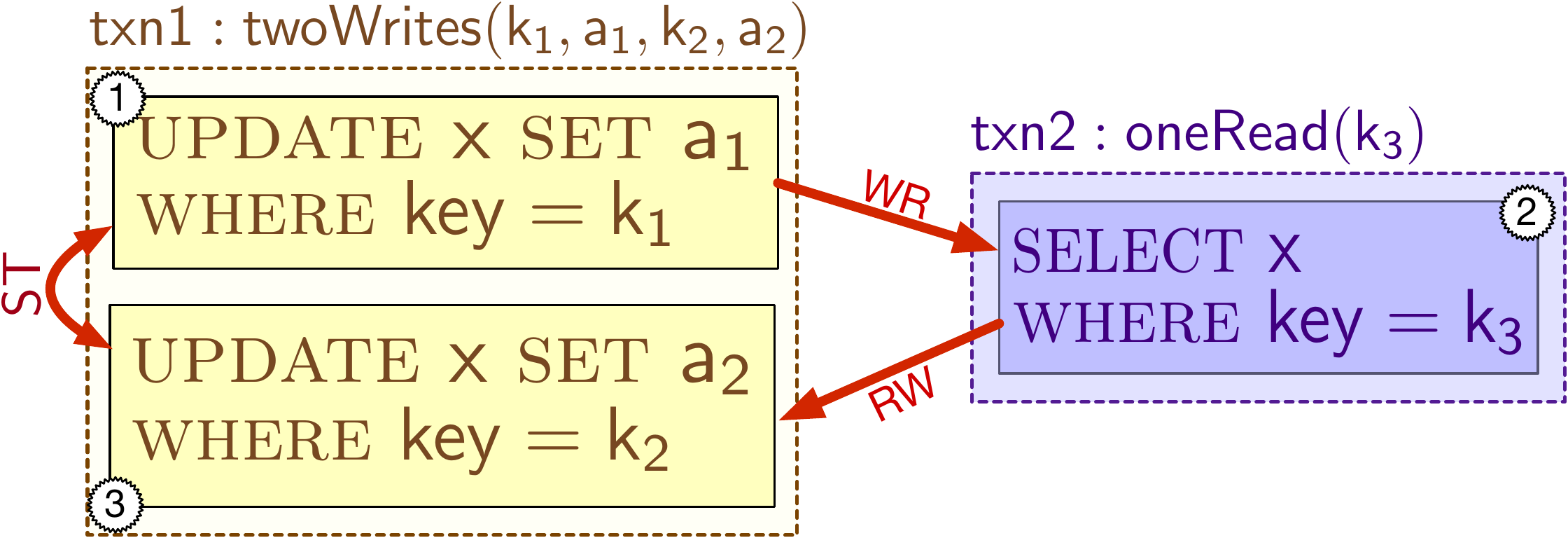}
\caption{Dirty Read}
\label{fig:abstract_anomaly_1}
  \end{subfigure}
  ~
  \begin{subfigure}[b]{.49\textwidth}
    \centering
\includegraphics[scale=0.3]{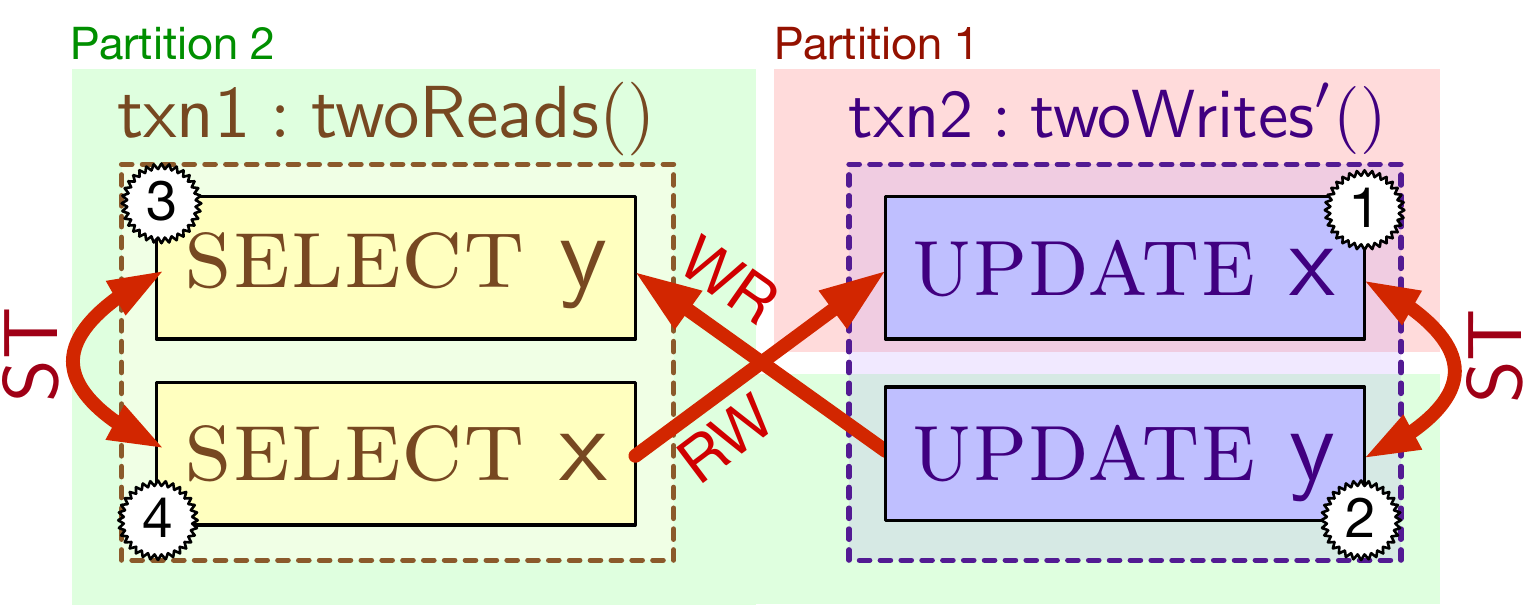}
\caption{Network Anomaly}
\label{fig:abstract_anomaly_2}
  \end{subfigure}
\begin{subfigure}{.49\textwidth}
  \vspace{1mm}
    \centering
\includegraphics[scale=0.26]{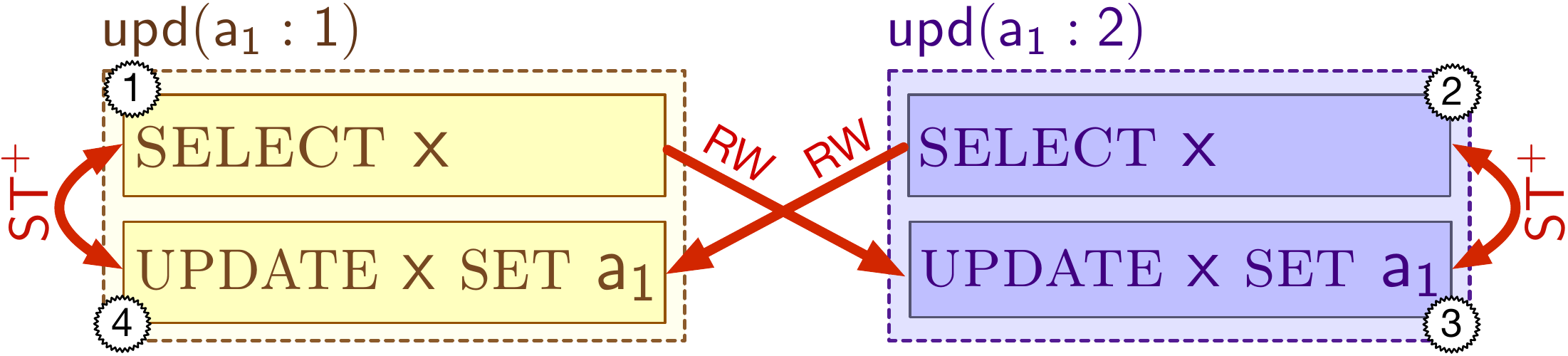}
\caption{External Lost Update}
\label{fig:abstract_anomaly_3}
  \end{subfigure}
  ~
  \begin{subfigure}{.49\textwidth}
  \vspace{1mm}
    \centering
\includegraphics[scale=0.26]{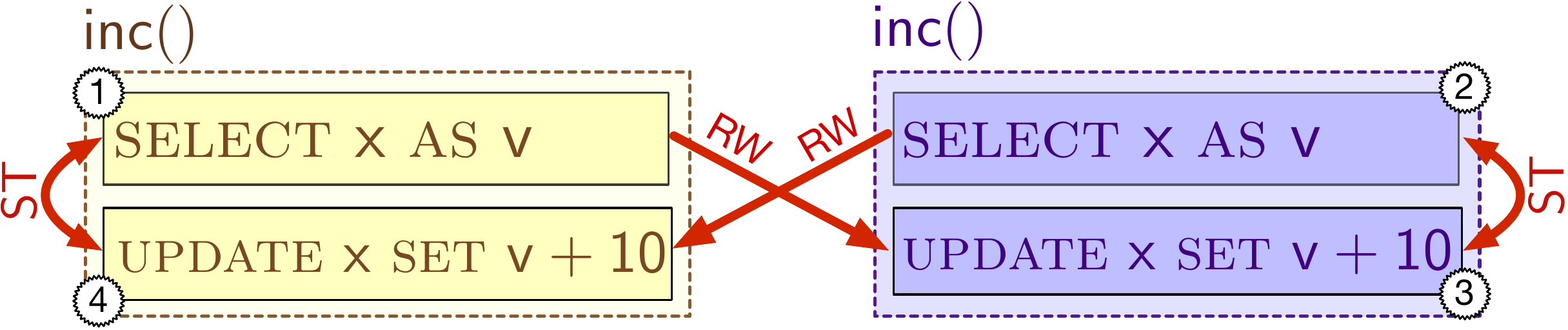}
\caption{Internal Lost Update}
\label{fig:abstract_anomaly_4}
  \end{subfigure}

 \end{mdframed}
\caption{Examples of dependency cycles generated by \tool}
\label{fig:abstract_anomaly}
\end{minipage}
\end{figure}

%% file: figures/dep_then_rule_example.tex
  \begin{figure}[h]%{0.46\textwidth}
  \begin{minipage}[b]{.6\textwidth} 
    \begin{mdframed}[backgroundcolor=Main-Theme-Complement]
\vspace{0mm}
  \begin{flushleft}
    \hspace{-2mm}\fcolorbox{black}{Main-Theme-Complement-Dark}{\scriptsize
    \textrm{\sc{\color{black} rw-select-update}}}
  \end{flushleft}
  \vspace{-6mm}
$$
\FOLRULE{
  \stx{q \equiv \sql{select}f\; \sql{as}
  x\;\sql{where} \phi } \\
  \stx{q' \equiv \sql{update}\sql{set}
  f=v\;\sql{where} \phi'} \\
  \stx{
  \folparent(q)=t \qquad \folparent(q')=t'
  \qquad t\not\eq t'
  %\qquad 
  %\V{v}{t'}=(\foldcond,\_\,)
  }
  \qquad
}
%% ------------------------------------------------------------------------------------------------
  {
  \mu^{\folrw\rightarrow}_{q,q'} =
  \stx{
    \color{dark-red}
    \exists r.\;
    \folB{\phi}{t,r
    %,\folversion{rd}(r,c_1)
    } 
  \wedge \folB{\phi'}{t',r
  %,\folversion{rd}(r,c_2)
  }\color{black}
  \wedge 
  \color{dark-green}
  \folalive{}(r,q)
  \wedge  
  }
  }
  {
    \stx{
      \color{dark-green}
      \folalive{}(r,q')
  %\wedge 
  %(\fol{\rho}_{r}(f,\folversionsub{wr}{r,f}(c_2))\!=\!\foldval_2)
    %\wedge \foldcond'
      \color{black}
      \wedge 
      \color{dark-yellow}
    \folB{\Lambda(q)}{t}
    \wedge \folB{\Lambda(q')}{t'}
    }
  }
$$
\end{mdframed}
\end{minipage}
\caption{An example of necessary conditions for a dependency relation}
\label{fig:dep_then_rule_example}
\end{figure}

%% file: figures/delivery.tex
\begin{figure}[t]
  \begin{subfigure}{0.45\textwidth}
  \centering
    \includegraphics[width=0.88\textwidth]{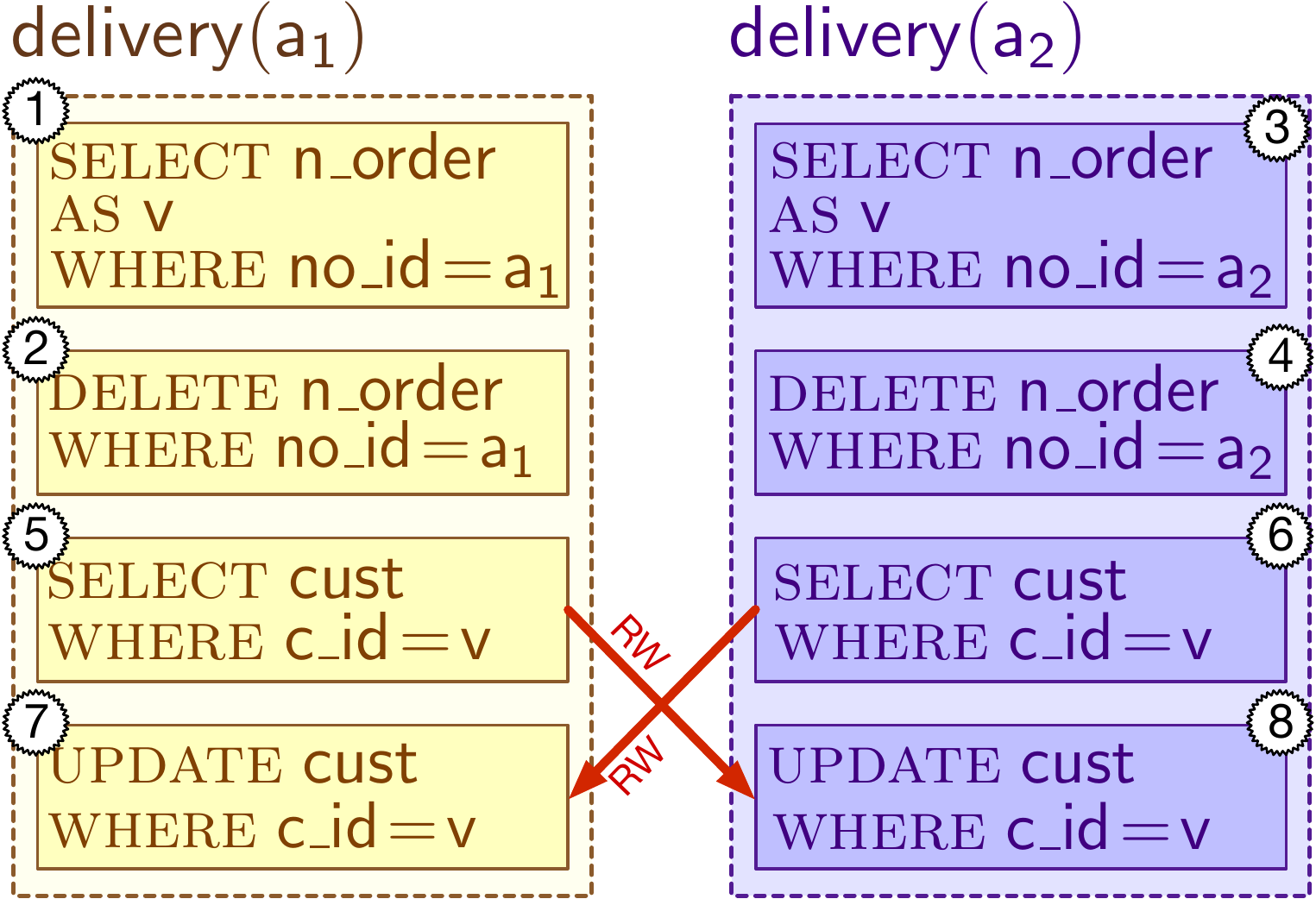}
\end{subfigure}
\quad
\begin{subfigure}{0.45\textwidth}
  \centering
  \includegraphics[width=0.88\textwidth]{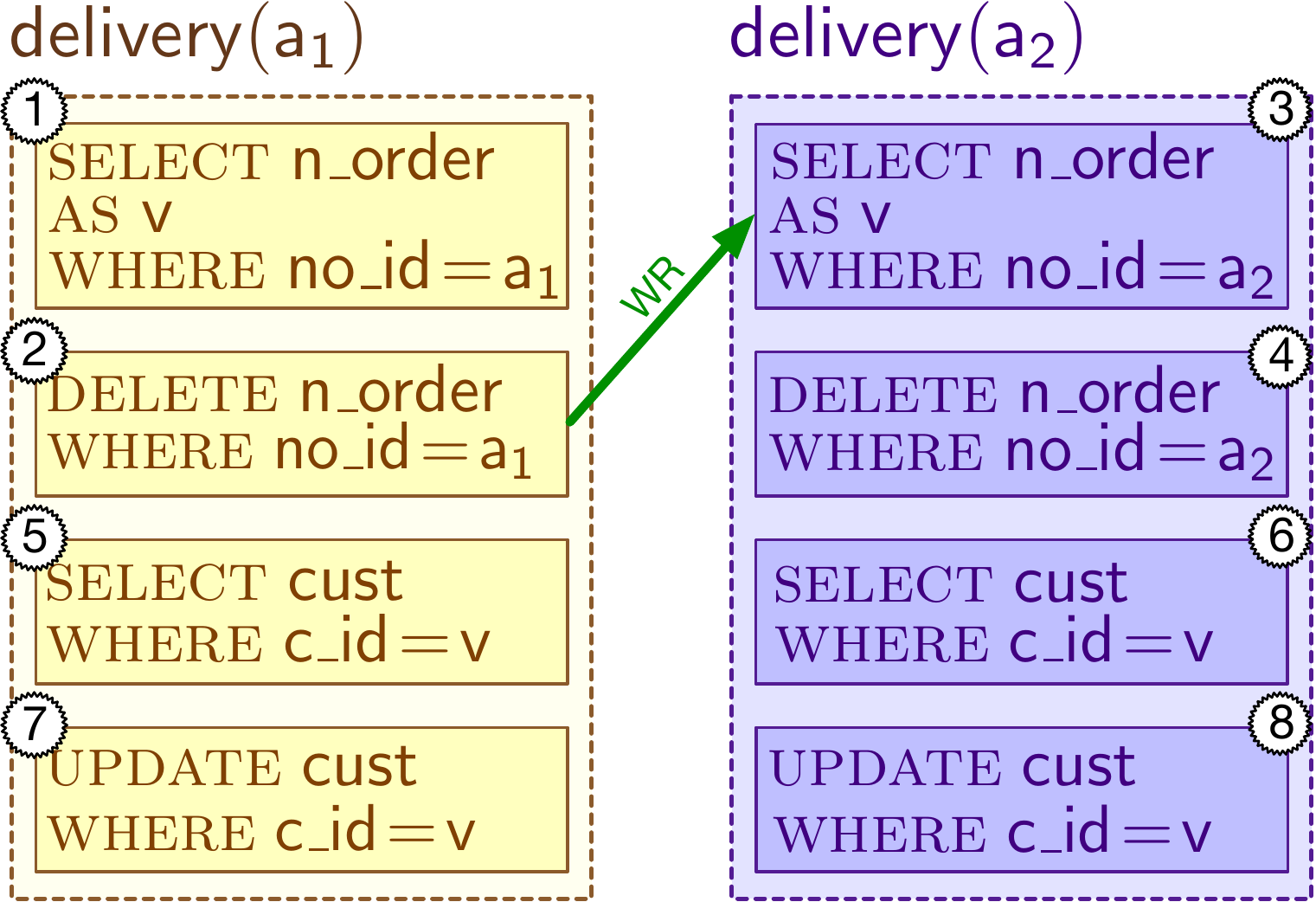}
  \end{subfigure}
  \caption{Lost update anomaly on $\mathsf{Cust}$ table in delivery transaction of
  TPC-C}
    \label{fig:delivery}
  \end{figure}

%% file: figures/then_dep_rule_example.tex
  \begin{figure}[h]
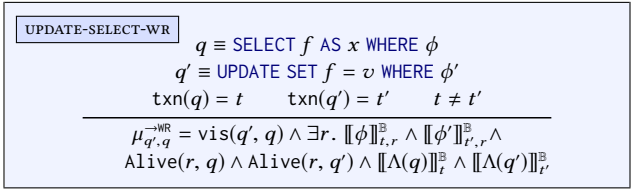
%{0.46\textwidth}
\begin{minipage}[b]{.6\textwidth} 
    \begin{mdframed}[backgroundcolor=Main-Theme-Complement]
\vspace{0mm}
  \begin{flushleft}
    \hspace{-2mm}\fcolorbox{black}{Main-Theme-Complement-Dark}{\scriptsize
    \textrm{\sc{\color{black} update-select-wr}}}
  \end{flushleft}
  \vspace{-6mm}
$$
\FOLRULE{
  \stx{q \equiv \sql{select}f\; \sql{as}
  x\;\sql{where} \phi } \\
  \stx{q' \equiv \sql{update}\sql{set}
  f=v\;\sql{where} \phi'} \\
  \stx{
  \folparent(q)=t \qquad \folparent(q')=t'
  %\qquad 
  %\V{v}{t'}=(\foldcond,\_\,)
  }
  \qquad 
  t\not\eq t'
}
%% ------------------------------------------------------------------------------------------------
  {
  \mu^{\rightarrow\folwr}_{q',q} =
  \stx{
    \color{dark-red}
    \vis(q',q)
    \color{black}
    \wedge
    \color{dark-green}
    \exists r.\;
    \folB{\phi}{t,r
    %,\folversion{rd}(r,c_1)
    } 
  \wedge \folB{\phi'}{t',r
  %,\folversion{rd}(r,c_2)
  }
  \wedge 
  }
  }
  {
    \stx{
      \color{dark-green}
  \qquad \folalive{}(r,q)
  \wedge  
      \folalive{}(r,q')
  %\wedge 
  %(\fol{\rho}_{r}(f,\folversionsub{wr}{r,f}(c_2))\!=\!\foldval_2)
    %\wedge \foldcond'
      \color{black}
      \wedge 
    \color{dark-yellow}
    \folB{\Lambda(q)}{t}
    \wedge \folB{\Lambda(q')}{t'}
    }
  }
$$
\end{mdframed}
\end{minipage}
\caption{An example of sufficient conditions for a dependency relation}
\label{fig:then_dep_rule_example}
\end{figure}

%% file: sections/impl.tex
In this section, we first present our algorithm for exhaustively 
finding the set of anomalous execution in a given application 
and then discuss details related to \tool's implementation.

\subsection{Search Algorithm}
\label{subsec:search}
\input{figures/search_algorithm}

The na\"ive algorithm described in the previous sections that 
iteratively queries a SMT solver to find all solutions to
$\varphi^{}_\mathbb{C}$ is unfortunately too inefficient to be effective in practice.
In our initial
experiments, we found that the SMT solver would often fail to find \emph{any} satisfying
assignment to $\varphi^{}_\mathbb{C}$ within a reasonable amount of time. We
hypothesized that the reason for this was that the solver had the
flexibility to instantiate $\folphi{anomaly}$ in (exponentially many) ways 
that did not force a cycle. Exploiting this intuition, we designed a two-stage search
algorithm, presented in \autoref{fig:search_algorithm}, which
iteratively guides the solver towards cycles via a series of successively  more
constrained SMT queries and then attempts to construct independent
anomalies based upon those cycles. 

The \textsf{FindAnomalies} algorithm takes as input 
a data schema $s$, an \ir\ application $a$, a set of 
user-defined constraints on the database and its consistency model
$\folphi{db}$, and $max_p$, $max_t$, $max_c$, which are bounds on 
the space of independent anomalies, as described in \autoref{sec:encoding}.
The algorithm also outputs $anoms$, the set
of all satisfying assignments to $\varphi_\mathbb{C}$.  
The algorithm
constructs $anoms$ by repeatedly querying a SMT solver, denoted by
\textsf{isSAT}. 
Satisfying assignments to $\varphi_\mathbb{C}$ are
constructed by first 
iteratively finding the set of all bounded cycles 
using the SMT query in 
\hyperref[l5]{line 5},
and then 
constructing independent anomalies 
atop of each cycle using the SMT query in 
\hyperref[l17]{line 17}.

The formula in the first query (\hyperref[l5]{line 5}) is a conjunction of four clauses
quantified over $t$ variables which represent the transactions on a
dependency cycle of length $c$. 
$\folphi{app}$ is constructed before the loop begins
using function \textsf{EncApp} and corresponds to $\folphi{dep$\rightarrow$}$ and
$\folphi{$\rightarrow$ dep}$ in the given program.
Finally, the formula also includes a clause $\folphi{neg}$, representing the negated
conjunction of all previously found assignments (stored in $cycles$)
so that the solver is forced to find new cycles at each iteration. 
The resulting satisfying assignment is
stored in the $new\_cyc$ variable, which is then added to 
$cycles$ in \hyperref[l7]{line 7}.
Similarly, the query in \hyperref[l17]{line 17}, encoded by function
\textsf{EncPath}, is quantified over $p$ serially executed transactions 
that are intended to update the initial state of the database 
such that the pre-conditions for the given cycle can be satisfied.

\begin{wrapfigure}{r}{0.45\textwidth}
    \centering
    \includegraphics[width=0.42\textwidth]{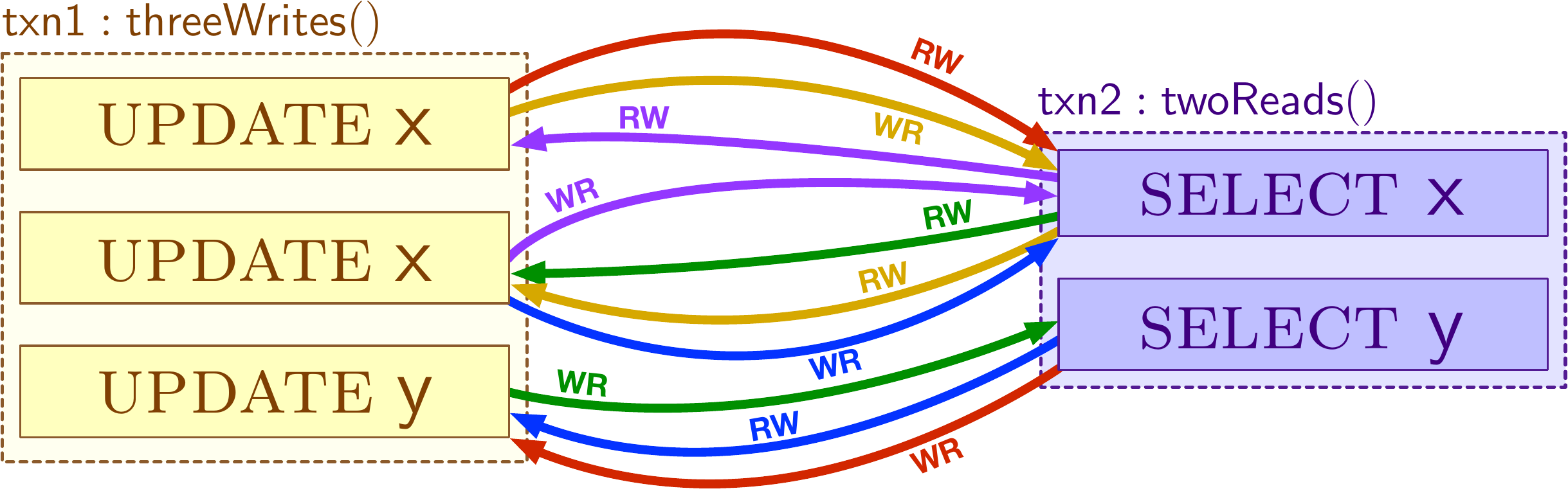}
    \caption{Structurally Similar Anomalies} %\KR{move to wrapfig}
\label{fig:similar_anomalies}

\end{wrapfigure}

%% intro to optimization
The above steps describe how to efficiently construct a satisfying
assignment to $\varphi_\mathbb{C}$. When experimenting with \tool\,, we
observed that many anomalies share a similar structure, in that they
all include the same sorts of dependency relations between the operations
in the transactions involved in the cycle. 
As an example, consider the
program in \autoref{fig:similar_anomalies} where five different
anomalies (depicted using dependency edges of different colors) can
form on the two transaction instance, all sharing the structure:
\tikzcdset{diagrams={arrows={shorten >=-1ex,shorten <=-1ex}}}
\begin{tikzcd}
  \mathsf{txn1} \ar[r,bend left=18, "\folwr" description]  & 
  \mathsf{txn2} \ar[l,bend left=18, pos=0.49,"\folrw" description]
\end{tikzcd}\!\!.

We refined our algorithm to exploit this observation by guiding the
solver to find all such structurally similar anomalies once one of them is
detected. 
It does so via
an inner loop (lines 9-13) which repeats the steps discussed above, but
first adds a clause, $\folphi{stct}$, that forces the solver to look for
cycles that are structurally similar to the one found on  \hyperref[l5]{line 5}.
$\folphi{stct}$ is constructed by the function \textsf{EncStruct} which
constrains the dependency edges and transactions that the solver
considers on a cycle.
The inner loop repeats until all such cycles are found. 
When the algorithm can no longer find any new anomaly with the shared structure,
it breaks the loop in \hyperref[l12]{line 12}, and the normal execution of the outer loop
is resumed. 
We 
empirically demonstrate
in \autoref{sec:eval} 
that this
optimized procedure is more performant than one that does not include
the inner loop.

\subsection{Prototype}
\tool\footnote{\url{https://github.com/Kiarahmani/CLOTHO}} is a complete directed testing framework that combines the
SAT-based approach of \autoref{sec:encoding} with a front-end compiler
and test replay environment.  The front-end compiler takes as input an arbitrary Java
class which manipulates a database through the standard JDBC API,
where each method is treated as a transaction. The compiler supports
queries written in the subset of SQL given in
\autoref{fig:syntax}. Consequently, it does not allow nested queries,
joins or update operations with implicit reads (e.g.  $\sql{UPDATE}
\mathtt{T} \;\sql{SET} \mathtt{f\!\!=\!f\!+\!1}$). These operations
are not typically supported by distributed, weakly consistent
databases, and translating application logic to avoid them is a necessary and standard
step of porting a database application to a distributed systems.

Our compiler is implemented as a pass on Shimple programs which are
produced by the Soot static analysis
framework~\cite{Vallee-Rai:1999:Soot}.
The compiler initially performs a simple goto analysis to determine
loops and conditional structures and marks all variables that are
modified or read within them.
It then detects all database access points, e.g.
{\lstinline[basicstyle=\ttfamily\small]{executeUpdate()}} function
calls. For each access point, it then constructs an \ir\ query that
may contain typed holes.
The compiler then walks backward through all assignments of variables
used in the query and tries to fill the holes with \ir\ values.
In the case that a variable is assigned a value produced by an unknown
function call, the compiler creates a new abstract variable to use
in place of the variable in the original query. \tool\, then uses
the algorithm presented in
\autoref{fig:search_algorithm} to search for anomalies, using
Z3~\cite{Moura:2008:Z3} as a backend.

For each satisfying assignment found, \tool\ extracts a set of test
configuration folders, each of which contains a database
initialization file and a set of Java class files annotated with
execution orders and input parameters. \tool's test administration
system uses this information to replicate anomalous executions on a
concrete environment. The test environment is initialized by creating
a cluster of Docker containers running replicated database instances
that have been instantiated according to the test configuration. The
replayer supports any distributed or centralized database that has JDBC drivers. The
replayer also spawns a set of Java worker containers executing
transactions from the test configuration with the specified
parameters. Partitioning requirements are implemented using the
open-source \citet{Blockade} tool which artificially induces network
partitions between replica or worker containers as required by the
test configuration. Transaction schedules are enforced by a custom
JDBC driver that receives requests from application workers and acts
as a \emph{managed wrapper} to the driver for the underlying
database. Requests are blocked until the driver communicates with the
central scheduling unit using the RMI functionality in Java, and are
only passed on to the actual driver once the permission to proceed has
been acquired.

%% file: figures/search_algorithm.tex
\begin{figure}[t]
  %{0.53\textwidth}
\vspace{-3mm}
  \begin{minipage}{0.87\textwidth}
  \begin{mdframed}[innertopmargin=5pt, innerbottommargin=5pt, 
    innerleftmargin=0.2cm, innerrightmargin=0.1cm,backgroundcolor=light-gray]

\SetKwInOut{Function}{function\!\!\!\!\!}
\SetKwInOut{Init}{Initialize\!\!\!}
\begin{algorithm}[H]
  \small
  \Function{$\mathsf{FindAnomalies}(s,a,\folphi{db},max_p,max_t,max_c)$}
  \KwIn{\hspace{0.4mm}$s$: data schema,
 \hspace{2mm} $a$: \ir\ application, 
 \hspace{2mm} $\folphi{db}$: database constraints, 
\\ \hspace{9mm} $max_p$: maximum serial transactions, 
 \hspace{2mm} $max_t$: maximum concurrent transactions,
\\ \hspace{9mm} $max_c$: maximum cycle length}
 \KwOut{$anoms$: a set of satisfying assignments}
 \Init{
   \hspace{1mm}
   $\folphi{app} \leftarrow \mathsf{EncApp}(s,a)$, 
   \hspace{1mm}
   $anoms \leftarrow \emptyset$,
   \hspace{1mm}
   $cycles \leftarrow \emptyset$
 }
 \For{$t\in[2,max_t]$}{
  $c\leftarrow 3$ \\
  \While{$c\leq max_c$}{
    $\folphi{neg}\leftarrow \mathsf{EncNeg}(cycles) $\\
    $new\_cyc \leftarrow \mathsf{isSAT}(\exists t_1,...,t_t.\;\folphi{cycle}^c(t_1,...t_t)\wedge\folphi{db} \wedge
    \folphi{app}\wedge \folphi{neg})$\\
    \label{l5}
    
    \lIf{$new\_cyc=\mathtt{UNSAT}$}{  
      $c\leftarrow c+1;$ \hspace{0.5mm}
        {\bf continue}}

    $cycles\leftarrow cycles\cup \{new\_cyc\}$ \\
    \label{l7}
    $\folphi{stct} \leftarrow \mathsf{EncStruct}(new\_cyc)$

      \Do{true}{
        $\folphi{neg}\leftarrow \mathsf{EncNeg}(cycles) $\\
        $new\_cyc \leftarrow \mathsf{isSAT}(\exists t_1,...t_t.\;
                          \folphi{cycle}^c(t_1,...t_t)\wedge
                          \folphi{db} \wedge
                          \folphi{app}\wedge 
                          \folphi{neg} \wedge
                          \folphi{stct}
                          )
        $ \\
        \leIf{$new\_cyc = \mathtt{UNSAT}$}{
          {\bf break}
        }{
          $cycles\leftarrow cycles\cup \{new\_cyc\}$
        }
        \label{l12}
      }
  }
 }
 \For{$cyc\in cycles$}{
   \For{$p\in[0,max_p]$}{
     $\folphi{path} \leftarrow \mathsf{EncPath}(cyc)$\\
    \label{l17}
     $new\_anml\leftarrow \mathsf{isSAT}(\exists t_1,...,t_p.\;\folphi{path} )$ \\
     \lIf{$new\_anml \not\eq \mathtt{UNSAT}$}{
      $anoms \leftarrow anoms \cup \{new\_anml\}$; \hspace{1mm}{\bf break}}
    } 
 }

\end{algorithm}
  
  \end{mdframed}
  \caption{Search Algorithm}
\label{fig:search_algorithm}
\end{minipage}
\end{figure}

%% file: sections/eval.tex
The goal of our experimental evaluation was to investigate the
following questions:
\begin{itemize}
\item[\textbf{(R1)}] Is \tool\ \emph{effective}, i.e. capable of detecting serializability
  anomalies in real weakly consistent database applications?
\item[\textbf{(R2)}] Is \tool\ \emph{practical}, i.e. how many anomalies does it find
  and how long does it take to find them?
\item[\textbf{(R3)}] Is our FOL encoding rich enough for cycles to be
  mapped to unserializable executions on a real datastore?
\item[\textbf{(R4)}] How does \tool\, compare to undirected or random
  testing at finding bugs?
\end{itemize}

Local evaluation of \tool\, was carried out on a Macbook Pro with a
2.9\textsc{Ghz} Intel Core\,i5 CPU and 16GB of memory, running Java 8
and equipped with Z3 v4.5.1.

%%%%%%%%%%%%%%%%%%%%%%%%%%%%%%%%%
%%%%%%%%%%%%%%%%%%%%%%%%%%%%%%%%%
\subsection{Detecting Serialization Anomalies}
%%%%%%%%%%%%%%%%%%%%%%%%%%%%%%%%%
%%%%%%%%%%%%%%%%%%%%%%%%%%%%%%%%%
%%
To answer questions \textbf{(R1)-(R3)}, we carried out three
experiments on the set of benchmarks shown in
\autoref{table:applicability_results}. These applications were drawn
from the OLTP-Bench test suite \cite{Difallah:OLTP_Bench}, and were
designed to be representative of the heavy workloads seen by typical
transaction processing and web-based applications that can also be 
deployed on weakly consistent data stores~\cite{Bailis:2014:RAMP, Shasha:2002:Database}.

A few of these benchmarks have distinctive features that deserve
special mention.
SEATS is an online flight search and ticketing system
with both read and write intensive transactions.
TATP simulates a caller location system with non-conflicting
transactions that are prone to be rolled back, which complicates
compilation.
%
%TPC-C is the industry standard benchmark for evaluating
%OLTP systems, modeling a warehouse and customer
%management system.
%
SmallBank simulates a financial account management system
comprised of transactions that repeatedly access the same
objects and therefore heavily depend on transactional isolation guarantees.
%
%Voter is derived from a talent show vote collecting software and
%includes a single read/write transaction.
%
Twitter and Wikipedia are inspired from the popular online services
featuring micro-blogging and collaborative editing facilities.
The data model and the transactional logic of Wikipedia is notably
complex, where transactions manipulate a large set of backed-up tables
for user-authentication and concurrent edit history management. To
apply \tool\ to this test suite, we had to manually replace any
unsupported database operations with equivalent operations, per
\autoref{sec:impl}. Out of the 131 distinct queries found in these
benchmarks, only 11 used $\sql{join}\!$, which we replaced with
other supported queries and application-level loops. While we made these
edits by hand, this process has been shown to be 
mechanizable~\cite{Tahboub:2018:How}.

Our first experiment investigated questions \textbf{(R1)} and
\textbf{(R2)} by applying \tool\, to detect serializability anomalies
of a fixed length in the benchmark applications. The first eight
columns of \autoref{table:applicability_results} presents the results
of this experiment.  For each benchmark the table lists the number of
transactions (\textsf{\#Txns}) and number of tables
(\textsf{\#Tables}) it includes, the maximum bound on cycle length
considered (\textsf{Max. Length}), the total number of anomalies found
(\textsf{\#Anmls}), unique number of structures (\textsf{\#Dist
  Strcts}), the average time needed to detect a single anomaly
(\textsf{Avg Anlys}), and the total analysis time (\textsf{Total
  Anlys}).  For each benchmark, we initially limited the length of
cycles to be at most 4, as all the canonical serializability
anomalies, e.g. dirty reads and lost
updates are of this length~\cite{Berenson:1995:Critique}. As section~\ref{sec:blackbox} 
will discuss in more detail, this bound is
also sufficient to discover executions violating every correctness
criteria in the specification of the TPC-C benchmark. For the two
benchmarks in which no anomaly of length 4 was found, TATP and Voter,
we iteratively increased the bound but were unable to find any longer
anomalies. These results demonstrate that \tool\ can effectively
detect anomalies in a reasonable amount of time on our benchmark
applications, needing about an hour to analyze our most complex
application.

\input{tables/applicability_results}

Our next experiment tackled \textbf{(R3)} by using the test
configuration information generated by \tool\ to manifest potential
anomalies on the popular Cassandra database storage engine. We chose Cassandra
because it only guarantees eventual consistency and provides no
transactional support, allowing for all serializability anomalies to
potentially occur and be caught.

Column (\textsf{Auto Replay}) in \autoref{table:applicability_results}
lists the number of unserializable executions \tool's test
administration framework was able to exhibit from the test
configurations generated by \tool, accounting for over 80\% of
detected anomalies. There were two main reasons this process failed
for the other anomalies: underspecified loop boundaries in the
encoding and the mismatch between the order of fetched rows in the
encoding and at the runtime.  Nevertheless, we were able to edit the
generated test configurations to include enough information to exhibit
an unserializable executions in all the remaining cases. In
particular, by specifying the loop boundaries in the encoding and
modifying the order at which initial database rows are
inserted. Column (\textsf{Man. Replay}) lists the number of anomalies
that required such interventions.

\input{figures/eval_results.tex}

To test the effectiveness of the optimized search algorithm presented
in section~\ref{subsec:search}, we created an unoptimized variant of
\textsf{FindAnomalies} by removing its inner do-while loop. To compare
the two algorithms, we applied the unoptimized version to the three
benchmarks with the most anomalies, using the total analysis of the
optimized version as a timeout.
\autoref{fig:eval_results} presents the number of anomalies found
by the basic version against the optimized.
We can see that the optimized algorithm performs better than the
unoptimized version, reporting more than double the anomalies in the
same amount of time for each benchmark.

%% file: tables/applicability_results.tex
% settings
\setlength{\arrayrulewidth}{0.2mm}
\setlength{\tabcolsep}{4pt}
\renewcommand{\arraystretch}{1}
\renewcommand\theadalign{cc}
\renewcommand\theadfont{  \scriptsize}
\setlength\cellspacetoplimit{0pt}
\setlength\cellspacebottomlimit{0pt}

% table
\begin{table}[h!]
\centering
\begin{footnotesize}
{\rowcolors{2}{Main-Theme-Light}{Main-Theme-Lighter}
  \begin{tabular}{|l|l|l|l|l|l|l|l|l|l|}
    \hline
 \rowcolor{Main-Theme-Dark}
 \multicolumn{1}{|c}{\thead{ Benchmark \;}}
     &  \multicolumn{1}{l} {\thead{ \#Txns}}
    &  \multicolumn{1}{l} {\thead{ \#Tables}}
    &  \multicolumn{1}{l} {\thead{   Max.  \\ Length}}
     &  \multicolumn{1}{l} {\thead{ \#Anmls}}
     &  \multicolumn{1}{l} {\thead{ \#Dist. \\ Strcts}}
     &  \multicolumn{1}{l} {\thead{   Avg   \\ Anlys}}
     &  \multicolumn{1}{l} {\thead{   Total \\ Anlys}}
     &  \multicolumn{1}{l} {\thead{   Auto  \\ Replay}}
     &  \multicolumn{1}{l|} {\thead{   Man.  \\ Replay}}
   \\[-0.25ex]
    \hline
    SEATS       & 6  & 10  & 4  & 32  & 18  & 26s   & 1970s   & 26 & 6  \\
    TATP        & 7  & 4   & 15 & 0   & 0   & -     & 55s     & -  & -  \\
    TPC-C       & 5  & 9   & 4  & 22  & 13  & 118s  & 3270s   & 18 & 4  \\
    SmallBank   & 6  & 3   & 4  & 60  & 15  & 3s    & 264s    & 50 & 10 \\
    Voter       & 1  & 3   & 20 & 0   & 0   & -     & 8s      & -  & -  \\
    Twitter     & 5  & 5   & 4  & 2   & 2   & 19s   & 211s    & 2  & 0  \\
    Wikipedia   & 5  & 12  & 4  & 3   & 3   & 227s  & 4343s   & 2  & 1  \\
    [0.2ex]
 \hline
\end{tabular}
}
\end{footnotesize}
  \vspace{4mm}
  \caption{Serializability anomalies found and manifested in OLTP
    benchmark suite. }
\vspace{-8mm}
\label{table:applicability_results}
\end{table}

%% file: figures/eval_results.tex
\begin{wrapfigure}[9]{r}{0.4\textwidth} 
  \vspace{-5mm}
  \centering
  \includegraphics[width=0.38\textwidth]{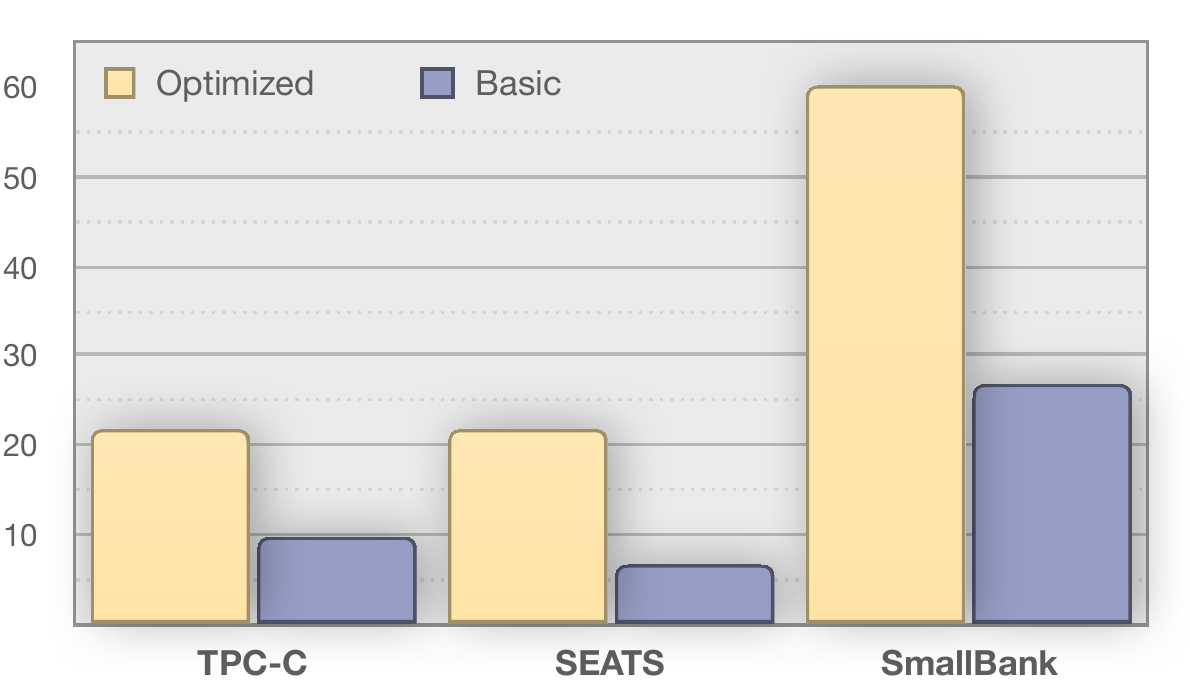}
%\vspace{-5mm}
\caption{Optimized vs basic analysis}
%\vspace{4mm}
\label{fig:eval_results}
\end{wrapfigure}

%% file: sections/blackbox.tex
%
% intro to section
%
Our final experiment was designed to compare \tool's ability to find
``real'' bugs over automated blackbox testing. To do so, we used the
TPC-C benchmark from the previous section, as it specifies a set of
twelve \emph{consistency requirements} (\crr{}) that must be preserved
by any sound database system. These \crr{}s give us objective notions
of buggy behaviors that can arise in realistic OLTP applications,
which we used to compare the relative effectiveness of \tool\ and
random testing.  The rows ($\crr{1}-\crr{12}$) of
\autoref{table:jepsen} list all of TPC-C consistency requirements. We
have broken "if and only if" requirements into separate rows.

 \input{tables/jepsen_results_tpcc}

In addition to \crr{}s, we identified seven new consistency requirements
(\ncr{1}-\ncr{7}) that should also be preserved by any serializable
execution of TPC-C. We identified these invariants after examining the
serializability anomalies generated by \tool.  These requirements
capture more subtle safety properties of the application, descriptions
of each of which is also shown in \autoref{table:jepsen}.  As an example,
\ncr{4} captures that a customer's $\fld{c\_payment\_cnt}$ field is
uniquely determined by its initial value and the number of times
\txn{payment} transaction has been called on that customer.

To the best of our knowledge, no random testing tool exists that can
be deployed on weakly consistent databases and that can also handle  our targeted
benchmark suit.
Thus, we were forced to develop our own tool for
testing high-level application properties. Our framework builds upon
\citet{jepsen}, a popular open source library for testing distributed
databases and systems. In addition to test automation and logging
infrastructure, Jepsen is capable of systematically injecting faults
(e.g. network partitions or node clock skews) during a test. Our
framework extends Jepson to:
\begin{enumerate*}[label=(\roman*)]
  \item manage concurrent Java clients that share a JDBC connection pool,
  \item support enhanced logging of client-side generated information and
  \item perform a set of customized database-wide safety checks
\end{enumerate*}, each of which is required for targeted testing of
application-level properties in TPC-C.
We evaluated our testing framework on a fully replicated Apache
Cassandra cluster running on a configurable number of \texttt{t2.xlarge}
EC2 instances. % We chose the popular Cassandra database because it
% guarantees eventual consistency with no transactional support,
% allowing for all serializability anomalies to potentially occur and be
% caught.
%

\input{figures/jepsen_testing_framework}
%

%
% strengths of our tool
%
In order to ensure a fair comparison and mimic how applications are
tested in practice, we ensured that:
\begin{enumerate*}[label=(\roman*)]
\item  the initial state of the database was realistic   \label{item:init}
\item each transaction call's arguments were realistic
  \label{item:load}
\item the maximum Cassandra throughput is achieved and reported
  for each test.   \label{item:throughput}
\end{enumerate*}
To achieve \ref{item:init} and \ref{item:load}, all our tests used
initial database states and transaction parameters taken directly from
OLTP-Bench. In addition, we used the smallest available tables in
order to make sure that the probability of conflicts was
maximized. Our tool utilizes Cassandra's sstableloader tool to
automatically and efficiently bulk-load snapshots into the cluster
before each test. Lastly, we made sure that none of our design
decisions, e.g. the choice of Cassandra's secondary indices would
compromise throughput, resulting in an unrealistically low load that
would minimize the chance for conflicts.
%To do so, \BD{We did something to determined maximum throughput on our
%  cluster, in order to show that our tests came close to that.}

%\BD{We need more details on how this tool actually works: what is
%  being randomized here?}

% \KR{--->>}
%We then tested whether our random testing tool which was capable of finding
%executions leading to violations of each of TPC-C standard requirements, as well
%as those we devised.
%\KR{<<---}

 We repeated each experiment with 10, 50 and 100 concurrent clients,
 each submitting 6 transactions per second, where the frequency of
 transactions adhered to TPC-C's workload requirements.  These
 settings achieved approximately 20\%, 80\% and 100\% of the maximum
 throughput we witnessed on the deployment.  Each experiment was
 repeated with a 10 minute and then 30 minute timeout.
 \autoref{table:jepsen} presents the results of these experiments,
 with \passed{} and \failed{} recording when an invariant was violated
 and was not violated, respectively. As the table shows, our random
 testing framework was only able to violate 14 out of 21 (67\%)
 requirements.

\input{tables/anomalies_table_tpcc}

Alternatively, we mapped each of TPC-C's internal serializability anomalies
found by \tool\ to the set of invariants that could be broken in
scenarios arising from those anomalies.
\autoref{table:anomalies_table_tpcc} lists the details of the internal
serializability anomalies of length 4 \tool\ finds for TPC-C along
with the set of requirements each anomaly breaks.
We found that every single one of TPC-C's consistency requirements was
broken as a result of at least one serializability anomaly.

These results evidence the importance of (internal) serializability
anomalies for detection of application-level bugs and showcase \tool's
ability to efficiently construct many of such anomalies that are
present in a realistic database.  \tool\ reports these anomalies as
slowed-down interleaved execution steps in a form comprehensible to
the developers.

%% file: tables/jepsen_results_tpcc.tex
\def\arraystretch{0.95}
\begin{table}[h!]
\centering
\begin{footnotesize}
{\rowcolors{2}{Main-Theme-Light}{Main-Theme-Lighter}
  \begin{tabular}{|l | l | l | c | c | c |} 
 \hline
 \rowcolor{Main-Theme-Dark}
        \multicolumn{1}{|c}{\scriptsize Inv.} 
     &  \multicolumn{1}{c}{\scriptsize Source} 
     &  \multicolumn{1}{c}{\scriptsize Description} 
     &  \multicolumn{3}{c|}{\thead{\vspace{-0.3mm}\Tiny Broken by rand testing?\,\\\tiny
     10c\;\;\;\;\;~\;50c\;\;~\;\;100c \vspace{-0.5mm}}} 
     %&  \multicolumn{1}{c}{\scriptsize 50c} 
     %&  \multicolumn{1}{c|}{\thead{\scriptsize \!100c\!}} 
     \\ [0.1ex] 
 \hline
    %1
    \crr{1} & spec & \Tiny W\_YTD = sum(D\_YTD) & \,\,\passed{} \,&
    \;\passed{}\;  & \, \passed{} \,\\

    %2
    \crr{2} & spec & \Tiny D\_NEXT\_O\_ID-1=max(O\_ID)=max(NO\_O\_ID)& \failed{} & \failed{} &  \passed{} \\
   
    %3
    \crr{3} & spec & \Tiny max(NO\_O\_ID)-min(NO\_O\_ID)+1=[number of corresponding rows in NEW-ORDER] & \failed{} &\failed{} &  \passed{} \\

    %4
    \crr{4} & spec & \Tiny sum(O\_OL\_CNT) = [number of corresponding rows in ORDER-LINE] &  \passed{} &  \passed{} &  \passed{} \\

    %5
    \crr{5a} & spec & \Tiny For any row in NEW\_ORDER there is a row in ORDER
    with O\_CARRIER\_ID set to null & \failed{}  & \failed{}  & \failed{}  \\
    
    %6
    \crr{5b} & spec & \Tiny For any row in ODER with O\_CARRIER\_ID set as null there is row in NEW\_ORDER  &  \failed{} & \failed{}  & \failed{} \\
    
    %7
    \crr{6} & spec & \Tiny  O\_OL\_CNT is equal to the number of rows in ORDER-LINE for that order& \passed{} & \passed{} & \passed{}  \\
    
    %8
    \crr{7a} & spec & \Tiny O\_CARRIER\_ID=null if all rows in ORDER\_LINE have OL\_DELIVERY\_D=null & \failed{} &\failed{}  & \failed{} \\
    
    %9
    \crr{7b} & spec & \Tiny  all rows in ORDER\_LINE have OL\_DELIVERY\_D=null if O\_CARRIER\_ID=null&\failed{} &\failed{} &\failed{}  \\
    
    %10
    \crr{8} & spec & \Tiny W\_YTD = sum(H\_AMOUNT) & \passed{} & \passed{} &  \passed{}\\
    
    %11
    \crr{9} & spec & \Tiny D\_YTD = sum(H\_AMOUNT) & \passed{} & \passed{}  &  \passed{} \\
    
    %12
    \crr{10} & spec & \Tiny C\_BALANCE = sum(OL\_AMOUNT) - sum(H\_AMOUNT)& \failed{} & \passed{}  & \passed{} \\

    \crr{11} & spec & \Tiny (count(*) from ORDER) - (count(*) from NEW-ORDER) = 2100 & \passed{} & \passed{}  & \passed{} \\

    %13
    \crr{12} & spec & \Tiny C\_BALANCE + C\_YTD\_PAYMENT = sum(OL\_AMOUNT)& \failed{}& \passed{} & \passed{} \\
    
    %14
    \ncr{1} & anlys & \Tiny s\_ytd=sum\_exec(ol\_quantity) & \failed{} & \passed{} &  \passed{} \\
    
    %15
    \ncr{2} & anlys & \Tiny sum(s\_order\_cnt) = sum\_exec(o\_ol\_cnt) &  \failed{}&  \passed{} &  \passed{}\\
    
    %16
    \ncr{3} & anlys & \Tiny In the absence of DELIVERY,
    (c\_ytd\_payment+c\_balance) must be constant for each cust & \failed{}& \failed{}& \failed{}\\
    
    %17
    \ncr{4} & anlys & \Tiny c\_payment\_cnt = initial\_value + 
    \#PAYMENT on that customer in execution log  & \failed{}& \failed{}&  \failed{}\\
    
    %18
    \ncr{5} & anlys & \Tiny sum(C\_DELIVERY\_CNT) = initial\_value + number of times DELIVERY is called & \passed{} &  \passed{}  & \passed{}\\
    
    %19
    \ncr{6} & anlys & \Tiny ratio of different o\_carrier\_id must
    respect the corresponding ratio on DELIVERY's inputs & \failed{}& \passed{} & \passed{}\\
    
    %20
    \ncr{7} & anlys & \Tiny H\_AMOUNT for each customer = sum(payment\_amounts) for that customer in log  & \failed{} & \failed{}& \failed{}\\
    [0.3ex] 
 \hline
\end{tabular}
}
\end{footnotesize}
  \vspace{4mm}
  \caption{Targeted random testing results}
  \vspace{-6mm}
  \label{table:jepsen}
\end{table}

%% file: figures/jepsen_testing_framework.tex
\begin {wrapfigure}[12]{r}{0.37\textwidth}
\centering
\vspace{-2mm}
\includegraphics[width=0.37\textwidth]{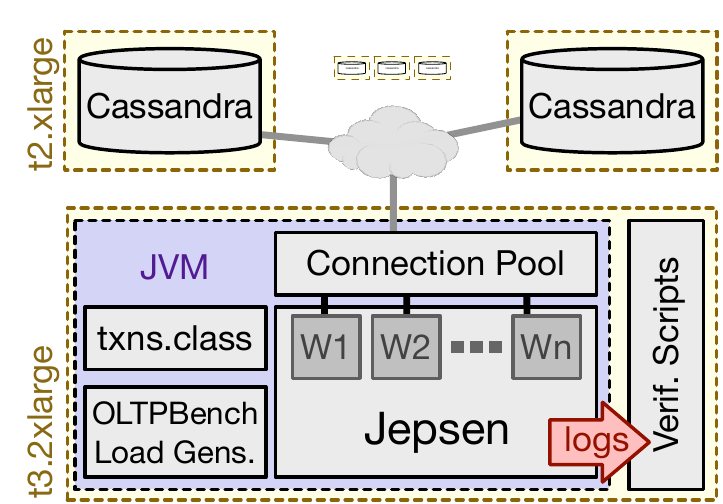}
  %\vspace{-4mm}	
  \caption{Testing framework}
  \label{fig:jepsen}
\end{wrapfigure}

%% file: tables/anomalies_table_tpcc.tex
% settings
\setlength{\arrayrulewidth}{0.2mm}
\setlength{\tabcolsep}{4pt}

\def\arraystretch{0.95}
% table
\begin{table}[h!]
\centering
\begin{footnotesize}
{\rowcolors{2}{Main-Theme-Light}{Main-Theme-Lighter}
  \begin{tabular}{|l | l | l | l | l | l | l | l |} 
 \hline
    \rowcolor{Main-Theme-Dark}
        \multicolumn{1}{|c}{\scriptsize\#} 
     &  \multicolumn{1}{c}{\scriptsize Len} 
     &  \multicolumn{1}{c}{\scriptsize Txns} 
     &  \multicolumn{1}{c}{\scriptsize Tables} 
     &  \multicolumn{1}{c}{\scriptsize Type} 
     &  \multicolumn{1}{c}{\scriptsize Broken Invariants} 
     &  \multicolumn{1}{c}{\scriptsize  Anlys} 
     &  \multicolumn{1}{c|}{\scriptsize Rply} \\ [0.1ex] 
 \hline
    1	& 4 & \txn{pm}/\txn{pm}	          & \tabl{warehouse}	& \scriptsize lost update  & \crr{1}, \crr{8}                       & 5s    & 7s \\ 
    2	& 4 & \txn{pm}/\txn{pm}	          & \tabl{district}	  & \scriptsize lost update  & \crr{1}, \crr{9}                       & 19s   & 12s \\
    3	& 4 & \txn{no}/\txn{no}	          & \tabl{district}	  & \scriptsize
lost update  & \crr{2}, ..., \crr{7}, \crr{11}, \ncr{1}         & 15s   & 13s \\
    4	& 4 & \txn{dv}/\txn{dv}	          & \tabl{customer}	  & \scriptsize lost update  & \crr{10}, \crr{12}, \ncr{5}            & 171s  & 12s  \\
    5	& 4 & \txn{dv}/\txn{dv}	          & \tabl{new\_order}	& \scriptsize lost
    update  & \ncr{6}                                & 16s    & 6s  \\
    6	& 4 & \txn{pm}/\txn{pm}	          & \tabl{history}	  & \scriptsize lost update  & \crr{8},\crr{9}, \crr{10}, \ncr{7}     & 125s  & 11s  \\
    7	& 4 & \txn{no}/\txn{no}/\txn{no}	& \tabl{stock}	    & \scriptsize unclassified & \ncr{1}, \ncr{2}                       & 697s   & 13s  \\
    8& 4 & \txn{no}/\txn{no}	          & \tabl{stock}	    & \scriptsize lost update & \ncr{1}, \ncr{2}                        & 70s   & 12s  \\
    9& 4 & \txn{dv}/\txn{os}/\txn{no}	& \tabl{order\_line}& \scriptsize unclassified & none                                   & 11s  & 12s  \\
    10& 4 & \txn{dv}/\txn{os}/\txn{no}	& \tabl{order\_line}& \scriptsize unclassified & none                                   & 9s    & 15s \\
    11& 4 & \txn{dv}/\txn{pm}	          & \tabl{customer}	  & \scriptsize lost update  & none                                   & 231s  & 9s  \\
    12& 4 & \txn{dv}/\txn{pm}	          & \tabl{customer}	  & \scriptsize lost update  & \ncr{3},\ncr{4}                        & 12s   & 11s  \\
    13& 4 & \txn{pm}/\txn{pm}	          & \tabl{customer}	  & \scriptsize lost update  & \ncr{3},\ncr{4},\crr{12}               & 108s   & 8s  \\
    [0.3ex] 
 \hline
\end{tabular}
}
\end{footnotesize}
  \vspace{3mm}
  \caption{Internal serializability anomalies found on TPC-C}
\label{table:anomalies_table_tpcc}
\end{table}

%% file: sections/related.tex
%% -------------------------------------------------------------------------------------------
%% -------------------------------------------------------------------------------------------

Serializability is a well-studied problem in the database community,
and there have been a number of efforts over the years to develop static techniques to
discover serializability anomalies in concurrent database systems.
For example, early works by 
%Fekete \emph{et al.} 
\citet{FE05a} and 
%Jorwekar \emph{et al.} 
\citet{JO07} proposed lightweight
syntactic analyses to check for serializability anomalies in centralized
databases, by looking for dangerous structures in the
over-approximated static graph of all possible dynamic dependency 
conflicts.
Several recent
works \cite{BE16, CE15, CE16, CE17, ZH13, WA17} have continued along
this line, by deriving different types of problematic structures in
dependency graphs that are possible under different weak consistency
mechanisms, and then checking for these structures on static
dependency graphs.

However, as we have also demonstrated in this paper, using just static
dependency graphs yields highly imprecise representations of actual
executions, and any analysis reliant on these graphs is likely to
yield a large number of false positives. 
Indeed, recent efforts in
this space \cite{BE16, CE16, CE17}
recognize this and propose
conditions to reduce false positives for specific consistency
mechanisms, but these works do not provide any automated methodology
to check those conditions on actual programs. Further, application
logic could prevent these harmful structures from manifesting in
actual executions.

Like our work,~\cite{KJ18} and \cite{KR18} also model application logic and
consistency specifications using a decidable fragment of first-order
logic (FOL), so that an underlying solver could automatically derive
harmful structures which are possible under the given consistency
specification and search for them in actual dependency graphs taking
application logic into account.  
Although their core language supports
SQL operators, they do not incorporate their techniques into a full
test-and-reply environment that allows mapping anomalies identified
in abstract executions to be translated to concrete inputs that
can be executed in a test environment.

\citet{BR17,BR18} recently proposed an analysis technique to find
serializability violations using a dependency graph-based approach.
Their notion of dependencies is very
conservative and is reliant upon accurate characterization of
commutativity and so-called ``absorption'' relations between operations.
Further, in a bid to make dependency characterizations local i.e. relying only on
the operations involved in the dependency, they over-approximate the
impact of other operations on commutativity and absorption (resulting
in what they call as ``far-commutativity'' and ``far-absorption'').

The localization strategy presented in \cite{BR18}, however, does not suit
query-based models where
dependences between two operations cannot be decided locally, but are reliant on other operations.
For example, following their approach, there will always be a dependence between 
$\sql{update}$queries
and $\sql{SELECT}$queries that access the same field \emph{on any row},
because there is always the possibility that 
%there is a copy
%from 
the values from  the row chosen by the $\sql{update}\!$ is copied into the row chosen by the 
$\sql{SELECT}\!\!$.
It is unclear how to accurately express far-commutativity and
far-absorption between SQL statements performing predicate read and
writes. Handling such fine-grained dependencies forms an important
contribution of our work.

%In order to explain the impact of locality on commutativity and
%absorption, \citet{BR18} use the following example involving key-value store
%operations {\tt put(k,v)}, {\tt get(k)}, and {\tt copy(k,k')} where
%{\tt copy(k,k')} copies the value from key {\tt k} to key {\tt
%k'}. Even though {\tt put(k,v)} and {\tt get(k')} (where {\tt k}
%$\neq$ {\tt k'}) clearly commute, they do not ``far-commute''
%according to their definition, because of the presence of the
%operation {\tt copy(k,k')}, which could occur between them and result
%in an indirect dependency. However, even in executions where this
%operation does not occur, they would still assume a dependency between
%{\tt put(k,v)} and {\tt get(k')}, because of their requirement of
%making dependences local. 

There have been several recent proposals to reason about programs
executing under weak consistency~\cite{bailisvldb, alvarocalm,
gotsmanpopl16,redblueatc, redblueosdi, ecinec}. All of them assume a
system model that offers a choice between a \emph{coordination-free}
weak consistency level (\emph{e.g.}, eventual
consistency~\cite{redblueosdi, redblueatc, ecinec, alvarocalm,
bailisvldb}) or causal consistency~\cite{lbc16,gotsmanpopl16}).  In
contrast, our focus is agnostic to the particular consistency
guarantees provided by the underlying storage system, and is concerned
with statically identifying violations of a particular class of
invariants, namely those that hold precisely when transactions exhibit
serializable behavior.

\cite{WA17} presents a dynamic analysis for weak isolation that
attempts to discover weak isolation anomalies from SQL log files.
Their solution, while capable of identifying database attacks due to
the use of incorrect isolation levels, does not consider how to help
ascertain application correctness, in particular determining if
applications executing on storage systems that expose guarantees
weaker than serializability are actually correct.

In addition to the mentioned threads of work on database applications, 
in the recent years several proposals have been made addressing similar
challenges in the context of shared memory concurrent programming models.
For example, static~\cite{CS03} and dynamic~\cite{CS04} approaches have been introduced
to determine if a particular locking policy in such programs actually enforces
a desired non-interference property.
Using code annotation and custom memory
allocations, \citet{RR09} extended these works and proposed a framework which allows detection of 
violations even in the  presence of ill-behaved threads that disobey the
locking~discipline.

In order to liberate users from writing low-level synchronization mechanisms,
a whole-program analysis  is offered by \citet{MZ06}
which compiles user-declared pessimistic atomic sets and blocks, similar to
transactional memory models, into performant locking policies.

While all above approaches rely on the programmer to annotate atomic regions in
their code, as a solution to an orthogonal problem, \cite{LP07} presents MUVI, a
static analysis tool for inferring  
correlated variables which must be accessed together, i.e. atomically. 
Several dynamic frameworks also attempted to detect potentially erroneous program executions
without depending on user annotations and by looking for dangerous access
patters at runtime,
either for a single shared variable~\cite{XB05} or multiple variables~\cite{HD08}.

Similar to our approach, \cite{HZ13} and \cite{MB15} rely on 
symbolic constraint solving to construct full, failing, multithreaded schedules
that manifest concurrency bugs. However, unlike these approaches which depend on dynamic path
profiling to detect conflicting operations, we analyze programs in an intermediate language
which can be generated from any source language using a proper compiler.
Our approach has the additional  completeness guarantee that no anomaly within
the given bounds is missed.

\begin{wrapfigure}[8]{r}{0.29\textwidth}
  \centering  
  \vspace{-2mm}
  \includegraphics[width=0.29\textwidth]{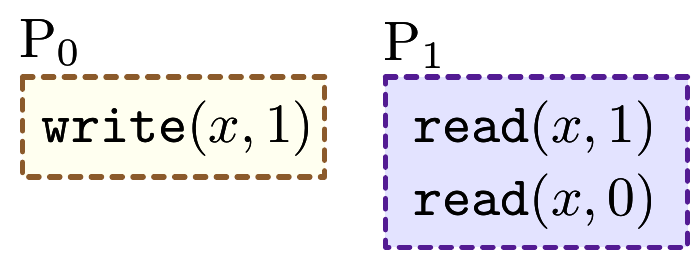}
  \caption{An impossible execution in a shared memory system}
  \label{fig:memory}
\end{wrapfigure}

It is worth noting that all shared-memory systems assume a coherence order on 
writes (a property that is not enforced in weakly consistent systems) 
rendering any existing analysis framework for such systems
inapplicable to replicated database models.
Such shared-memory systems do not permit certain executions 
which can occur in eventually consistent data-stores
while \tool's semantics is rich enough to model these sorts of executions.
As an example, consider the simple execution in \autoref{fig:memory} with two concurrent programs. 
Assuming $x$ is initially 0, models of shared-memory systems disallow this execution, as they force the
second read event in $P_1$ to witness $P_0$'s write.
No such restriction applies in a weakly consistent environment, which may allow
the second read to be routed to a replica that has not yet received the update
from $P_0$'s write event.

Several violation detection mechanisms of developers' atomicity assumption for
code regions in shared memory programs
have been recently introduced where certain conflicting access patterns have
been deemed safe due to lack of data flow between the read and write operations involved in such
conflicts~\cite{LT06,Lucia:ColorSafe:2010}.
Unlike their purely structural approach which only considers patterns within
individual operations, our notion of internal serializability
(to the best of our knowledge) is the first formulation of a 
semantic property of whole programs which takes into account actual
data-dependencies between arbitrary read and write operations that form a conflicting
cycle.
For example, following the approach presented in these works,
the external serializability anomaly presented 
\autoref{fig:abstract_anomaly_3} and
the internal serializability anomaly presented in
\autoref{fig:abstract_anomaly_4} are both considered harmful, as they
include conflicting read and writes from distinct operations.

There have been a number of testing frameworks developed for
distributed systems over the years.  Some notable examples include
MaceMC~\cite{KA+07} a model-checker that discovers liveness bugs in
distributed programs, and~\citet{jepsen}, a random testing tool that
checks partition tolerance of NoSQL distributed database systems with
varying consistency levels to enable high-availability. \tool\
differs from these in a number of important ways.  In particular,
MaceMC does not consider safety issues related to replication, while
Jepsen is purely a dynamic analysis that does not leverage semantic
properties of the application in searching for faulty executions.
Both of these considerations have played an important role in
shaping \tool's design and implementation.

%% file: sections/conclusion.tex
This paper presents \tool, an automated test generation and replaying tool
for applications of weakly consistent distributed databases.
\tool\ is backed by a precise encoding of database applications 
allowing it to maintain fine-grained dependency relations between
complex SQL query and update operations.
Notably, the encoding enforces special relationships on operations
that capture concrete execution details required for an automated and complete replay
of anomalous scenarios without any input from the users.
We applied \tool\ on a suite of real-world database benchmarks 
where it successfully constructed and replayed a wide range of
bugs, providing strong evidence for its applicability. 
Additionally, we compared our approach to a state-of-the-art random testing tool 
where \tool\ performed more efficiently and more reliably.

Concurrency anomalies detected by \tool\ can be straightforwardly fixed by using
stronger database-offered consistency/isolation guarantees (e.g. by using
serializable transactions). Prescribing more optimal fixes, based on the
structure of dependency cycles that \tool\ generates is an intriguing research
problem which we leave for the future works.